\documentclass[twocolumn,showpacs,preprintnumbers,amsmath,amssymb]{revtex4}

\usepackage{graphicx}
\usepackage{dcolumn}
\usepackage{bm}
\usepackage{subfigure}
\begin{document}

\preprint{APS/123-QED}

\title{Alternative statistical-mechanical descriptions of decaying two-dimensional turbulence in terms of ``patches'' and ``points''}

\author{Z. Yin}
 \email{z.yin@tue.nl}
 \author{D.C.Montgomery}
 \altaffiliation[Also at ]{Dept. of Physics \& Astronomy, Dartmouth College, Hanover, NH 03755, U.S.A.}
 \author{H.J.H.Clercx}
 
\affiliation{Fluid Dynamics Laboratory, Applied Physics Department, Eindhoven University of Technology, \\ P.O.Box 513, 5600 MB, Eindhoven,The Netherlands}

\begin{abstract}
Numerical and analytical studies of decaying, two-dimensional (2D) Navier-Stokes (NS) turbulence at high Reynolds numbers are reported. The effort is to determine computable distinctions between two different formulations of maximum entropy predictions for the decayed, late-time state. Though these predictions may be thought to apply only to the ideal Euler equations, there have been surprising and imperfectly-understood correspondences between the long-time computations of decaying states of NS flows and the results of the maximum-entropy analyses. Both formulations define an entropy through a somewhat ad hoc discretization of vorticity to the ``particles'' of which statistical mechanical methods are employed to define an entropy, before passing to a mean-field limit. In one case, the particles are delta-function parallel ``line'' vortices (``points'' in two dimensions), and in the other, they are finite-area, mutually-exclusive convected ``patches'' of vorticity which in the limit of zero area become ``points.'' The former are taken to obey Boltzmann statistics and the latter, Lynden-Bell statistics. Clearly, there is no unique way to reach a continuous differentiable vorticity distribution as a mean-field limit by either method. The simplest method of taking equal-strength points and equal-strength, equal-area patches is chosen here, no reason being apparent for attempting anything more complicated.  In both cases, a non-linear partial differential equation results for the stream function of the ``most probable'' or maximum-entropy state, compatible with conserved total energy and positive and negative vorticity fluxes. These amount to generalizations of the ``sinh-Poisson'' equation which has become familiar from the ``point'' formulation. They have many solutions, which must be obtained numerically, and only one of which maximizes the entropy on the basis of which it was derived. These predictions can differ for the point and patch discretizations. The intent here is to use time-dependent, spectral-method direct numerical simulation of the Navier-Stokes equations to see if initial conditions which should relax to different late-time states under the two formulations actually do so. 
\end{abstract}

\maketitle

\section{Introduction}

It has been known for several years that two-dimensional Navier-Stokes (2D NS) turbulence with periodic boundary conditions and at high Reynolds numbers (in excess of a very few thousand) will relax to long-lived, quasi-steady states whose topology is preserved and whose energy decays more or less inversely proportionally to the Reynolds number, computed with respect to a box dimension and an rms initial turbulent velocity. When this energy decay time is large compared to the initial eddy turnover time, the quasi-steady state can be reached at a time when the total energy decay is fractionally small, though the enstrophy decay can be fractionally large. It came as a surprise when, a decade or more ago, a series of such computations ~\cite{kn:q1, kn:q2, kn:q3} revealed that the late-time quasi-steady state for an initially turbulent run at a Reynolds number above 14,000 showed a pointwise hyperbolic sinusoidal dependence between stream function and vorticity.

	The prediction of such a dependence, in the context of a mean-field treatment of ideal line vortices (or guiding-center plasma rods) had been given thirty years ago ~\cite{kn:q4, kn:q5} and has since been extended and refined in a series of investigations by several groups ~\cite{kn:sinh2, kn:q6, kn:q7, kn:q8, kn:q9, kn:q10, kn:q11, kn:q12, kn:q13, kn:q14, kn:q15, kn:q16, kn:q17, kn:q18, kn:sinh1}: in every case referring to ideal, non-viscous systems. The system is Hamiltonian with a finite phase space, and it is natural to apply Boltzmann statistics to its dynamics, as originally suggested by Onsager ~\cite{kn:q19} (see also Lin ~\cite{kn:q20}). The surprise came in the extent to which the ideal Euler mean-field predictions fit the Navier-Stokes results. At least one attempt was made to define an entropy for the case of finite viscosity ~\cite{kn:q21} but generated new puzzles of its own.

	In the late 1980s and early 1990s, an alternative formulation was given by Robert, Sommeria and Chavanis ~\cite{kn:q22, kn:q23, kn:q24, kn:q25}, by Miller and colleagues ~\cite{kn:q26} and later explored by Brands, Maassen and Clercx \cite{kn:pq26}. The principal difference was that the vorticity field was discretized not in terms of delta-functions, but rather in terms of finite-area, mutually-exclusive ``patches'' of vorticity, to which Lynden-Bell statistics ~\cite{kn:q27} could be applied. The choice of parameters in the patch formulation is wider than that of equal-strength point vortices, in that one must decide in advance the size of the patches, and the number of ``levels,'' or strengths, that the patches carry. There seems to be no deductive mechanism for making such choices, and so we stick with the simplest, taking equal-area patches and  never more than three levels, including zero.

	For readers not inclined to pursue the original references for these formulations, which in some cases require considerable mathematical sophistication, we offer in Appendix A a summary treatment of the derivation of the ``patch'' dependence of vorticity on stream function in the ``most probable,'' or maximum-entropy, state. It is attempted to do this using the least forbidding and most transparent mathematics available to the purpose. We show how in the limit of zero-area patches, the point dependence is recovered. Thus the reader is being referred to Appendix A for a derivation of the nonlinear partial differential equations to be introduced in Section \ref{sec:s2} as candidates for the predictor of the most-probable state of late-time decaying 2D NS turbulence.  
	
	Section \ref{sec:s2} will summarize both the numerical method (now well-known in its essentials), due originally to Orszag and Patterson ~\cite{kn:q28, kn:q29}, to which various sets of initial conditions will be subjected. We will also describe the motivation for choosing these initial conditions, in terms of what we expect from them as possible most-probable states. We will concentrate on initial conditions for which it appears that the predicted outcomes using point and patch entropies will be as different as possible. We will refer the reader to Appendix B for a description of how the sinh-Poisson relation and its patch generalizations are dealt with numerically (a non-trivial task).
	
	Section \ref{sec:s3} contains the main body of the results that we have found. Among other unexpected features, we have found a series of one-dimensional (1D) solutions to the most-probable state equations, whereas the literature up to now has dealt only with 2D solutions (``dipoles,'' ``quadrupoles,'' ``octupoles,'' etc.). There are cases in which the entropies of the 1D solutions (which we call ``bars'') are extremely close to those of the dipole over a considerable range of energy. Which  is greater depends on seemingly arbitrary choices, in the patch formulation, such as the areas of the patches used. Classes of initial conditions are found (not essentially turbulent ones) for which relaxation to the ``bar'' states is observed. It will be easier, in the context of the details of the solutions, to illustrate the variety of behavior we have been able to catalogue.

	Section \ref{sec:s4} will be a summary of what we think we have learned, and of what remains to be learned. One of the more radical, if incidental, conclusions which we believe may be extracted from these runs (to be discussed in more detail later) is that in the context of the initial value problem, all 2D NS turbulence may result only from its having been there initially. Otherwise, it can come into existence only through random external ``stirring.'' 
	
        We should caution the reader that the word ``turbulence'' in this manuscript will be used in a somewhat looser sense than is customary. It will not always mean a state of the fluid in which the kinetic energy is widely shared by many Fourier modes. We have concentrated on initial conditions which seemed to us most relevant to producing effects peculiar to the ``patch'' description: namely, initial vorticity distributions in which a few flat levels of nearly constant vorticity could be readily identified. We expected, because of instabilities in the shear flows these represent, that turbulence of a typical broad-band modal structure would soon be generated, as it is in three dimensions.  We found that instabilities could indeed be activated, but often, broad-band turbulence of the conventional kind was not the result. Instead, energy spectra dominated by only a few low-lying (in Fourier space) modes were nonlinearly dynamically converted to Fourier spectra dominated by only a few (but different) low-lying modes. We have in fact found it nearly impossible to generate genuinely broad-band 2D NS turbulence through instabilities, rather than through initial conditions or externally-imposed stirring. Perhaps surprisingly, the dynamical evolution involved sometimes, but not nearly always, led to late-time quasi-steady states that seemed to correspond to most-probable ``patch'' predictions: most interestingly, the one-dimensional ``bar'' state, as will be seen. Broad-band, initially-excited turbulence continued to lead to the classical dipolar late-time state, as it has in previous turbulence simulations.

\section{The Numerical Agenda}
\label{sec:s2}
	We address ourselves to the long-time dynamics in two dimensions (2D) of the Navier-Stokes equation in the usual vorticity representation,
\begin{equation}
  \frac{\partial \omega}{\partial t} + \mathbf{v} \cdot \nabla \omega = \nu \nabla^2 \omega,
 \label{eqn:navier}
\end{equation}
where the fluid velocity $\mathbf{v}$ is to be written in terms of the stream function $\psi$ as
$ \mathbf{v} = \nabla \times (\hat{\mathbf{e}} \psi)$, and $\hat{\mathbf{e}}$ is a unit vector in the $z$ direction. $\mathbf{v}$ has only $x$ and $y$ components, which depend only upon $x$, $y$, and the time $t$. In the natural dimensionless units of the problem, the kinematic viscosity $\nu$ may be interpreted as the reciprocal of the Reynolds number, which we will specify in more detail presently. The curl of ${\mathbf v}$ is the vorticity 
$\mbox{ \boldmath{$\omega$}} = (0,0,\omega)$, which has only one component, also in the $z$-direction. The stream  function $\psi$ and the non-vanishing component of vorticity $\omega$ are related by Poisson's equation,

\begin{equation}
 \nabla^2 \psi = - \omega.
 \label{eqn:stream1}
\end{equation}

We note that dropping the final viscous term in Eq. (\ref{eqn:navier}) leaves us with the 2D equations of an ideal Euler fluid. We note moreover that a time-independent solution of these Euler equations results any time $\omega$ is a differentiable function of $\psi$.

	We will work in a periodic box with sides $2 \pi$ in length, and will use the unit  length 1 to define the Reynolds number; thus our basic unit of length is roughly 1/6 of a box dimension.  The unit of velocity will typically be a root-mean-square value of the initial velocity, and we will attempt to make this equal to 1 whenever possible.  Thus $\nu$ in Eq. (\ref{eqn:navier}) may reasonably be identified with the reciprocal of an initial Reynolds number, Re. We are interested in values of Re of at least several thousand, so that the final term in Eq. (\ref{eqn:navier}) is formally very small, except in regions of steep vorticity gradients. The ``eddy turnover time,'' which we shall use as a unit of $t$,  is thus about 1/6 of a box dimension divided by an initial rms velocity. 
	It has been known for many years that under such circumstances, the kinetic energy $E$, defined by

\begin{equation}
 E = \frac{1}{2} \frac{1}{(2 \pi)^2} \int \int \mathbf{v}^2 dxdy
 \label{eqn:energy1}
\end{equation}
decays slowly proportionally to $\nu$. However, the decays of higher-order Euler-equation invariants constructed as integral moments of $\omega$, such as the enstrophy,

\begin{equation}
  \Omega = \frac{1}{2} \frac{1}{(2 \pi)^2} \int \int \omega^2 dxdy,
  \label{eqn:entrophy1}
\end{equation}			
appear to continue to decay at an $\mathcal{O}(1)$ rate; even a weak dependence of their decay rate upon $\nu$ has not been convincingly demonstrated. Note that both ideal invariants are referred to unit volume, in the dimensionless units.

	When high Reynolds numbers exist, then, the turbulent decay of energy is seen to be very slow, relative to any other identifiable ideal Euler invariant. When the energy decay time is large enough to be well-separated from the eddy turnover time, it is possible to observe in numerical computations ~\cite{kn:q1, kn:q2, kn:q3} that a quasi-steady state is reached in a time over which the fractional decay of energy is small (a few hundred eddy turnover times). One might expect to find a ``selectively decayed'' state, in which the enstrophy to energy ratio is minimal~\cite{kn:q30}, and indeed, if one waits long enough, it can be analytically proved that such a state must be approached ~\cite{kn:sinh3}. However, it was found ~\cite{kn:q3} that long before that time, the quasi-steady, slowly-decaying state that is reached has a rather sharp one-to-one pointwise correspondence between $\omega$ and $\psi$. The elucidation and testing of this correspondence is the principal purpose of this paper.

	Clearly, even though a slow viscous decay may be superimposed on such a state with $\omega$ approximately a function of $\psi$, the state is also closely approximated by a time-independent solution of the Euler equations. There is no a priori reason why this should be so. It is a fact we attempt to incorporate here into a coordinated set of direct numerical simulations of Eq. (\ref{eqn:navier}) and a combined analytical and numerical argument, based upon statistical mechanics, in pursuit of what the connection between $\omega$ and $\psi$ should be. 

	The reader is referred to Appendix A for a summary of the statistical mechanical arguments. They depend upon a discretization of vorticity in terms of delta-function line vortices (``points'') or in terms of mutually-exclusive, finite-area ``patches.'' Boltzmann statistics are applied to the former, to define an entropy, or logarithm of the probability of a state, and Lynden-Bell statistics are applied to the latter. The limit of zero-area, finite-vorticity  patches are points. Thus the patch formulation can be viewed as containing the point version of the theory, and either is or is not a useful generalization of it. In both cases, what results is a ``most probable'' dependence of vorticity upon stream function. A mean-field limit (infinitely-many points or patches, of arbitrarily small strength) is then taken, to yield continuous and differentiable functions $\omega$($\psi$). In Appendix A, it is shown that for points, the resulting function is:	
\begin{equation}
\nabla^2 \psi = - \omega = - e^{\alpha_+ - \beta \psi} + e^{\alpha_- + \beta \psi}.
\label{eqn:point1}
\end{equation}
For patches, the resulting function is:
\begin{equation}
\nabla^2 \psi = - \omega = - \sum_{j=1}^q \frac{M}{\Delta} K_j 
\frac{e^{\alpha_j - \beta \psi K_j}}{\sum\limits_{l=0}^q e^{\alpha_l - \beta \psi K_l}}.
\label{eqn:general1}
\end{equation}

The symbols $\alpha$, $\beta$ are Lagrange multipliers, which enter via a maximization of the appropriate entropy subject to the constraints of given energy and positive and negative fluxes of vorticity. They are determined in principle by demanding that the energies and vorticity fluxes calculated on the basis of Eqs. (\ref{eqn:stream1}) and (\ref{eqn:point1}) or (\ref{eqn:general1}) match specified values. In Eq. (\ref{eqn:general1}), the $K_j$ are the ``levels'' of vorticity corresponding to the different sized patches, and must be chosen somewhat arbitrarily.
$\Delta/M$ is a fixed size of a ``patch,'' defined in detail in Appendix A, and may be chosen arbitrarily, within wide limits.

	Both Eqs. (\ref{eqn:point1}) and (\ref{eqn:general1}) have infinitely many solutions. The physical ones are interpreted to be those which maximize the appropriate entropy from which they were derived. The others represent local maxima, which, as we shall see, may in some cases represent attainable states in the computations.

	The questions before us are: 
\begin{itemize}
\item
   How does one determine the entropy-maximizing time-independent solutions to Eqs. (\ref{eqn:point1}) and (\ref{eqn:general1})?
\item
   How close may we come to those solutions in a dynamical computation that solves Eqs. (\ref{eqn:navier}) and (\ref{eqn:stream1}) as a time-dependent initial-value problem with large but finite Reynolds number?
\item
  Are there noticeable differences between the predictions of Eqs. (\ref{eqn:point1}) and (\ref{eqn:general1}) and are there reasons for preferring one to the other as predictors of late-time turbulent decays?
\end{itemize}

	It is to be stressed that all three questions are answerable not in the abstract, but only as a result of somewhat demanding computations. There is no a priori reason why Navier-Stokes turbulent decays should be predicted at all by anything having to do with the Euler equations. The latter will never be soluble, for continuous initial conditions, over time intervals long enough to make ideal solutions of much interest, because it is in the nature of Euler codes, having no minimum physically-determined length scale, to overrun their own resolution in a very few eddy turnover times. If the predicted states had not shown some empirical relevance to Navier-Stokes solutions, there would be no justification for this activity.

	The dynamical code used here is of the now familiar Orszag-Patterson pseudospectral variety ~\cite{kn:q28, kn:q29}. The code is a parallelized MPI Fortran 90 version of an earlier Fortran 77 code provided by W.H. Matthaeus (private communication). It is fully de-aliased, using the shifted-grid method. It has been run, in the runs reported here,  on the SGI Origin 3800 at SARA Supercomputing Centre in Amsterdam. Some of the simulations (resulting in the ``bar'' final state in section \ref{sec:s3:b:1:bb}) have also been recomputed with a different pseudospectral Fourier code (provided by Dr. A.H. Nielsen, Riso National Laboratories, Denmark.) These computations yielded the same conclusions.  
	
	All runs resolve the Kolmogorov dissipation wave number based on enstrophy dissipation. Previously, initial conditions for turbulent decay runs have tended to use randomly loaded Fourier coefficients in the spectrum or vorticity fields up to some upper wave number. These have typically been chosen to match some cascade wavenumber $k$-spectrum, such as the ``$-3$'' direct enstrophy cascade ~\cite{kn:q31, kn:q14} spectrum predicted by Kraichnan ~\cite{kn:q31}.  The phases have been chosen from a random number generator. This has been done to achieve the most disordered (in some sense) initial field compatible with a particular power spectrum. Several of the runs to be reported later are in fact of this variety. Such an initialization can only yield a vorticity distribution that is analytic in $x$ and $y$, since only sinusoidal functions are involved, even though the spatial dependence may be wildly fluctuating. Such a vorticity distribution can take on any particular value of $\omega$ only over a set of measure zero. For this reason, one might imagine it would tend to de-emphasize features that corresponded to the ``levels'' in the patch formulation, over which the vorticity is supposed to take on a constant value inside a compact area. Therefore, we have also stressed initial conditions that have large, flat areas of vorticity, separated by thin regions, with as steep spatial gradients between them as the code will resolve. Our original intention was that these would correspond to unstable laminar shear flows whose instabilities would subsequently generate turbulence. Somewhat to our surprise, we have found it easy to produce shear-flow instabilities but not ones that led to turbulence, in the sense of broad $k$-spectra. We have come to suspect that this is a feature inherent to two-dimensional flows, and perusal of the literature has uncovered only three-dimensional turbulence as a consequence of unstable shear flows, but not two-dimensional; but this must be left as a conjecture rather than a demonstrated fact.

	In any case, in order to induce a large number of Fourier modes to participate in the subsequent dynamics, we have found it necessary, in the runs evolving from vorticity distributions with flat areas, to add significant amounts of random noise initially, in order to get a subsequent $k$-spectrum that could readily be called ``turbulent.'' Even then, some of the evolution we find might have its status as turbulence disputed.

	So referring to Appendix A for the derivations of Eqs. (\ref{eqn:point1}) and (\ref{eqn:general1}), and to Appendix B for the method of their numerical solution, we proceed in Sec. 3 to a description of initial conditions used in the dynamical runs, the states to which they evolved, and their comparisons with the maximum-entropy or ``most probable'' states that Eqs. (\ref{eqn:point1}) and (\ref{eqn:general1}) imply.

\section{NUMERICAL RESULTS}
\label{sec:s3}
\subsection{Extracting Maximum-Entropy (``Most Probable'') States}

\begin{figure*}[!htbp]
\begin{minipage}[c]{.31 \linewidth}
\subfigure[]{
 \label{fig:good:a}
 \includegraphics[width= \linewidth]{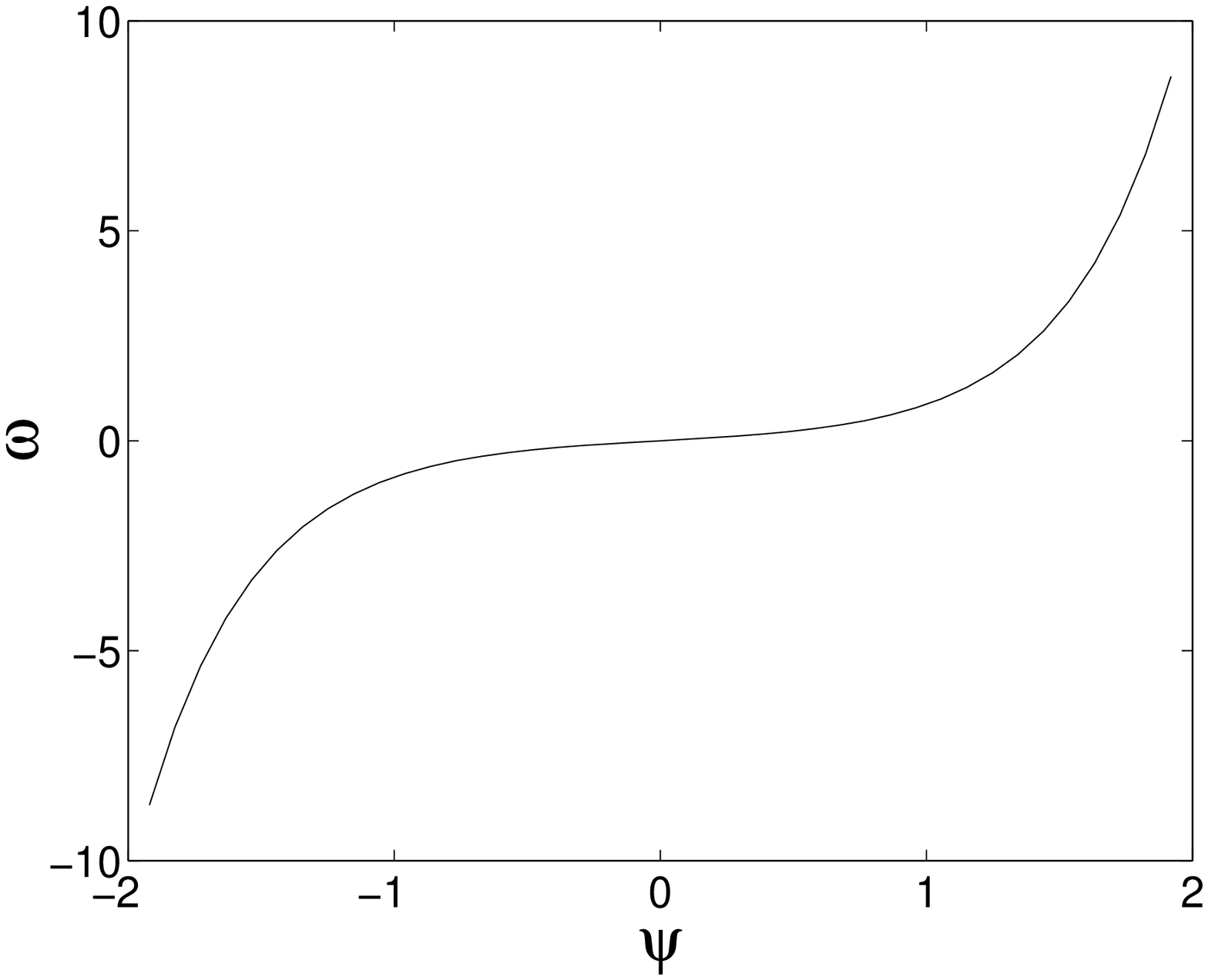}}
\end{minipage}
\begin{minipage}[c]{.31 \linewidth}
 \subfigure[]{
 \label{fig:good:b}
 \includegraphics[width= \linewidth]{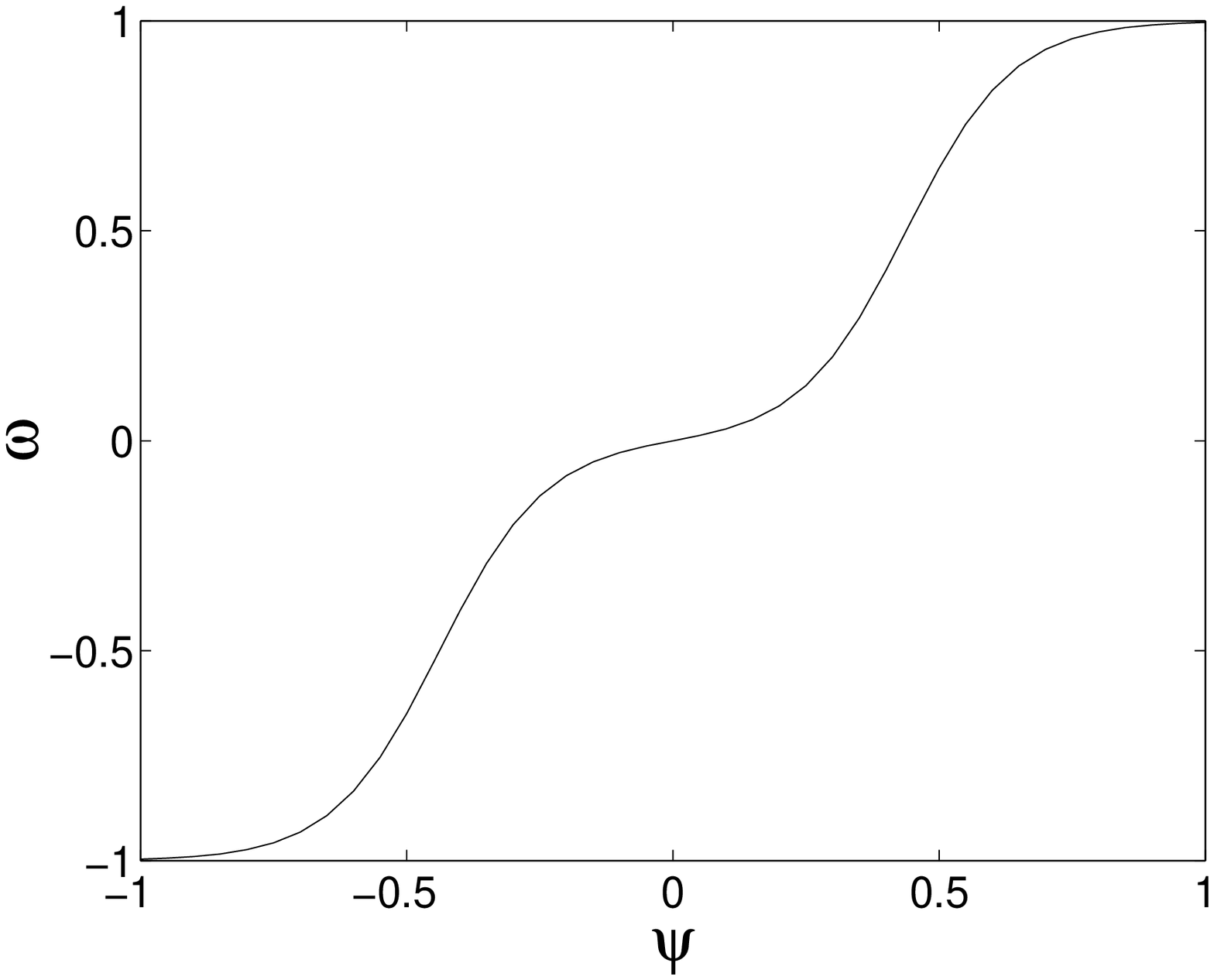}}
\end{minipage}
\begin{minipage}[c]{.31 \linewidth}
 \subfigure[]{
 \label{fig:good:c}
 \includegraphics[width= \linewidth]{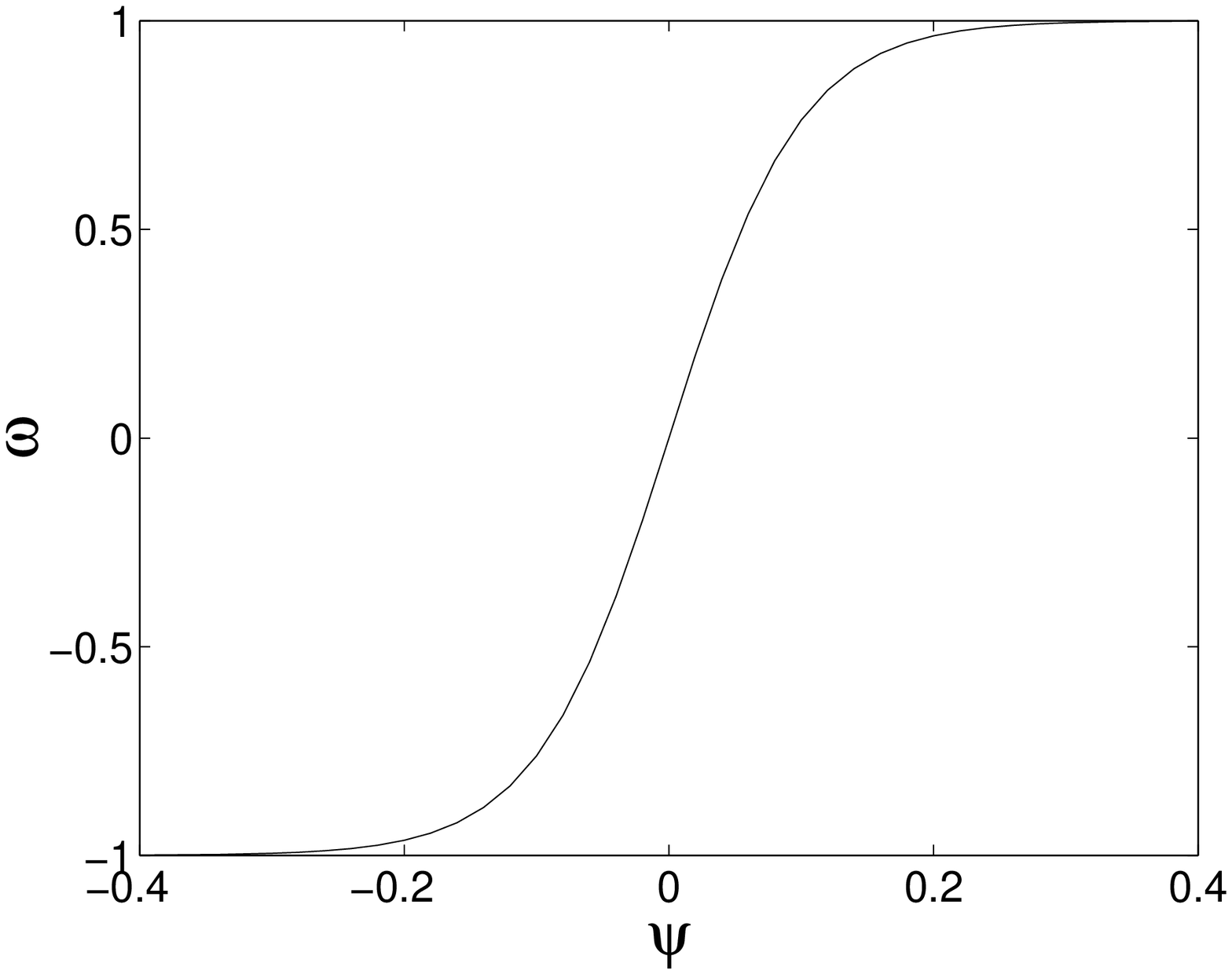}}
\end{minipage}
\caption{Typical $\omega - \psi$ relation from (a) sinh-Poisson, (b) 3-level Poisson equation and (c) tanh-Poisson equation (special case when $e^{-\alpha} \rightarrow 0$ in Eq. (\ref{eqn:3_level})).}
\label{fig:good}
\end{figure*}

\begin{figure*}[!htbp]
\begin{minipage}[c]{.31 \linewidth}
\subfigure[]{
 \label{fig:Dipole:a}
 \includegraphics[width= \linewidth]{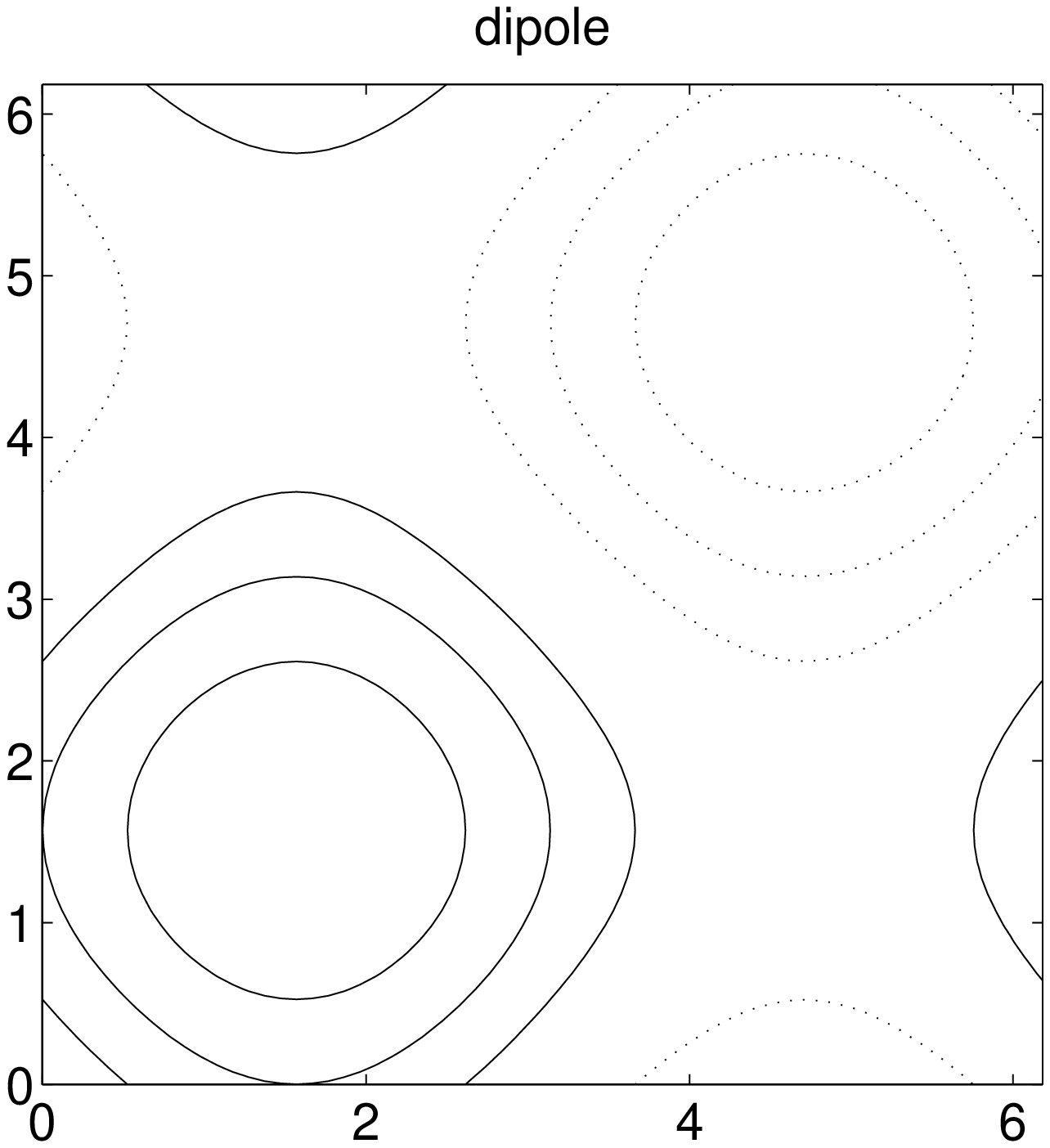}}
\end{minipage}
\begin{minipage}[c]{.31 \linewidth}
 \subfigure[]{
 \label{fig:Dipole:b}
 \includegraphics[width= \linewidth]{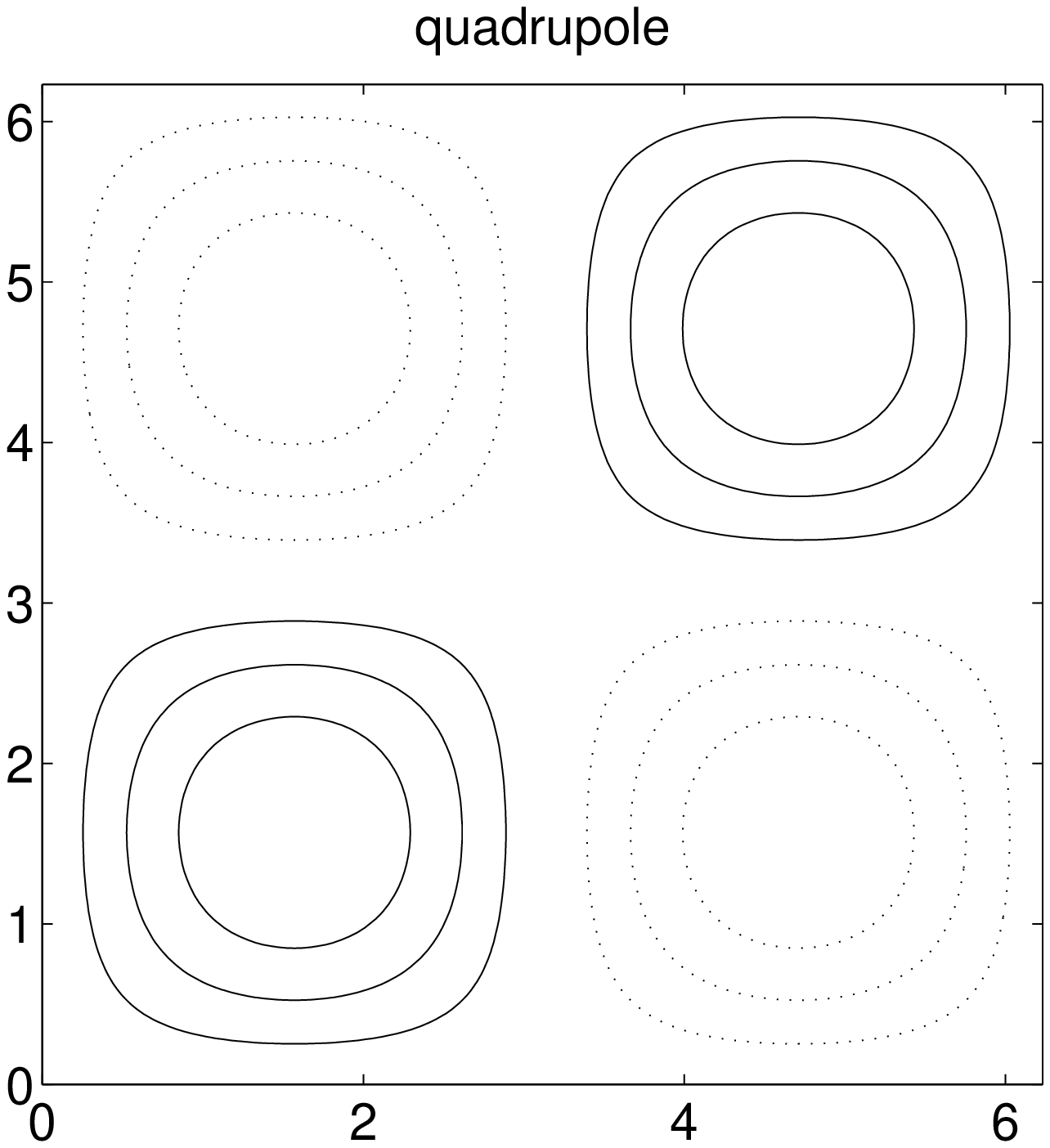}}
\end{minipage}
\begin{minipage}[c]{.31 \linewidth}
 \subfigure[]{
 \label{fig:Dipole:c}
 \includegraphics[width= \linewidth]{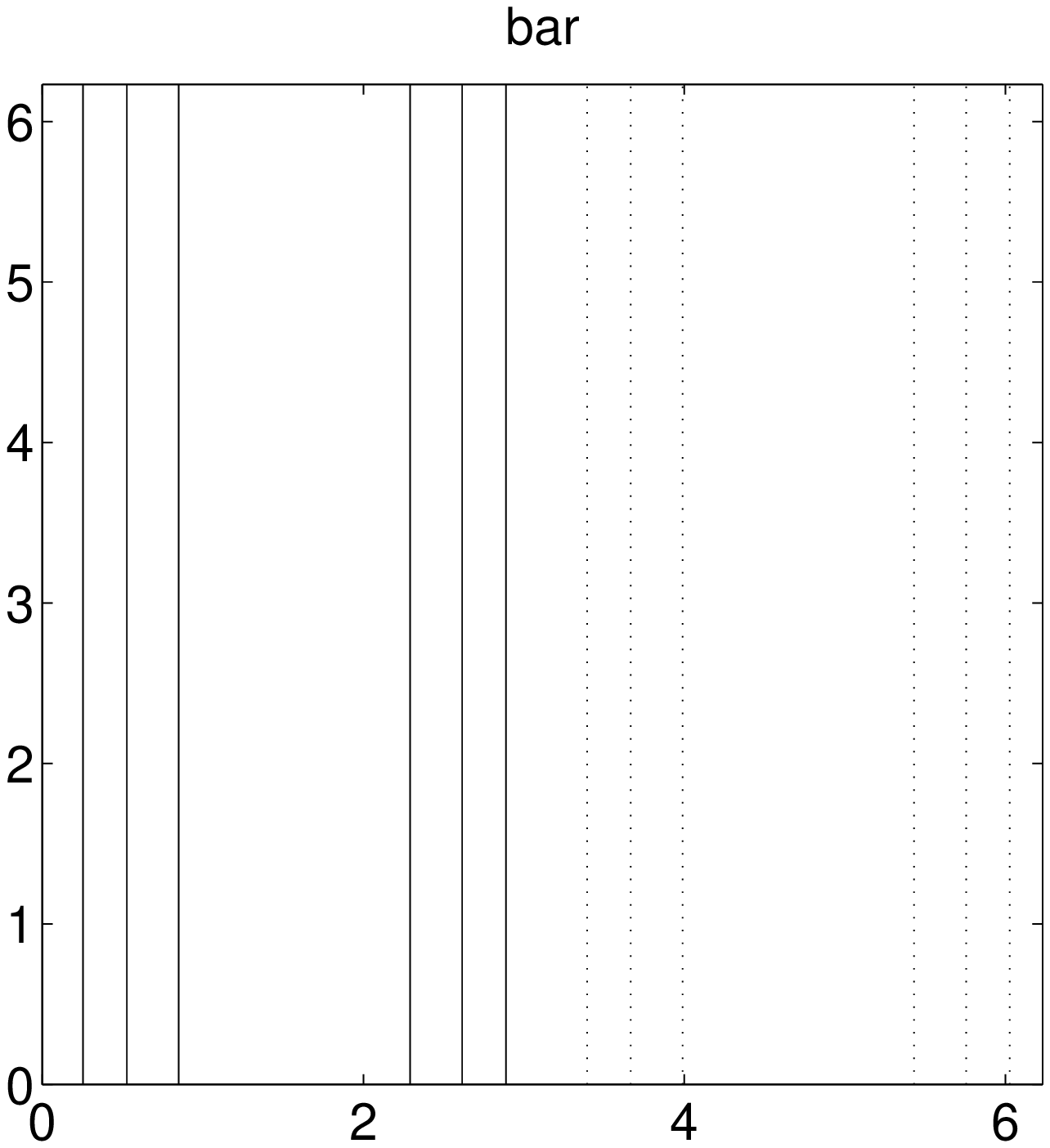}}
\end{minipage}
\caption{Contours of constant stream function $\psi$ for the solutions of Eq. (\ref{eqn:point1}). Negative values are shown as broken lines, throughout.}
\label{fig:Dipole}
\end{figure*}

	We first summarize the results of the solutions of the most-probable-state equations, derived in Appendix A; the method of their solution is described in Appendix B. The goal is not only to find numerical solutions but to determine which among the solutions has the highest entropy for given energy and vorticity fluxes. Generally speaking, the entropies of states with more maxima and minima have been seen to be less than those with fewer, and we can be guided by this. But there are instances where solutions of two different topologies can be found whose entropies lie very close to each other, as found by Pointin and Lundgren ~\cite{kn:q7} , and we must concentrate on those. 

	We consider the point relation, Eq. (\ref{eqn:point1}) in the absence of any conditions that would suggest asymmetry between positives and negatives, so that the two Lagrange multipliers $\alpha_+ = \alpha_- = \alpha$. The result is the sinh-Poisson equation (a typical $\omega - \psi$ dependence of which is shown in Fig. \ref{fig:good:a}),
	
\begin{equation}
 \nabla^2 \psi =2 e^{\alpha} \sinh(\beta \psi), \; \beta < 0.
 \label{eqn:points}
\end{equation}
 
We also consider the patch relation, Eq. (\ref{eqn:general1}), specialized to the three-level case of vorticity levels $-1$, $0$, and $+1$. Again assuming symmetry between positives and negatives (Fig. \ref{fig:good:b}), we have
 
\begin{equation}
\nabla^2 \psi = \frac{M}{\Delta} \cdot 1 \cdot \left[ \frac {2 \sinh(\beta \psi)}
  {e^{- \alpha} + 2 \cosh(\beta \psi) } \right]
\label{eqn:3_level}  
\end{equation}
where $\beta < 0$ and $\Delta/ M$ is the (arbitrary) size of a patch.

Eq. (\ref{eqn:points}) can be rewritten as
\begin{equation}
 \nabla^2 \Psi =- \lambda^2 \sinh \Psi
 \label{eqn:s_point}
\end{equation}
where we have defined
\begin{equation}
 \Psi = \left| \beta \right| \psi
 \label{eqn:re1}
\end{equation}
and
\begin{equation}
\lambda^2 = 2 \left| \beta \right| e^{\alpha}.
 \label{eqn:re2}
\end{equation}

The form of Eq. (\ref{eqn:s_point}) makes it most susceptible to numerical solution. Once it is solved, for a given value of $\lambda$, we must keep in mind that in order to evaluate the entropy associated with the solution,  we must revert to the parameters of Eq. (\ref{eqn:points}), while guaranteeing that $\alpha$ and $\beta$ satisfy Eq. (\ref{eqn:re2}). Since for each fixed value of $\lambda^2$, we have infinite sets of possibilities for $\alpha$ and $\beta$, a recipe is needed for choosing them. Since our goal is to plot the entropy vs. energy for a fixed value of the vorticity flux, 
\[
   \Omega_{\psi}= \frac{1}{L^2} \int \int e^{\alpha + \beta \psi} dxdy = \frac{1}{L^2} \int \int e^{\alpha - \beta \psi} dxdy
\]
for the domain $[0, L] \times [0, L]$ (here we let $L=2 \pi$ and $\Omega_{\psi} = 1$, for convenience), we must also be assured that the solution we get for $\psi$ satisfies this condition. 

Fortunately, these two conditions can be satisfied simultaneously.  Combining Eqs. (\ref{eqn:re1}), (\ref{eqn:re2}) and $\Omega_{\psi} = 1$ leads to the conclusion that $\left| \beta \right | = \Omega_{\Psi}$:
\begin{eqnarray} 
  \Omega_{\Psi} &=& \frac{1}{L^2} \int \int \frac{\lambda^2}{2} e^{-\Psi}dxdy \nonumber \\
                &=& \frac{1}{L^2} \int \int \left | \beta \right | e^{\alpha} e^ {- \left | \beta \right | \psi} dxdy =\left | \beta \right |.
\label{eqn:get_p2}
\end{eqnarray}

Because Eq. (\ref{eqn:s_point}) is the equation solved, the value of $\Omega_{\Psi}$ is readily obtained. Thus we obtain $\beta$ and from Eq. (\ref{eqn:re2}), $\alpha$. We have now all the parameters that correspond to fixed vorticity flux of either sign. From Eqs. (\ref{eqn:get_p3}) and (\ref{eqn:get_p4}),

\begin{equation}
   E_{\psi} = \frac{1}{2 L^2} \int \int \psi \omega_{\psi} dxdy = \frac{1}{2 L^2 \beta^2} \int \int \Psi \omega_{\Psi} dxdy,
\label{eqn:get_p3}
\end{equation}
\begin{equation}
   S_{\psi}= -2 \alpha \Omega_{\psi} -2 \left| \beta \right| E_{\psi} = -2 \alpha -2 \left| \beta \right| E_{\psi},
\label{eqn:get_p4}
\end{equation}
we can draw plots of the entropy ($S_{\psi}$) vs. energy ($E_{\psi}$) for fixed unit flux of positive and negative vorticity.

It is useful to introduce terms for the solutions of differing topology. For example, for a given energy and vorticity flux, the contours of stream function may look like Figs. \ref{fig:Dipole}. Fig. \ref{fig:Dipole:a} will be called the ``dipole'' solution. Fig. \ref{fig:Dipole:b} will be called the ``quadrupole'' solution, and so on through higher ``multipoles'' that correspond to successively higher (even) numbers of maxima and minima. We have also found one-dimensional solutions as illustrated in Fig. \ref{fig:Dipole:c}.  These one-dimensional solutions will be called ``bar'' solutions, and there are also an infinite sequence of them, with basic periodicities $2 \pi$, $\pi$,  $\pi/2$, ... These states are obtained, as explained in Appendix B, by iterating a trial solution with similar topology until convergence is obtained.

\begin{figure}[!htbp]
   \begin{center}
   \includegraphics[width= 0.85 \linewidth]{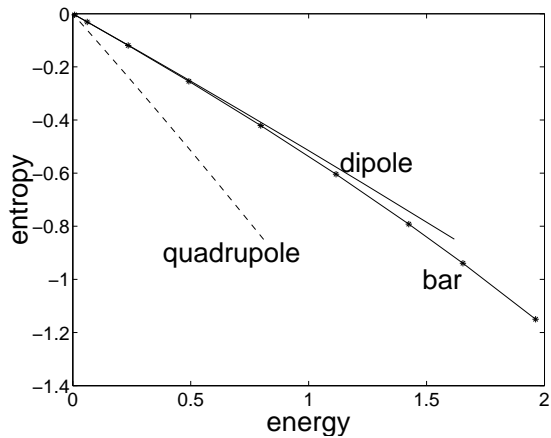}
   \end{center}
   \caption{Entropy vs. energy for the solutions of Eq. (\ref{eqn:point1}) for unit positive and negative vorticity flux, computed for the ``point'' discretization.}
   \label{fig:pentropy}
\end{figure}

Fig. \ref{fig:pentropy} shows a plot of the entropy vs. energy, for fixed unit positive and negative vorticity flux, for the quadrupole, bar, and dipole solutions. Observe the very small difference that makes the entropy of the dipole greater than that of the bar. It will turn out that it is possible to find either solution as a consequence of the development of certain initial conditions, as we shall see later. 

	Turning to the patch prediction, Eq. (\ref{eqn:3_level}) can be simplified as
 
\begin{equation}
 \nabla^2 \Psi = - \frac{\lambda^2 \sinh \Psi}{g + \cosh \Psi},
 \label{eqn:s3_level}
\end{equation}
by letting:
\begin{equation}
 \Psi = \left| \beta \right| \psi,
 \label{eqn:3re1}
\end{equation}
\begin{equation}
 g = \frac{1}{2} e^{- \alpha},
 \label{eqn:3re2}
\end{equation}
and
\begin{equation}
 \lambda^2 = \frac{M \left| \beta \right|}{ \Delta}.
 \label{eqn:3re3}
\end{equation}

Solving Eq. (\ref{eqn:s3_level}) is not as easy as solving Eq. (\ref{eqn:s_point}) because there is an extra parameter, $g$, which implictly reflects the arbitrary choice of the patch size. Another strategy is required:

\begin{enumerate}
\item We choose the size of the patch, so that 

\begin{equation}
\frac{M}{ \Delta}=\frac{\lambda^2}{\left| \beta \right|} = Const.;
 \label{eqn:3re4}
\end{equation}

\item Then a trial value of $g$ is chosen before solving Eq. (\ref{eqn:s3_level}), again iterating about a trial solution of desired topology.  Using Eq. (\ref{eqn:3re4}), we have the value of $\beta$, and $\alpha$ from Eq. (\ref{eqn:3re2}), $E_\psi$ and $\Omega_\psi$ are given by
   
\begin{equation}
  E_\psi = \frac{1}{2 L^2 \beta^2} \int \int \omega_{\Psi} \Psi dxdy,
 \label{eqn:3get1}
\end{equation}    

\begin{equation}
 \Omega_\psi = \frac{M}{L^2 \left| \beta \right| } \int \int \frac{e^{\Psi}}{2g + 2 \cosh \Psi} dxdy;
 \label{eqn:3get2}
\end{equation}    

\item  However, the parameters obtained in this way are not usable since the condition of unit positive and negative vorticity flux is not in general satisfied. We must return to the second step and change $g$ until the desired accuracy is obtained from Eq. (\ref{eqn:3get2}) with $\Omega_\psi =1$.  Once this has been done, we are then in a position to evaluate the entropy from the algebraic expression (see Appendix A):

\begin{eqnarray}
   S_\psi &=& -2 \alpha \Omega_\psi -2 \left| \beta \right | E_\psi +  \nonumber \\ 
          & &     \frac{\lambda^2}{\left| \beta \right| L^2} \int \int \ln (1+ e^{\alpha - \Psi} + e^{\alpha +
	           \Psi}) dxdy.
\label{eqn:3get3}
\end{eqnarray}

\end{enumerate}

\begin{figure}[!htbp]
\centering    
\subfigure[]{
   \label{fig:bbrag:3entropy4}
   \begin{minipage}[c]{.7 \linewidth}   
   \scalebox{1}[1.1]{\includegraphics[width= \linewidth]{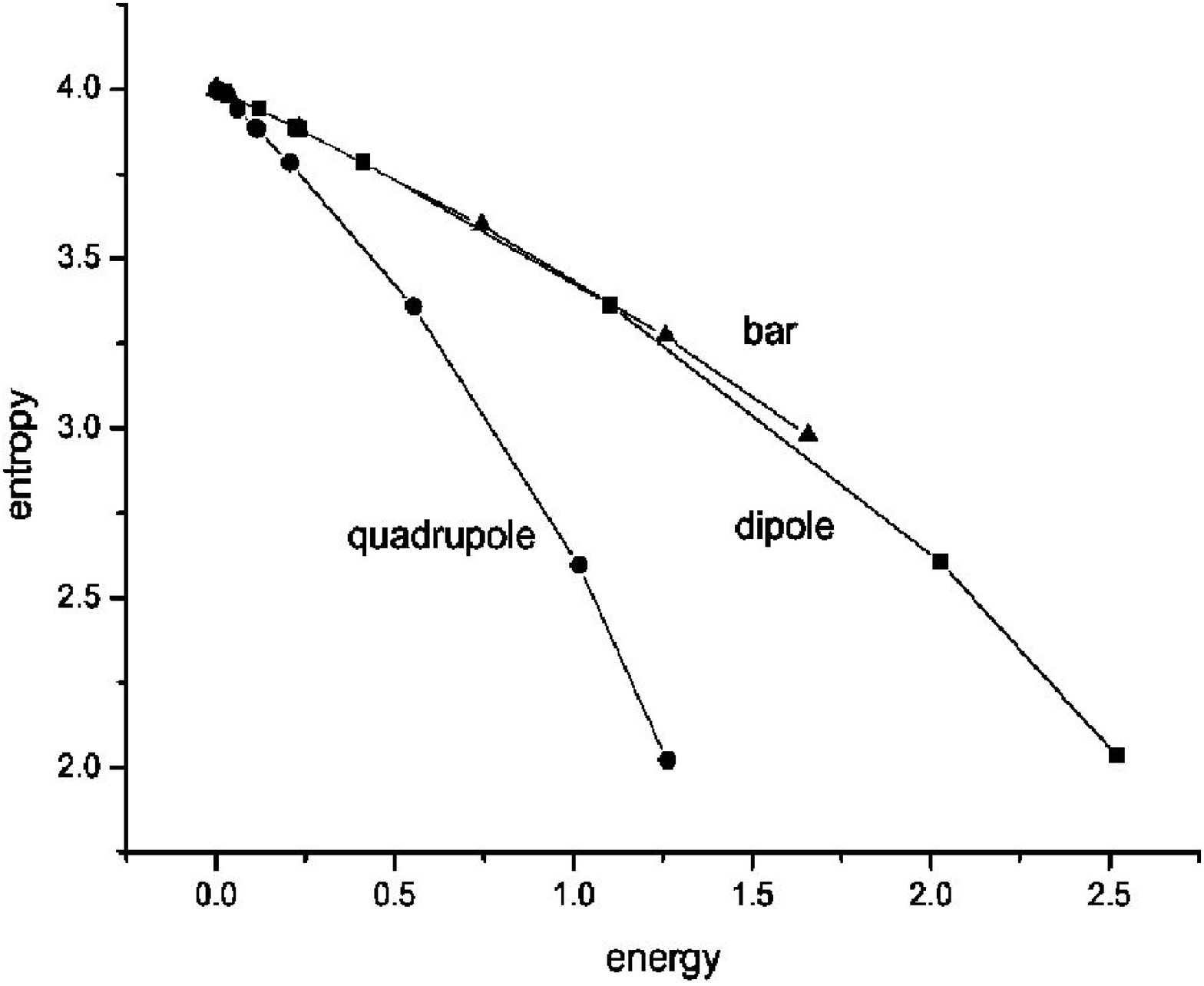}}
   \end{minipage}}
\subfigure[]{
   \label{fig:bbrag:3entropys1}
   \begin{minipage}[c]{.7 \linewidth}
   \scalebox{1}[1.1]{\includegraphics[width=\linewidth]{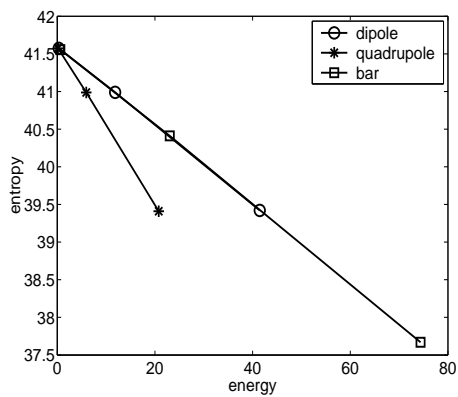}}
   \end{minipage}}
\subfigure[]{ 
   \label{fig:bbrag:3entropys2}
   \begin{minipage}[c]{.7 \linewidth}
   \scalebox{1}[1.1]{\includegraphics[width=\linewidth]{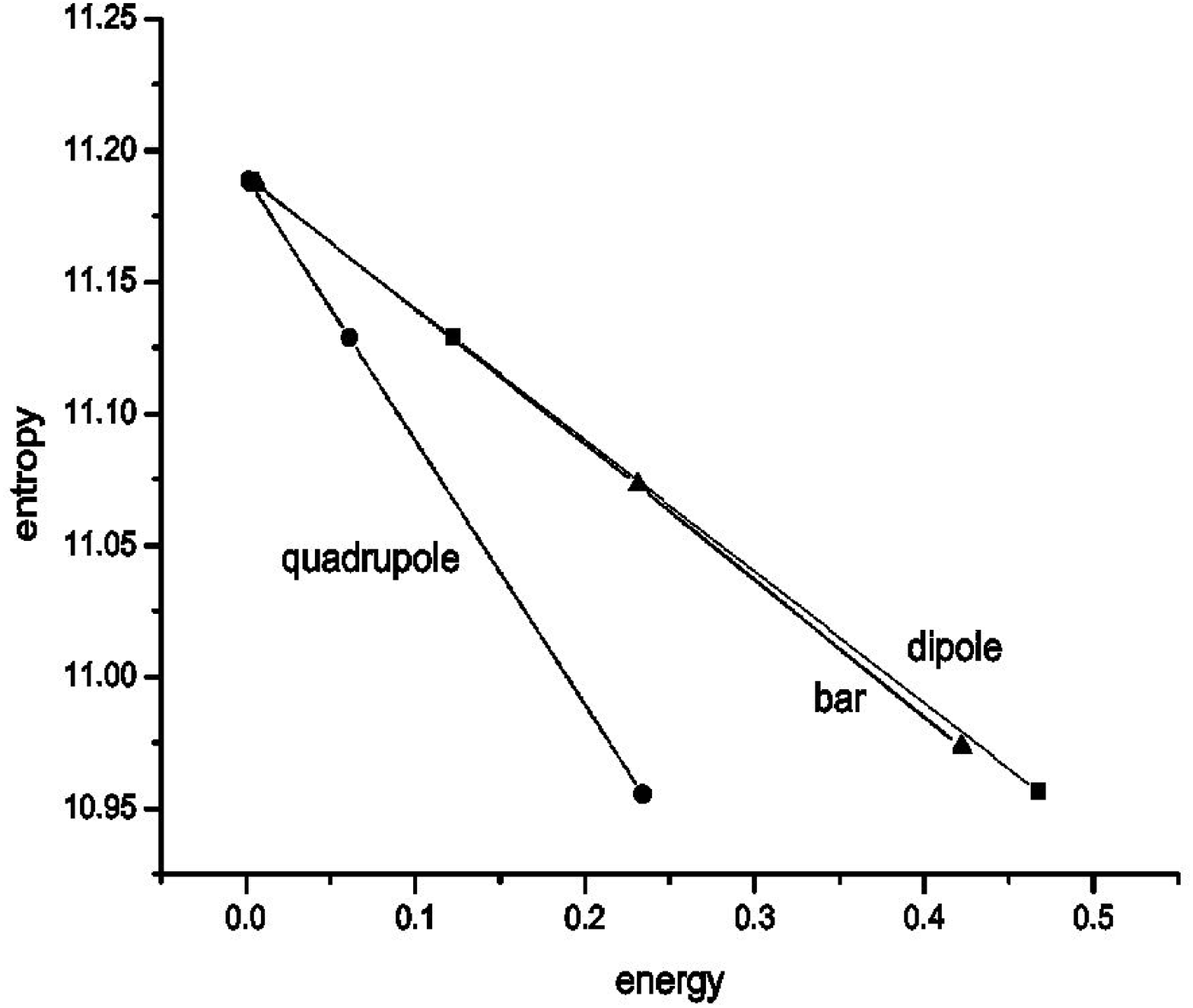}}
   \end{minipage}}

\caption{Entropy vs. energy at unit vorticity fluxes for the solutions of Eq. (\ref{eqn:3_level}): (a) with a relatively large  patch size ($M/ \Delta = 3.7814$); (b) with a somewhat smaller patch size ($M/ \Delta = 25$); (c) with a still smaller patch size ($M/ \Delta = 100$). Note that in (c) the dipole solution has become slightly more probable than the bar.}
\label{fig:bbrag} 
\end{figure}

We may plot the entropy of Eq. (\ref{eqn:3get3}) in Fig. \ref{fig:bbrag:3entropy4}, which is obtained by the choice of $M/ \Delta = 3.7814$, a relatively large ``patch'' size. It will be seen that for this large patch size, the entropy of the bar solution is greater than that of the dipole, a different conclusion than that of the point calculation. However, if we reduce the size of the patch, we find that the result approaches that of the point formulation.  In Fig. \ref{fig:bbrag:3entropys1}, we see the result when $M/ \Delta = 25$, and in Fig. \ref{fig:bbrag:3entropys2} the result when $M/ \Delta = 100$, where the maximum entropy state is again the dipole one. Considering the very small entropy differences between the dipole and the bar solutions shown in Figs. \ref{fig:pentropy} and \ref{fig:bbrag}, it might not be thought surprising if the fluid had difficulty making up its mind which state to relax into.

	Before turning to the results of the dynamical computations, we offer a few observations on the relations of the patch versions of the theory to the point version, and to each other. It is clear from the derivation that in the limit that the patch sizes become smaller and smaller, at fixed and finite separation, the point version of the theory is recovered. Depending upon the number of levels chosen for the patch formulation, there are many versions of the most-probable-state patch equation; the three-level version of Eq. (\ref{eqn:3_level}) is not the most general by any means. Each will have different solutions. There is, however, no apparent unique or practical prescription for how many levels, or what size patches, should be used to represent a particular initial analytic vorticity distribution with high accuracy. It is not clear that it can be done without requiring the patch size to shrink to zero, at which point it becomes indistinguishable from a point representation. In fairness, we should also say that there is another aspect to the point formulation that is also ambiguous: namely, there is no reason to choose the point vortices of the mean-field theory to be of equal strength. If some are of different strength, Eq. (\ref{eqn:points}) will also change. Keeping these in mind, we turn now to the effort to see some of the solutions as consequences of direct numerical solution of the Navier-Stokes equation.

\begin{figure}[!htbp]
\centering
\begin{minipage}[c]{.489 \linewidth}
\scalebox{1}[1.05]{\includegraphics[width=\linewidth]{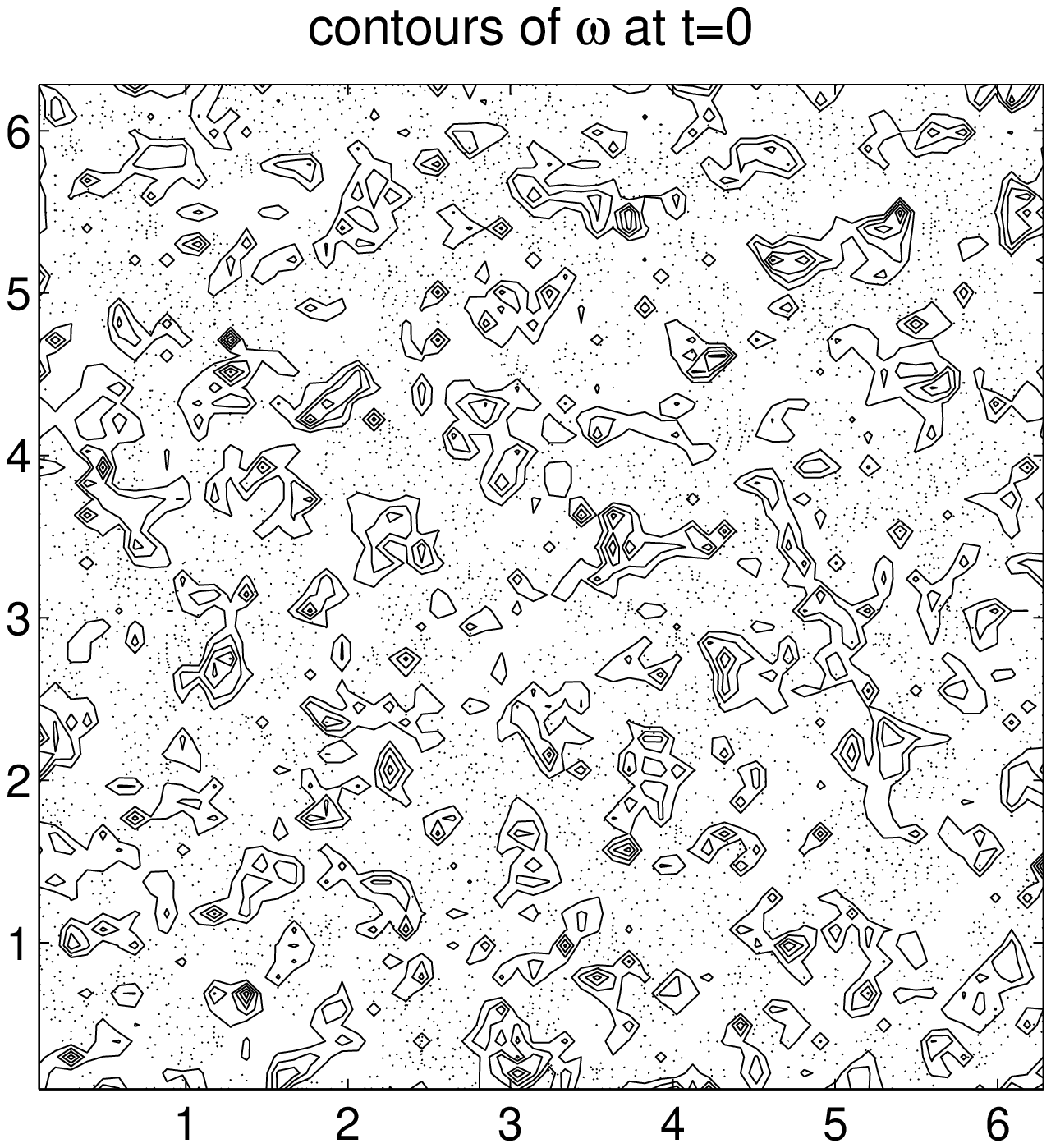}}
\end{minipage}
\begin{minipage}[c]{.489 \linewidth}
\scalebox{1}[1.05]{\includegraphics[width=\linewidth]{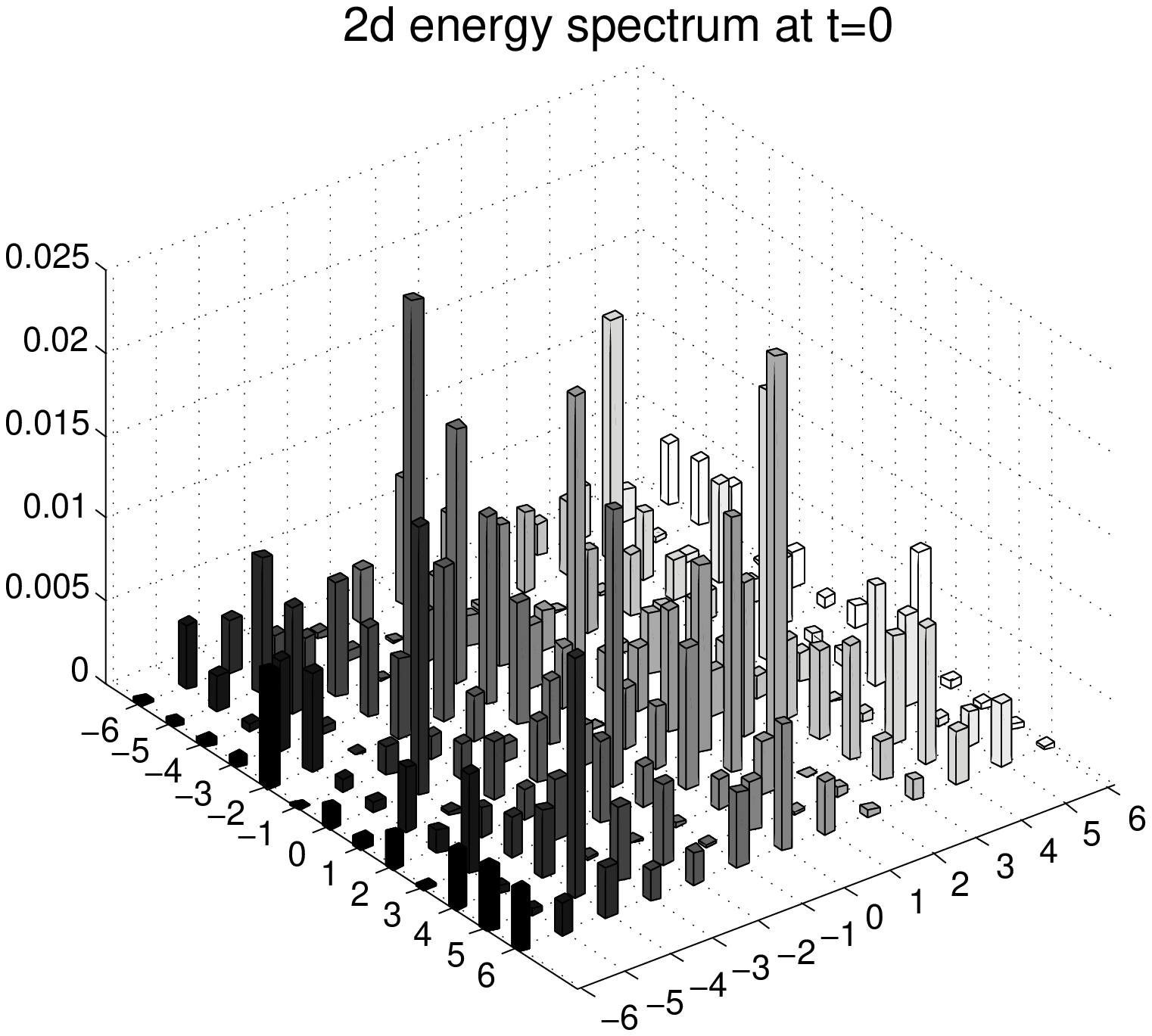}}
\end{minipage}
\begin{minipage}[c]{.489 \linewidth}
\scalebox{1}[1.05]{\includegraphics[width=\linewidth]{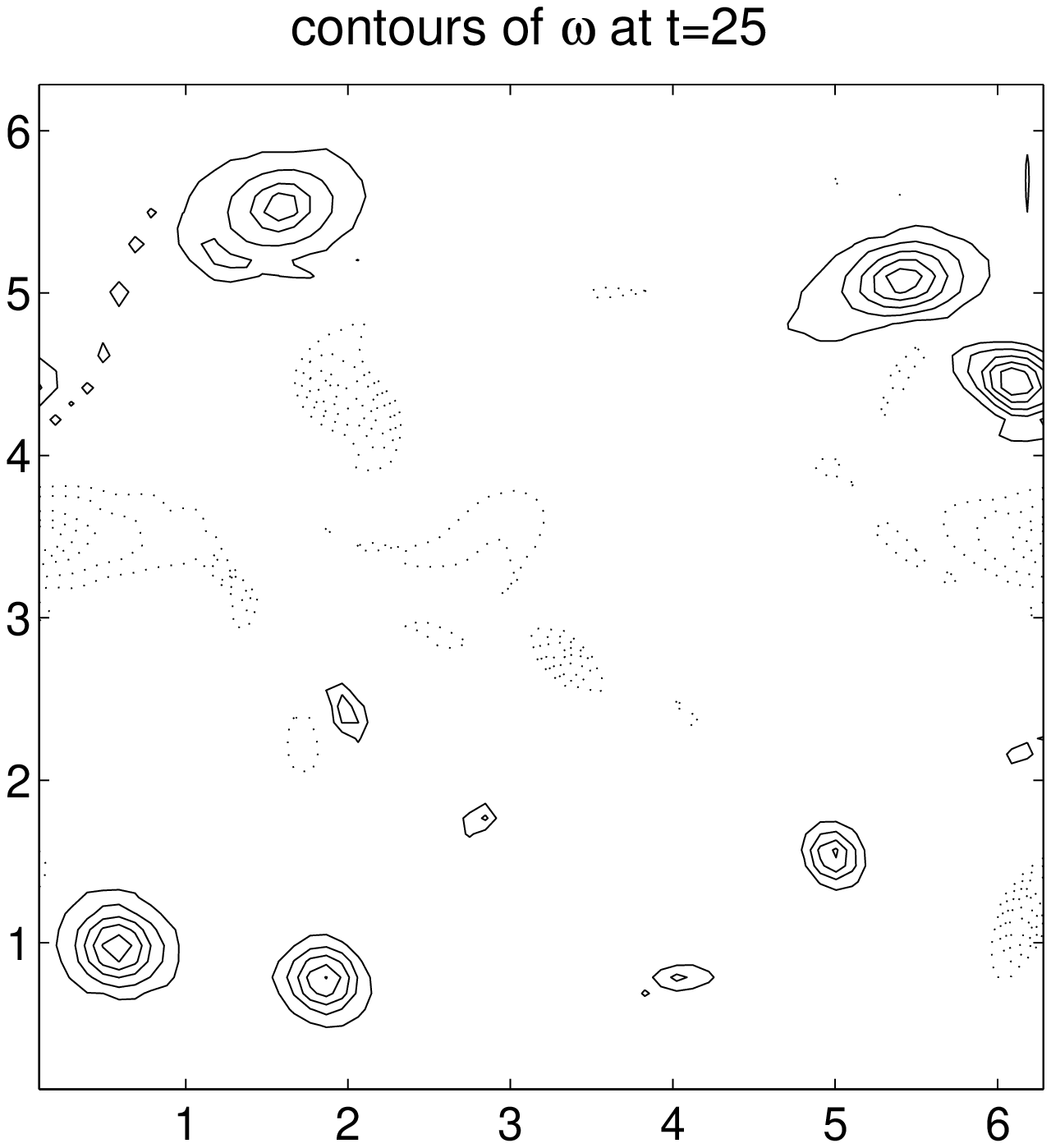}}
\end{minipage}
\begin{minipage}[c]{.489 \linewidth}
\scalebox{1}[1.05]{\includegraphics[width=\linewidth]{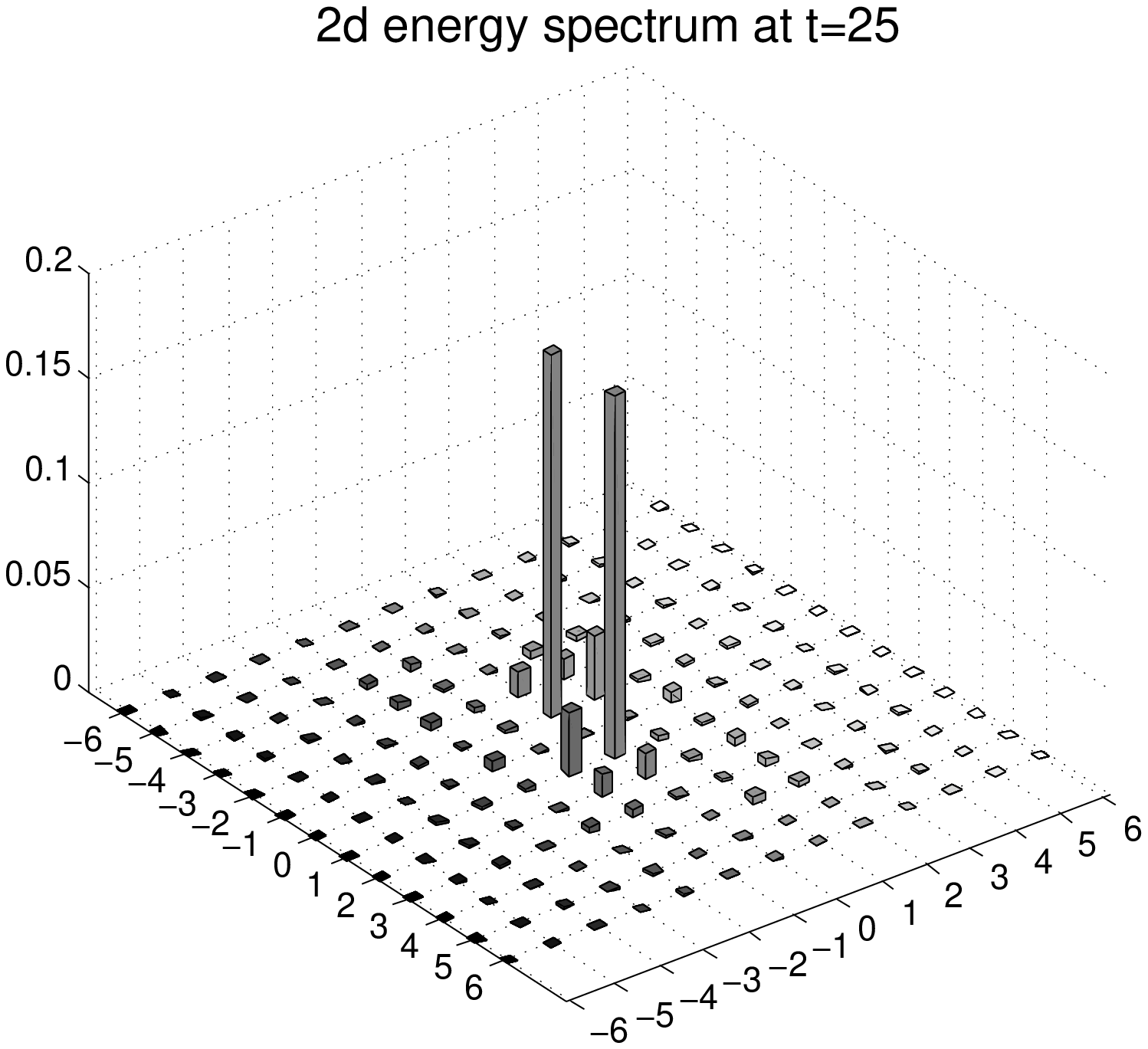}}
\end{minipage}
\begin{minipage}[c]{.489 \linewidth}
\scalebox{1}[1.05]{\includegraphics[width=\linewidth]{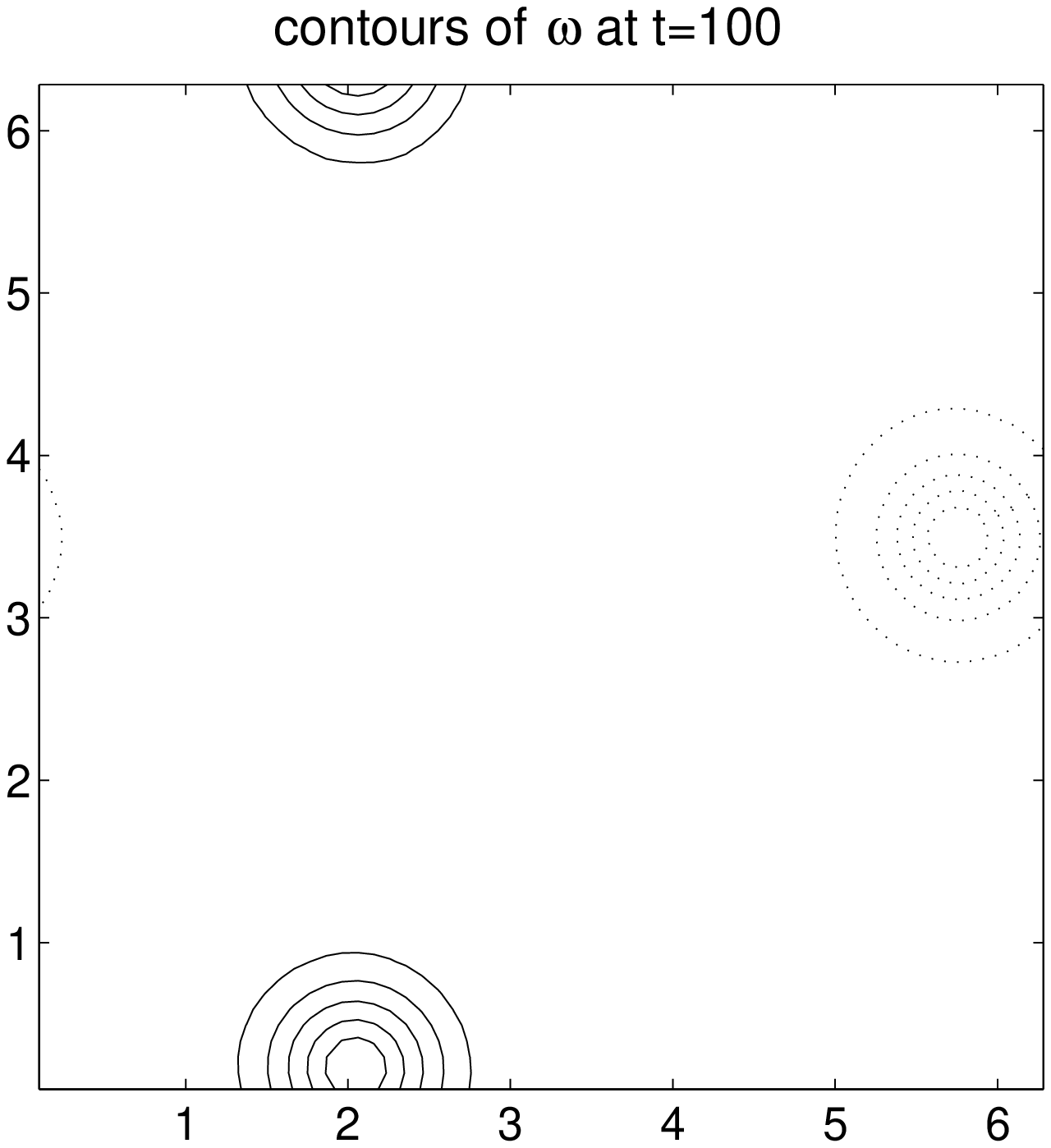}}
\end{minipage}
\begin{minipage}[c]{.489 \linewidth}
\scalebox{1}[1.05]{\includegraphics[width=\linewidth]{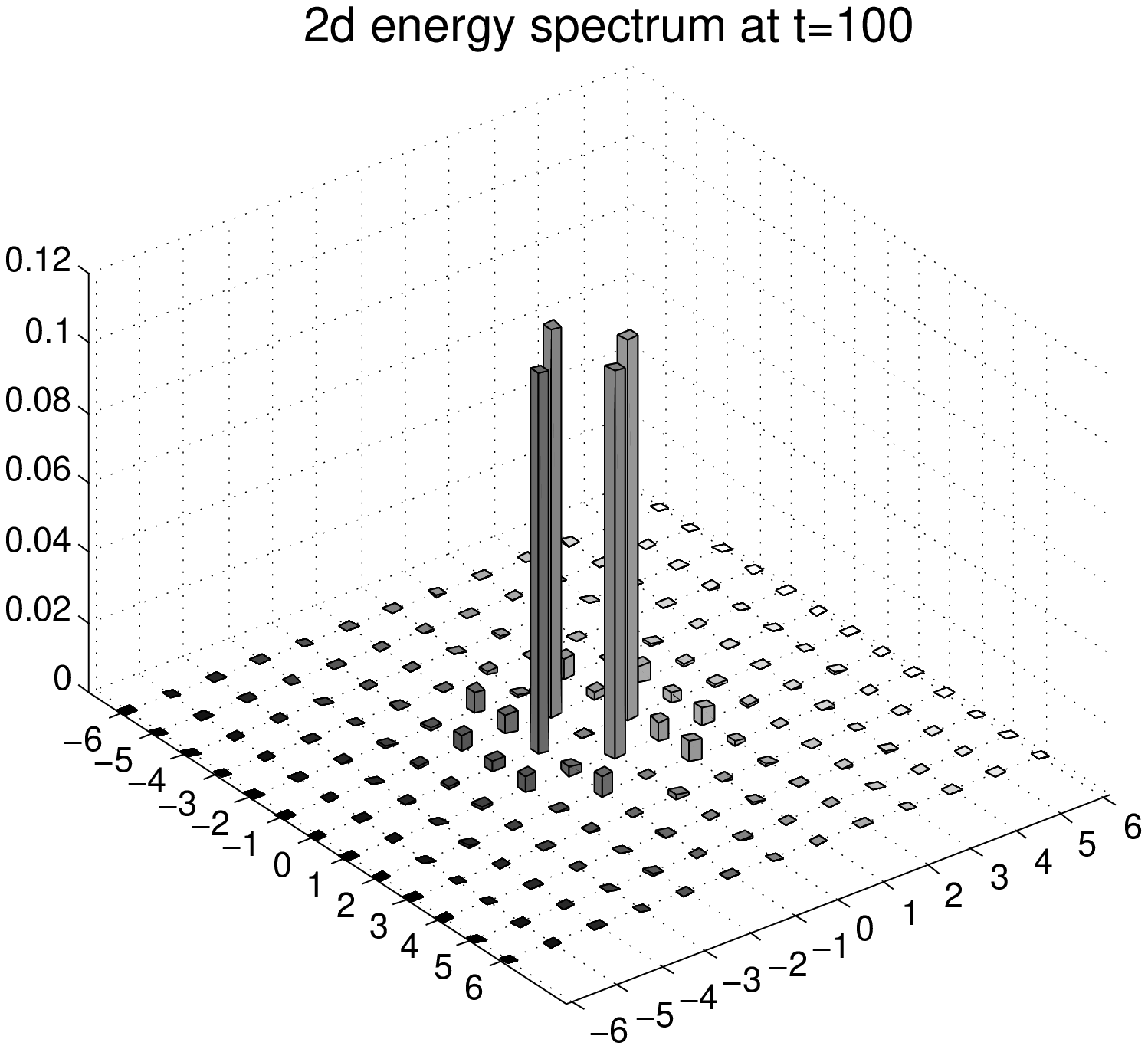}}
\end{minipage}
\caption{Equally spaced contours of constant vorticity at three different times (left column) and corresponding modal energies at the lower values of $k$ (right column), during the evolution of the McWilliams/Matthaeus initial conditions. These have no flat patches of vorticity initially, even approximately.}
\label{fig:noise1}
\end{figure}

\begin{figure}[!htbp]
\begin{minipage}[c]{.8 \linewidth}
\centering
\includegraphics[width=\linewidth]{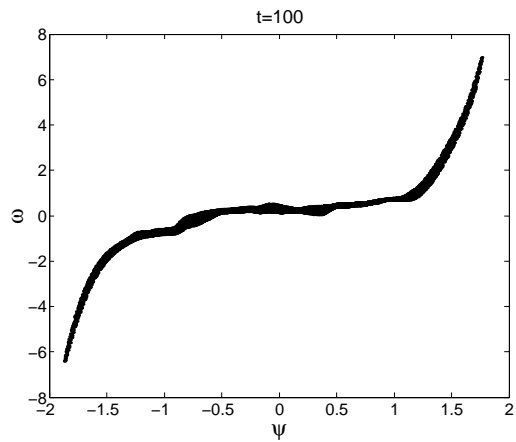}
\end{minipage}
\caption{The $\omega - \psi$ scatter plot for the run shown in Fig. \ref{fig:noise1} (which is close to Fig. \ref{fig:good:a}).}
\label{fig:noise2}
\end{figure}

\subsection{Dynamical Solutions and Comparisons}
\label{sec:s3:b}
We now present the results of dynamical, pseudo-spectral-method solutions of the 2D NS equation, using a resolution of $512^2$. The time step is 0.0005. The initial energy, using the normalization of Eq. (\ref{eqn:energy1}), is 0.5. There is no hyperviscosity or small-scale smoothing of any kind.

\subsubsection{``Dipole'' and ``bar'' final states.}
\label{sec:s3:b:1}
\paragraph{Random, broad-band initial conditions.}

The first run we report here is essentially a reproduction of the run of Matthaeus \textit{et al.} ~\cite{kn:q1, kn:q2, kn:q3}, but with a lower Reynolds number ($1/ \nu = 5000$), corresponding to an initial Taylor-scale Reynolds number 
\[
R_{\lambda} = \sqrt{\frac{10}{3}} \frac{E}{\nu \sqrt{\Omega}} \approx 558.
\]
$R_{\lambda}$ has increased to $3143$ by the end of the run.

 This run served as a benchmark for our parallelized code in the beginning, and the McWilliams initial conditions ~\cite{kn:q32} can also be used as noise, with variable amplitude, to be added to later simulations to break unwanted symmetries and accelerate the dynamical development. The run also provides an opportunity to introduce the types of data displays  that will be used throughout the rest of this Section.
The Fourier modal energy spectrum $E(\mathbf{k}) \equiv (1/2) \left| \mathbf{v}(\mathbf{k}) \right|^2$
is initialized according to 
\[
E(k) = \frac{C}{1 + (\frac{k}{6})^4}
\]
for $1 \leq k \leq 120$ (here, and for this purpose only, the wave number $k$ is binned in integer values by a standard Fortran routine, and the partition of energies among modes with the same integer $k$ is decided by a random number generator), and zero otherwise. The phases of the Fourier coefficients are chosen from a Gaussian random number generator.  $C$ is a constant to be adjusted to make the total initial energy equal to 0.5.

The results were quite similar in all respects to those of Matthaeus \textit{et al.} ~\cite{kn:q1, kn:q2, kn:q3}, so only a few figures (Figs. \ref{fig:noise1} and \ref{fig:noise2}) are shown here. The contour plots of vorticity and stream function relax to the familiar dipole final states. The left column of Figs. \ref{fig:noise1} shows the vorticity contours at three separate times.  The right column of Figs. \ref{fig:noise1} shows the modal energy spectra, for the lower part of $k$-space (the maximum $k$ is 241) at those same times, exhibited as three-dimensional perspective plots. They are initially broad-band, but evolve as expected to be concentrated in the lowest values of $k$. In Fig. \ref{fig:noise2}, a scatter plot of $\omega$ vs. $\psi$ shows a good agreement with the hyperbolic sinusoidal dependence of vorticity on stream function, as predicted in the point formulation (Fig. \ref{fig:pentropy}).

\begin{figure}[!htbp]
\subfigure[]{
\label{fig:banoise:a}
\begin{minipage}[c]{.43 \linewidth}
\scalebox{1}[1.2]{\includegraphics[width=\linewidth]{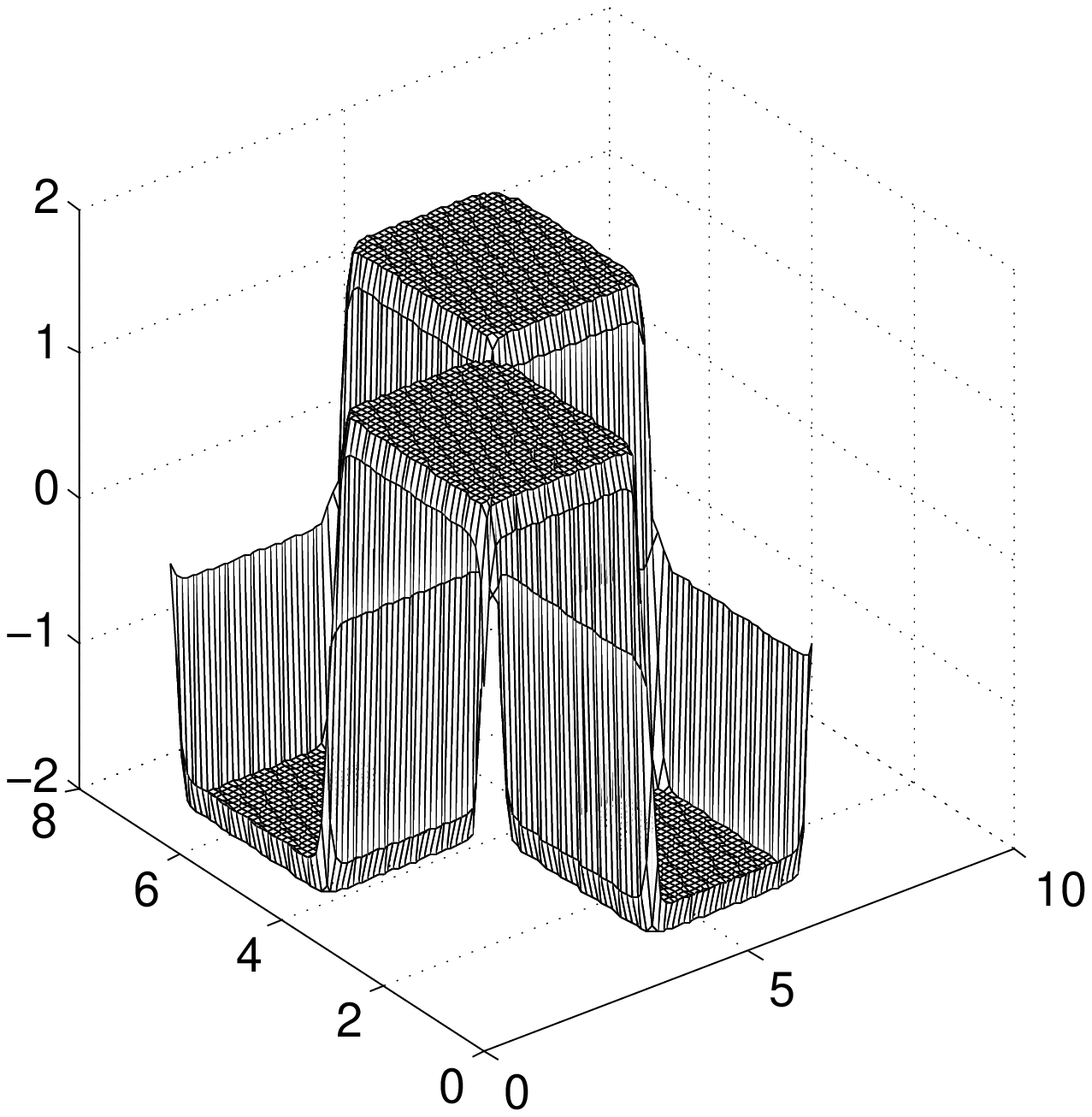}}
\end{minipage}}
\subfigure[]{
\label{fig:banoise:b}
\begin{minipage}[c]{.43 \linewidth}
\scalebox{1}[1.2]{\includegraphics[width=\linewidth]{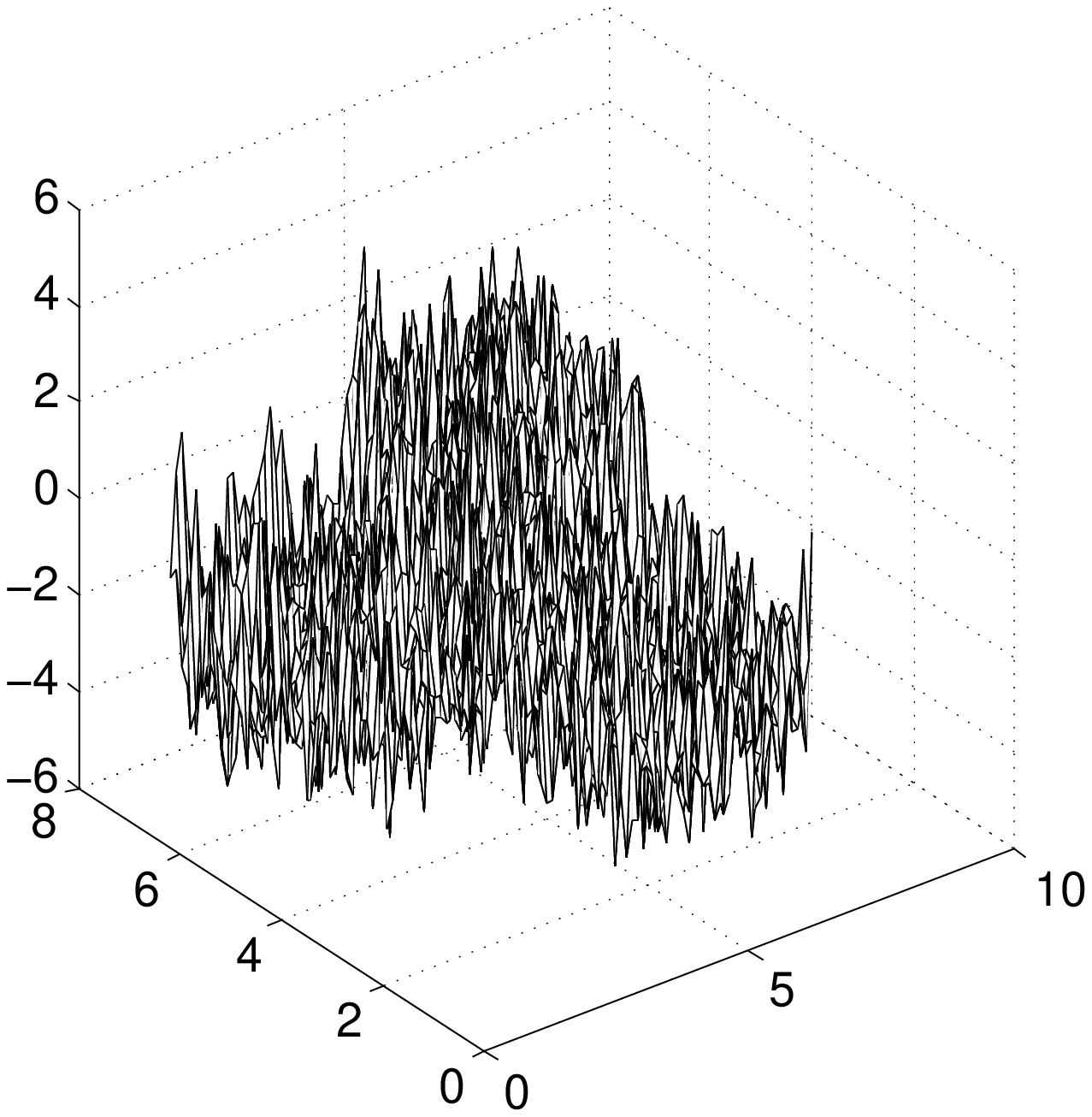}}
\end{minipage}}
\caption{The initial vorticity field as a function of $x$ and $y$ for a run intended to exhibit patch characteristics. (a) without random noise, (b) has had substantial random noise added to the vorticity field.}
\label{fig:banoise}
\end{figure}

\paragraph{``Bar''}
\label{sec:s3:b:1:bb}
A different, and interesting, evolution results ($1/ \nu = 5000$) if we initialize using the quadrupole solution from Eq. (\ref{eqn:s3_level}). In this run, $R_{\lambda}$ increases from $2391$ initially to $4920$ at the end. This  is not predicted to be the maximum entropy state from Eq. (\ref{eqn:3get3}), or any other equation, patch or point; and so should evolve when initialized with some random noise. The quadrupole solution was found to have a symmetry which persists in time, and the noise, it is hoped, will permit that symmetry to be broken.
It was found that the symmetry persisted until the noise was raised to quite substantial levels. Figs. \ref{fig:banoise} are perspective plots of the initial vorticity as a function of $x$ and $y$, with and without the noise added. The noise level in this case is about twice the quadrupole vorticity, and its randomness should break any symmetry that might be present.

\begin{figure}[!htbp]
\centering
\begin{minipage}[c]{.489 \linewidth}
\scalebox{1}[1.]{\includegraphics[width=\linewidth]{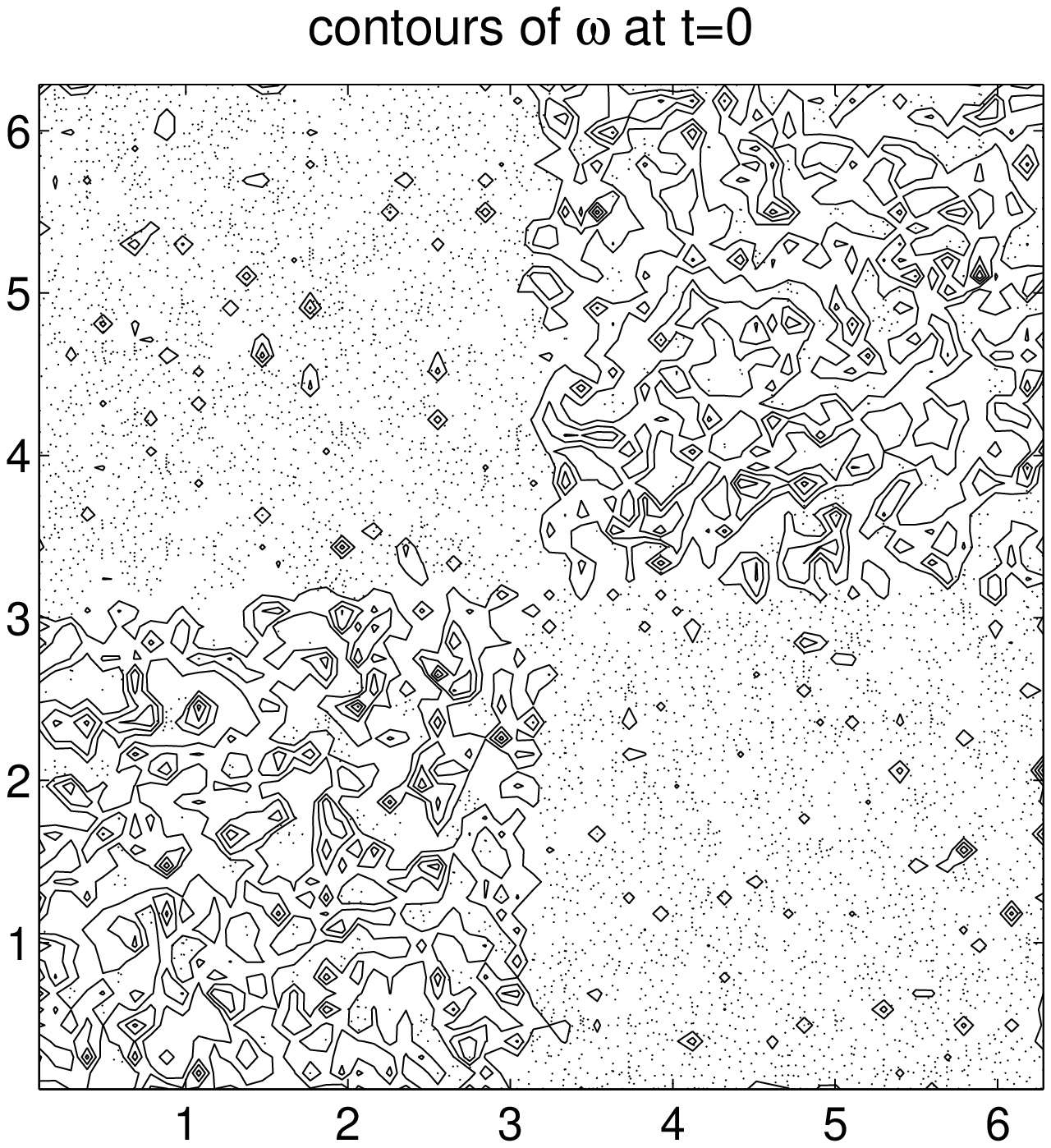}}
\end{minipage}
\begin{minipage}[c]{.489 \linewidth}
\scalebox{1}[1.]{\includegraphics[width=\linewidth]{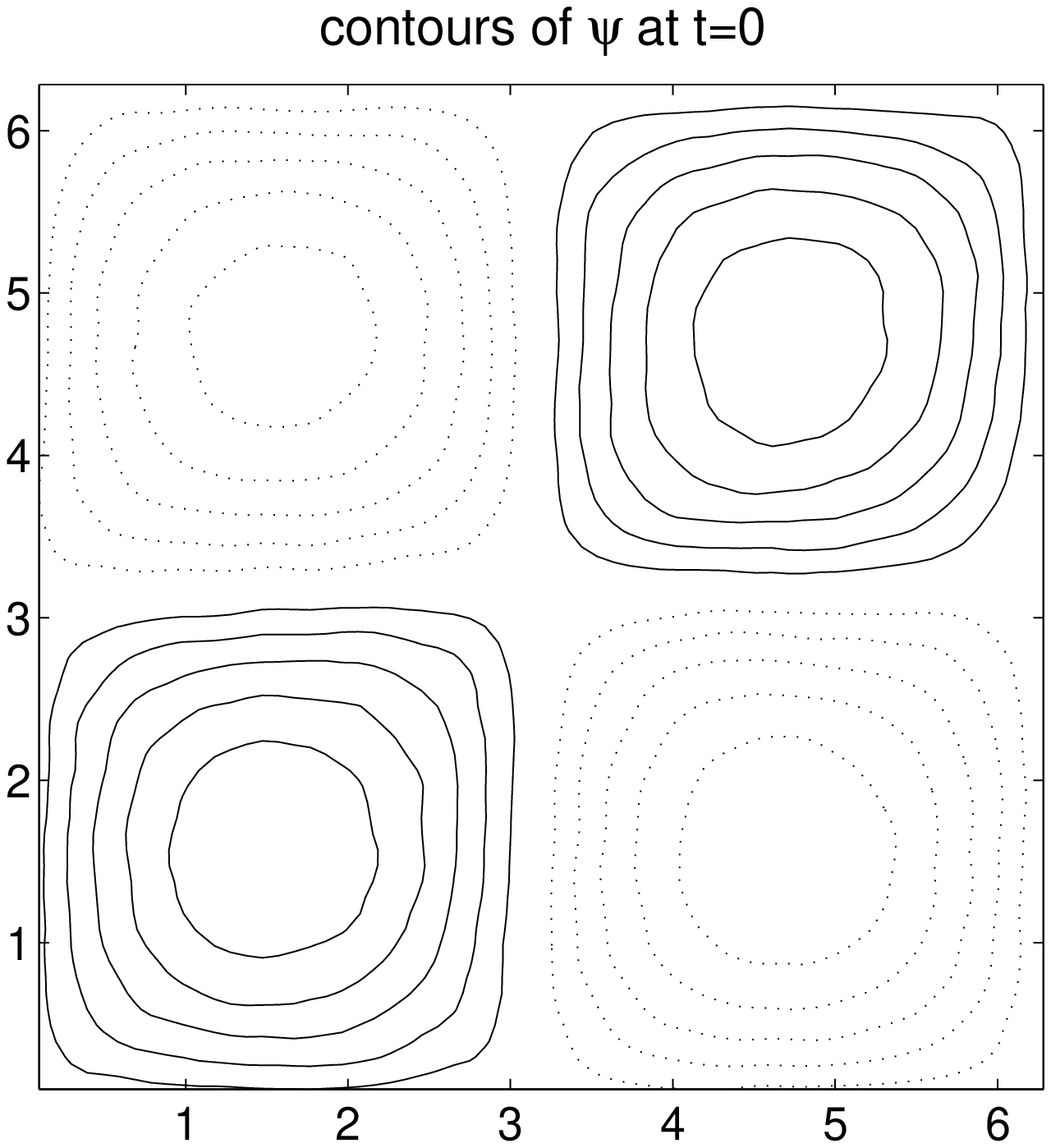}}
\end{minipage}
\begin{minipage}[c]{.489 \linewidth}
\scalebox{1}[1.]{\includegraphics[width=\linewidth]{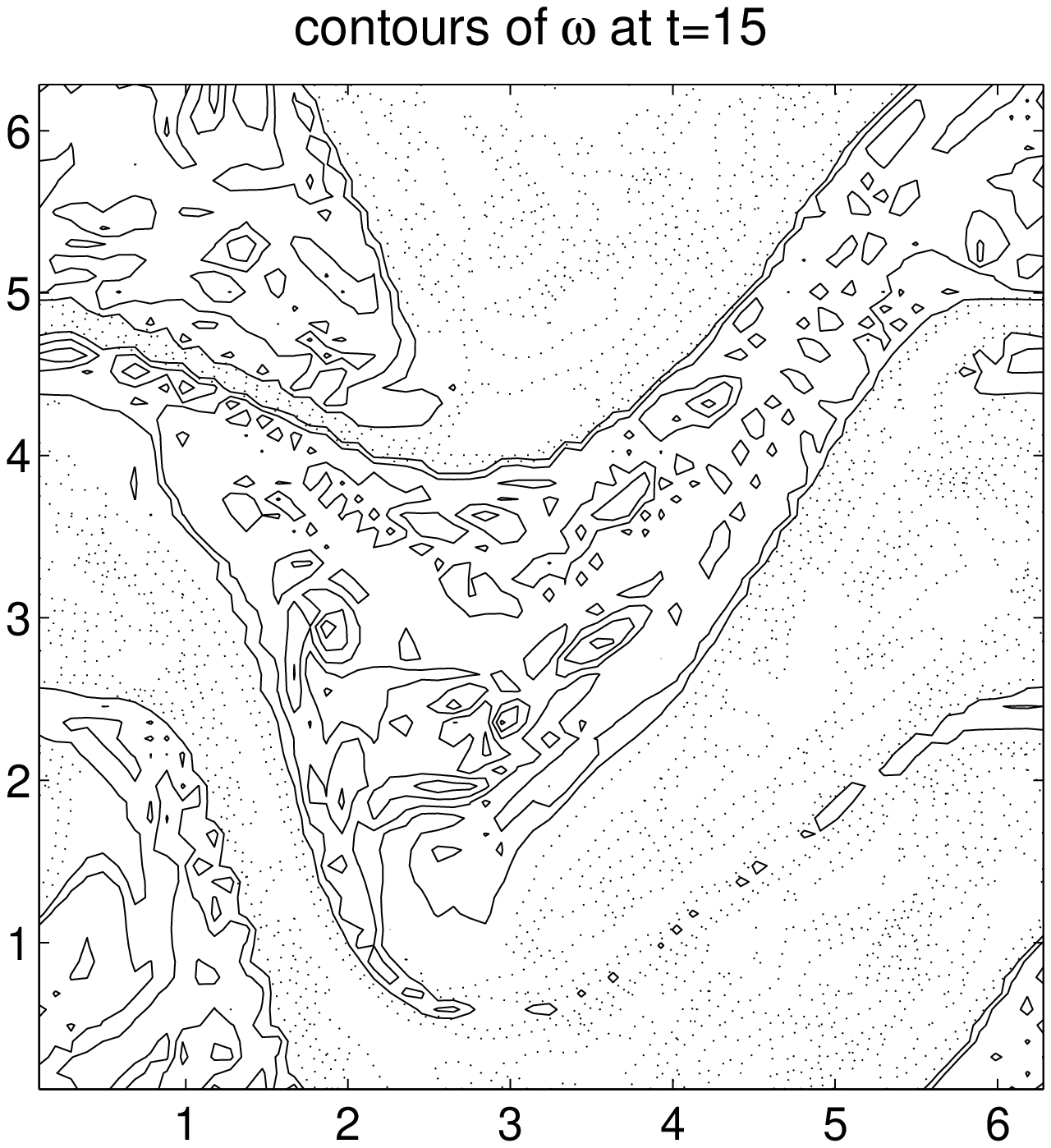}}
\end{minipage}
\begin{minipage}[c]{.489 \linewidth}
\scalebox{1}[1.]{\includegraphics[width=\linewidth]{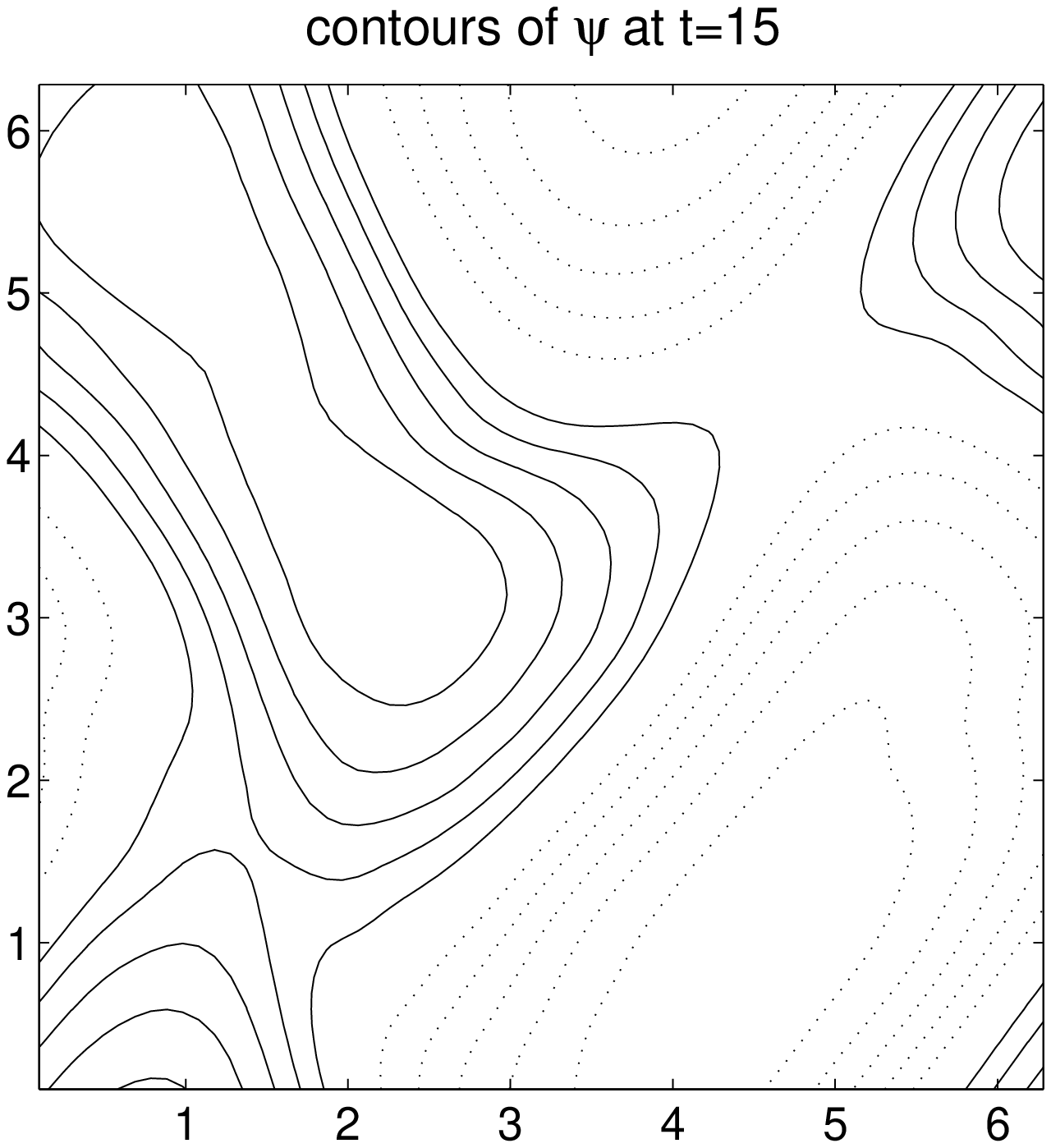}}
\end{minipage}
\begin{minipage}[c]{.489 \linewidth}
\scalebox{1}[1.]{\includegraphics[width=\linewidth]{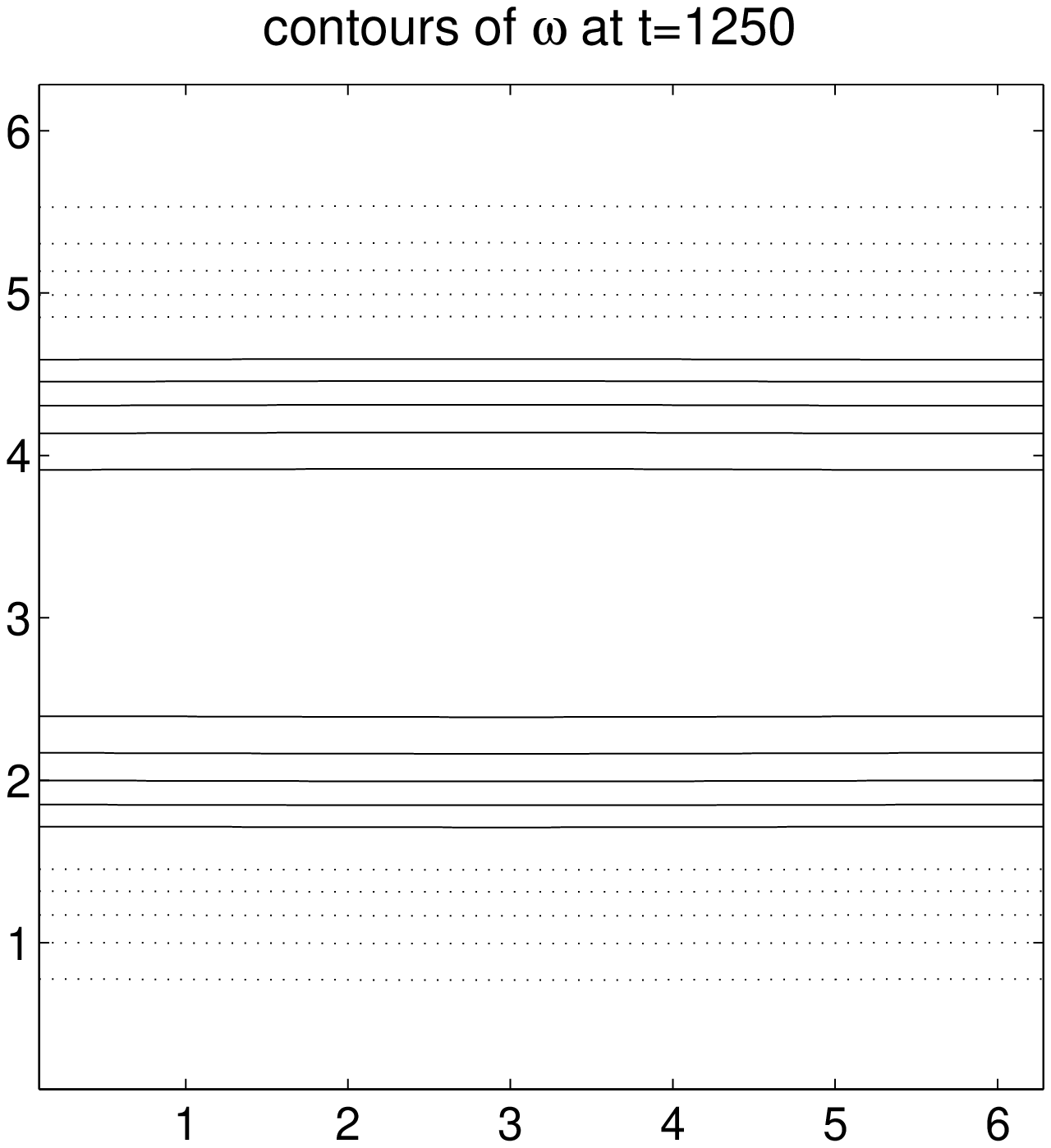}}
\end{minipage}
\begin{minipage}[c]{.489 \linewidth}
\scalebox{1}[1.]{\includegraphics[width=\linewidth]{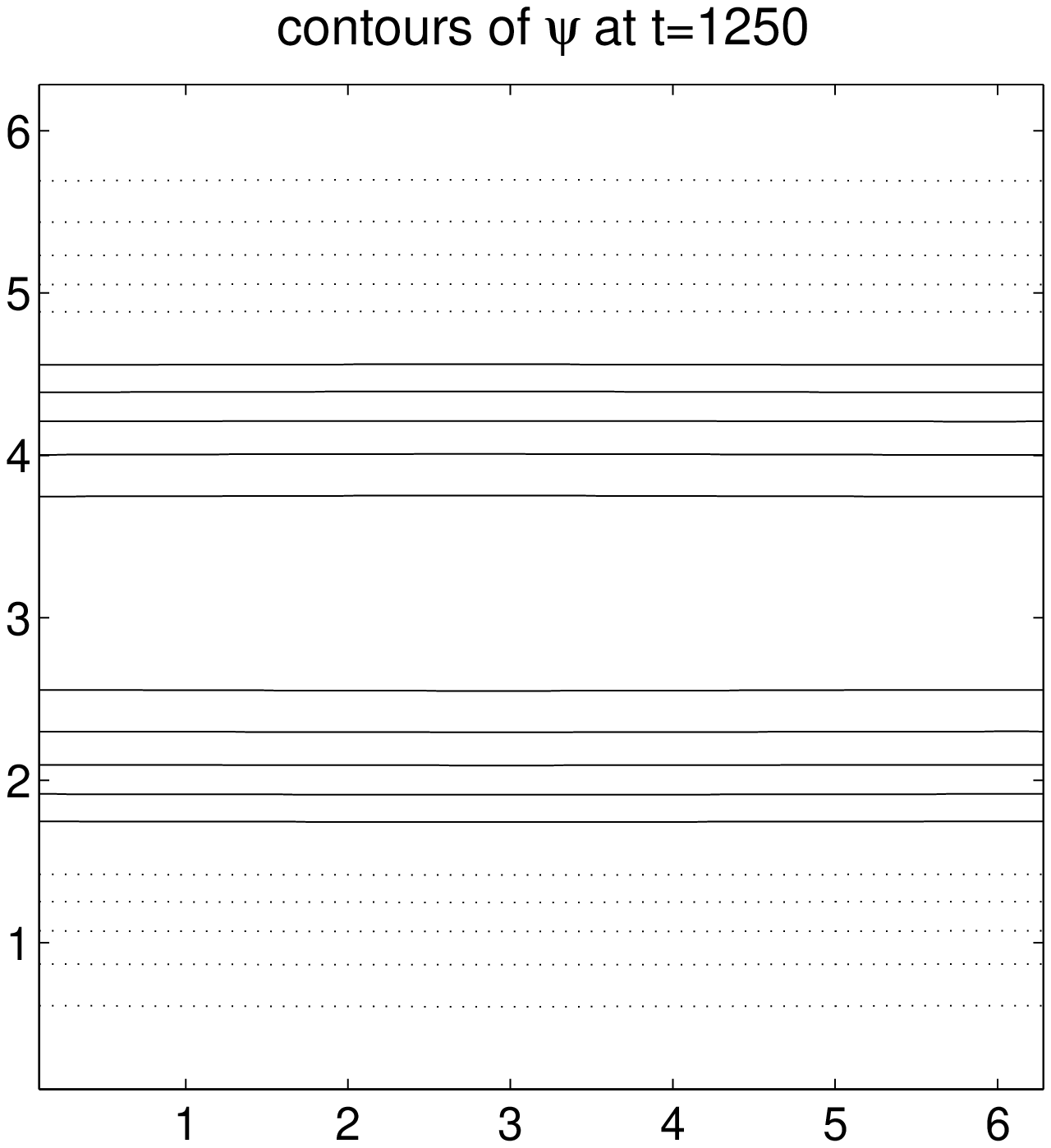}}
\end{minipage}
\caption{Contours of constant vorticity (left column) and constant stream function (right column) at three different times for the run originating from the intial vorticity distribution (with noise) shown in Fig. \ref{fig:banoise:b}.}
\label{fig:bar1}
\end{figure}

\begin{figure}[!htbp]
\centering
\begin{minipage}[c]{.7 \linewidth}
\includegraphics[width=\linewidth]{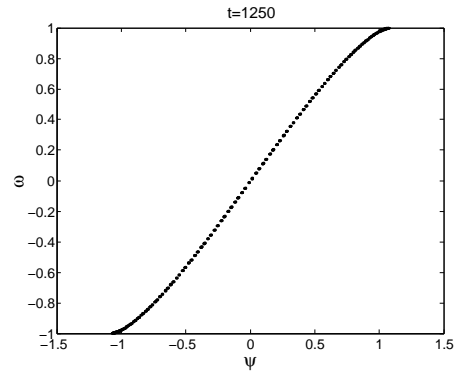}
\end{minipage}
\caption{The $\omega - \psi$ scatter plot of for the late-time state achieved in the run shown in Fig. \ref{fig:bar1} (which is close to Fig. \ref{fig:good:c}).}
\label{fig:barst}
\end{figure}

	Fig. \ref{fig:bar1} shows the evolution of the vorticity and stream function contours that result from the initial conditions shown in Fig. \ref{fig:banoise:b}. The intermediate panel ($t=15$) shows no obvious connection to the symmetry of Fig. \ref{fig:banoise:a}  or to that of the final panels (t=1250) of Figs. \ref{fig:bar1}  (the first is odd about the center lines of the basic square, but clearly, the t=15 state has no such symmetry). The relaxation to the one-dimensional ``bar'' state at t=1250 has been quite robust in several such runs. Fig. \ref{fig:barst} is a $\omega - \psi$ scatter plot for the bar state. There has been considerable decay of vorticity, as evidenced by the disappearance in Fig. \ref{fig:barst} of the high and low values visible in Fig. \ref{fig:banoise:b}. By $t=1250$, the energy ($E$) has decreased to $58.4\%$ of the initial value.

\begin{figure}[!htbp]
\centering
\begin{minipage}[c]{.489 \linewidth}
\scalebox{1}[1.]{\includegraphics[width=\linewidth]{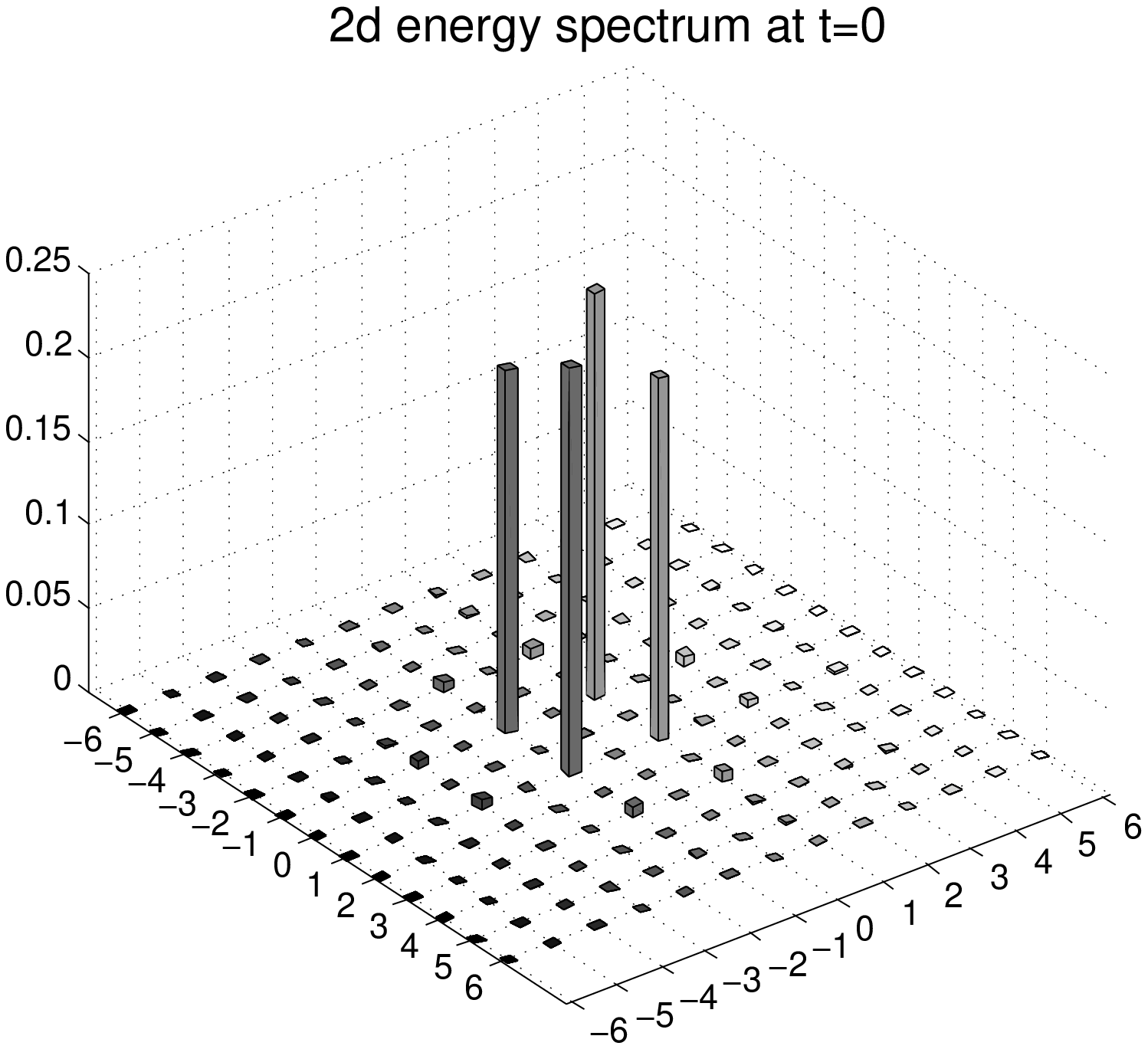}}
\end{minipage}
\begin{minipage}[c]{.489 \linewidth}
\scalebox{1}[1.]{\includegraphics[width=\linewidth]{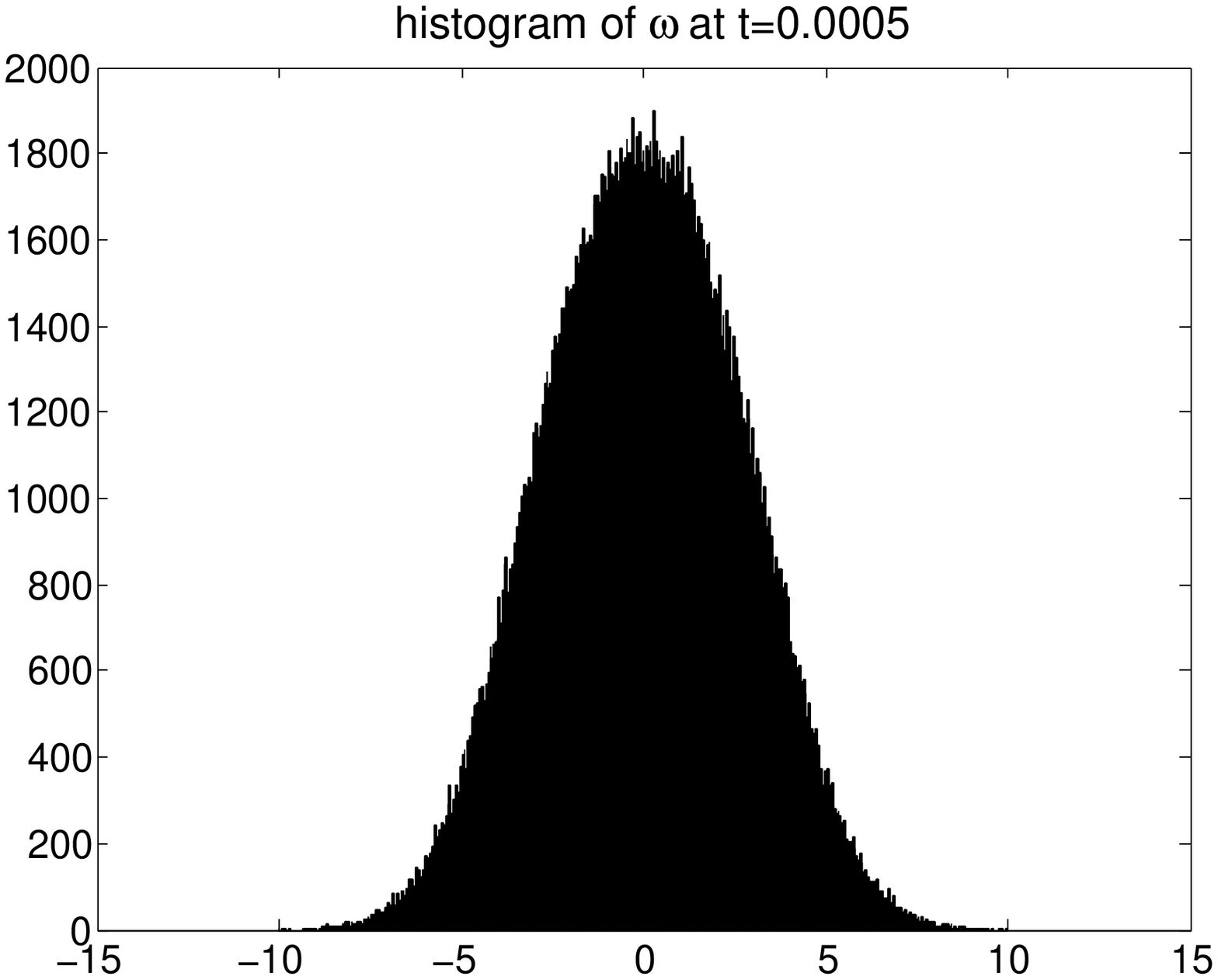}}
\end{minipage}
\begin{minipage}[c]{.489 \linewidth}
\scalebox{1}[1.]{\includegraphics[width=\linewidth]{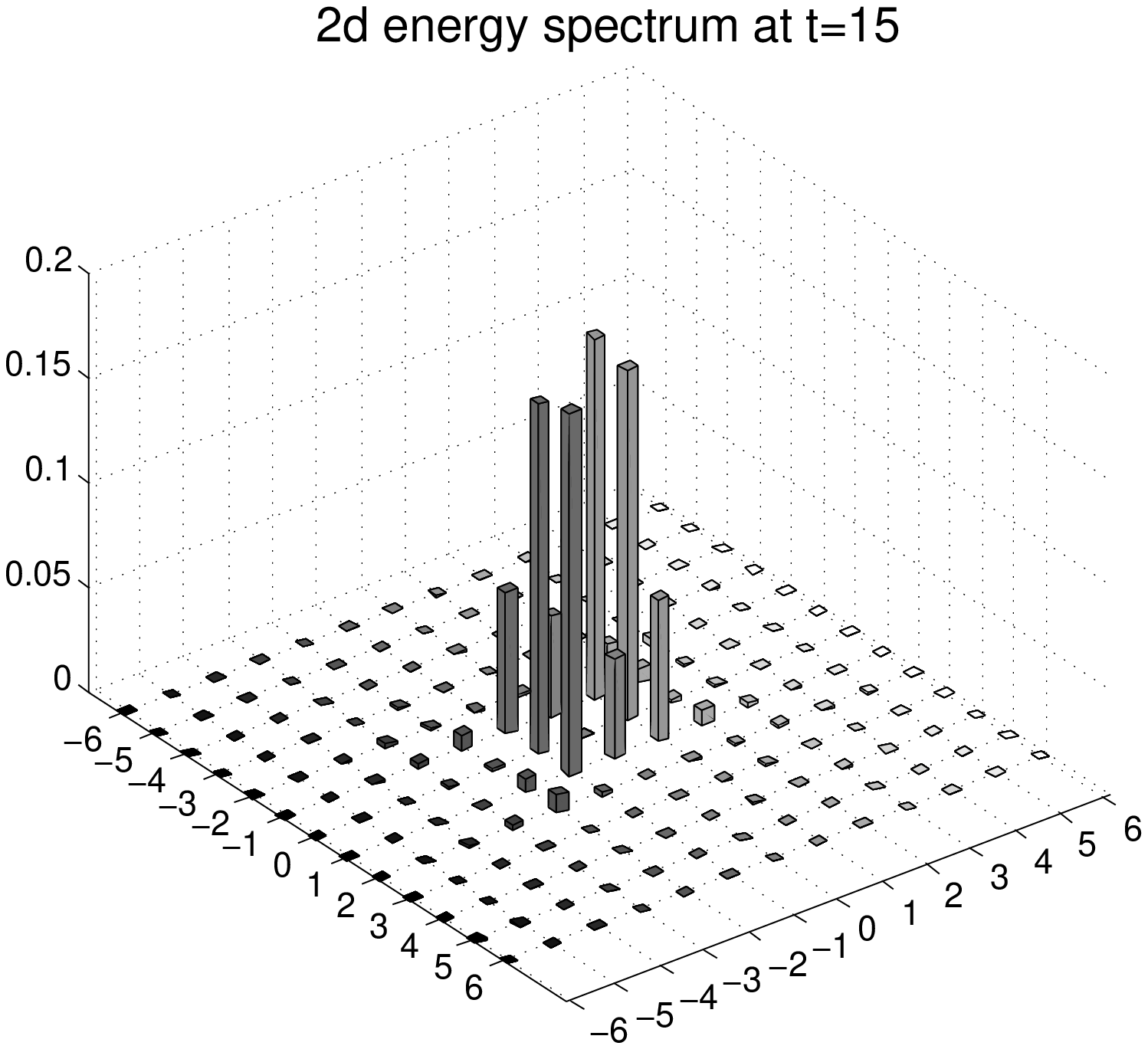}}
\end{minipage}
\begin{minipage}[c]{.489 \linewidth}
\scalebox{1}[1.]{\includegraphics[width=\linewidth]{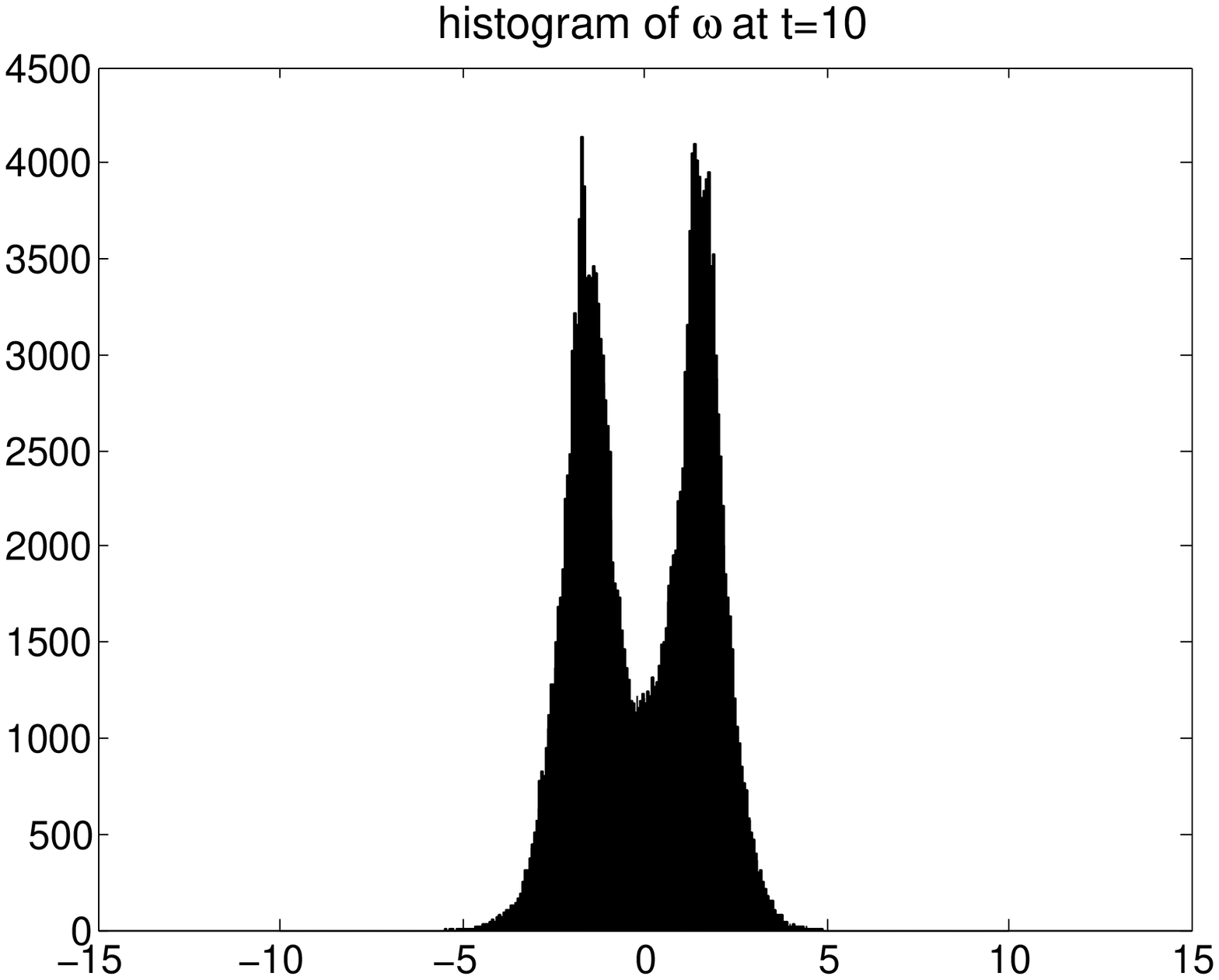}}
\end{minipage}
\begin{minipage}[c]{.489 \linewidth}
\scalebox{1}[1.]{\includegraphics[width=\linewidth]{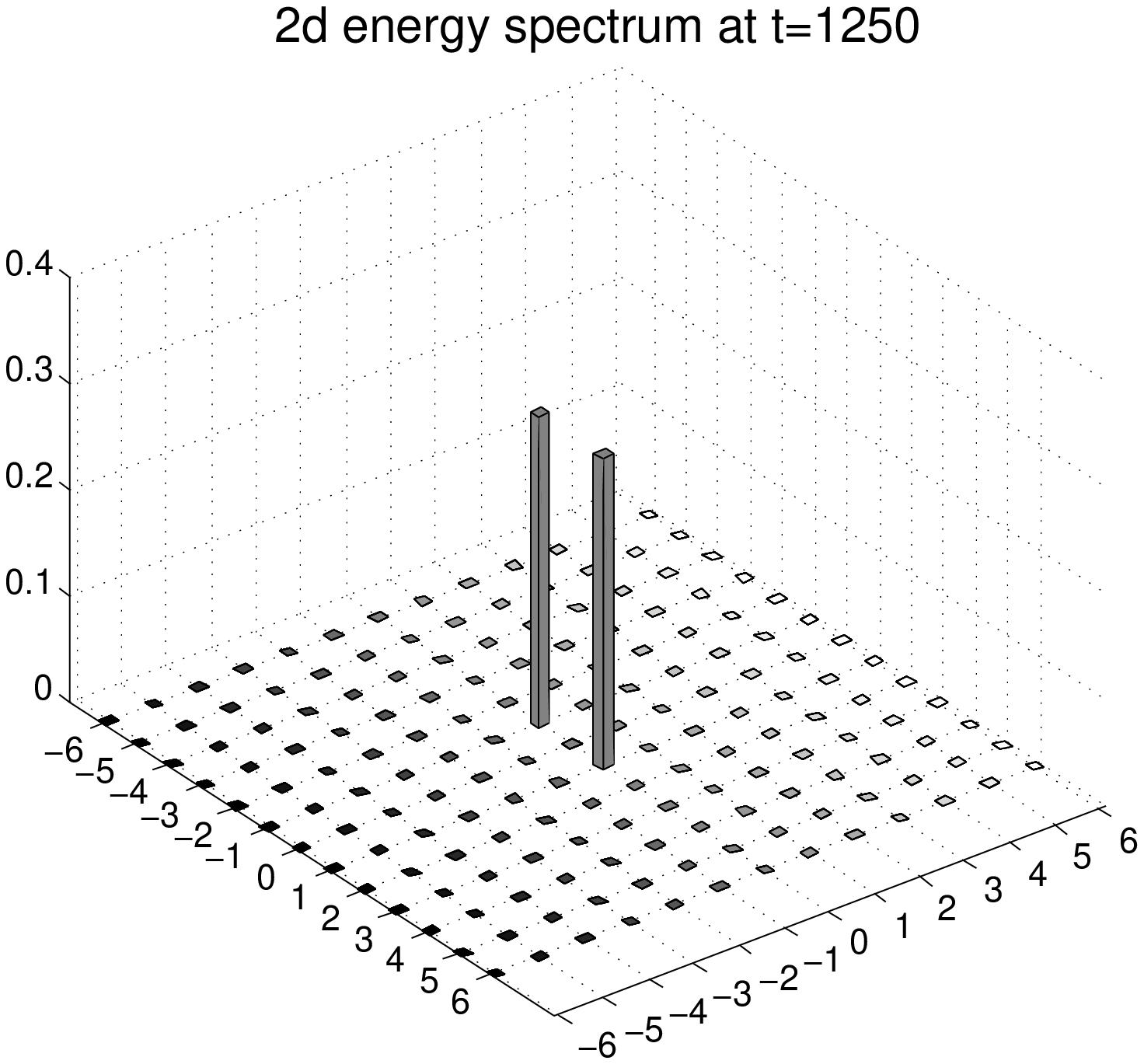}}
\end{minipage}
\begin{minipage}[c]{.489 \linewidth}
\scalebox{1}[1.]{\includegraphics[width=\linewidth]{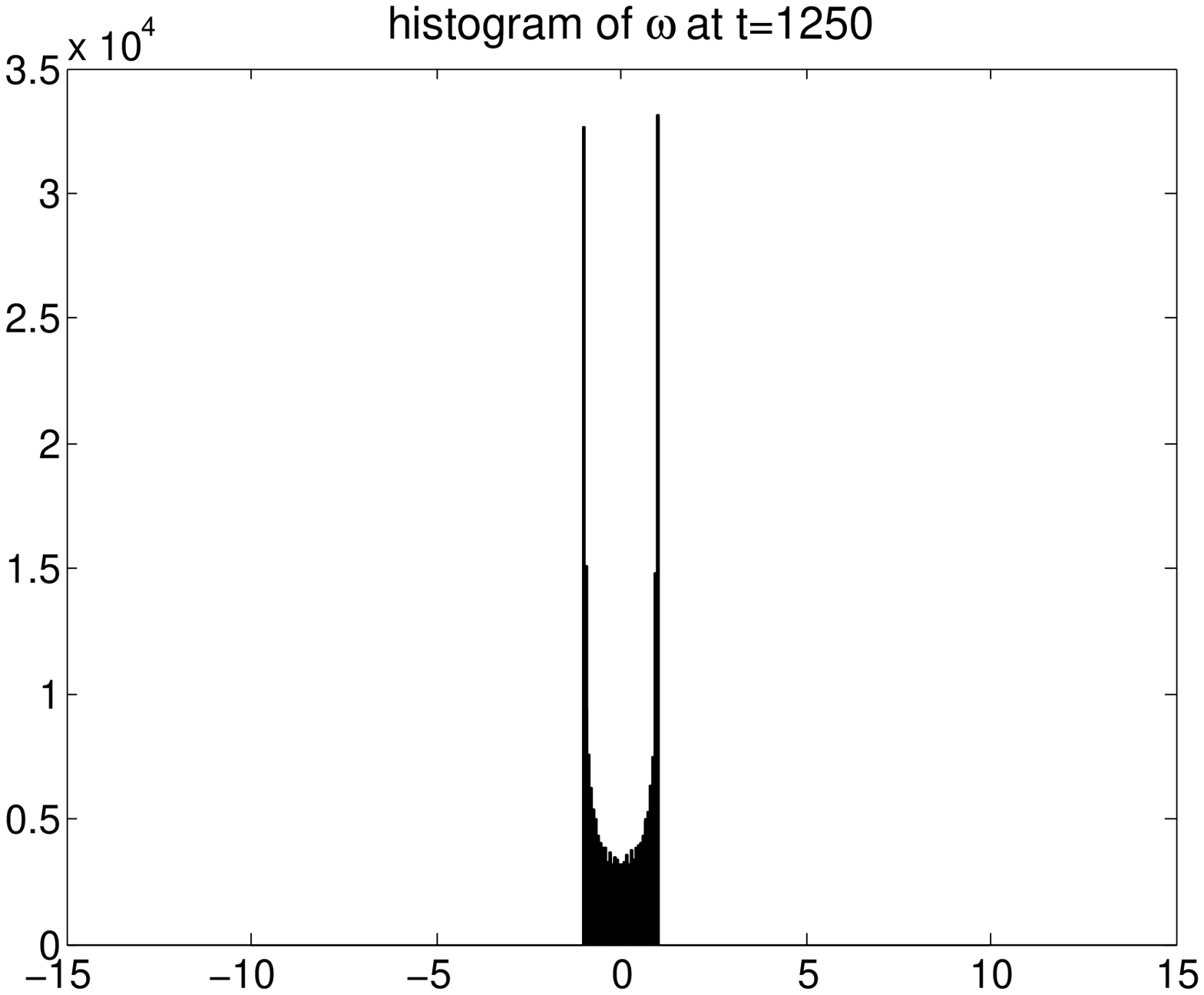}}
\end{minipage}
\caption{Low-k modal energy spectra (left column) and vorticity histogram (right column) at three different times for the run shown in Figs. \ref{fig:banoise:b} and \ref{fig:bar1}. The vorticity histogram shows the number of times a particular value of $\omega$ is recorded as one cycles through the computational grid. Notice that there is substantial variation from one time to another. (An Euler equation solution should preserve this histogram in time.)}
\label{fig:barmhhi}
\end{figure}

\begin{figure}[!htbp]
\centering
\begin{minipage}[c]{.323 \linewidth}
\scalebox{1}[1.]{\includegraphics[width=\linewidth]{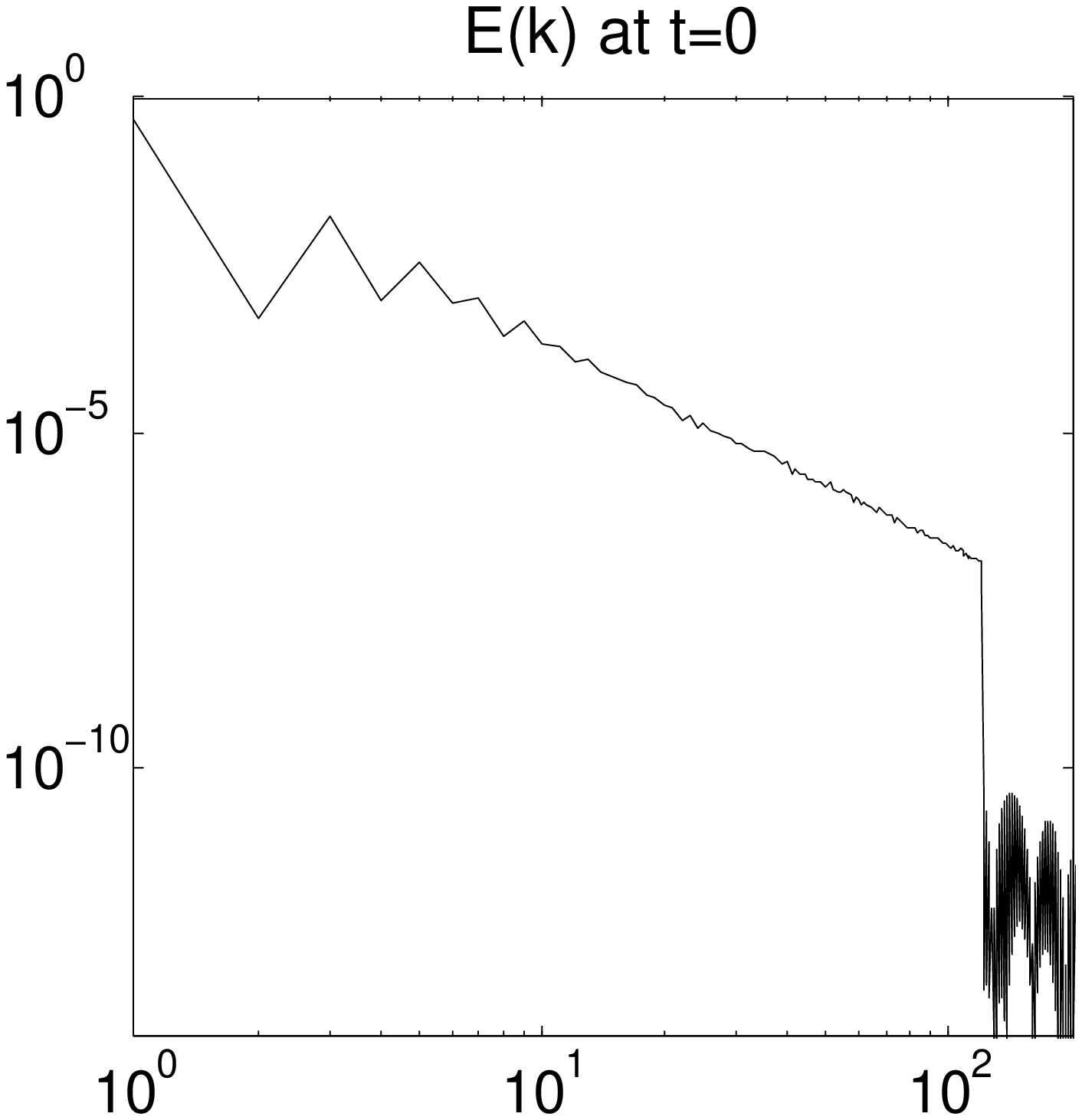}}
\end{minipage}
\begin{minipage}[c]{.323 \linewidth}
\scalebox{1}[1.]{\includegraphics[width=\linewidth]{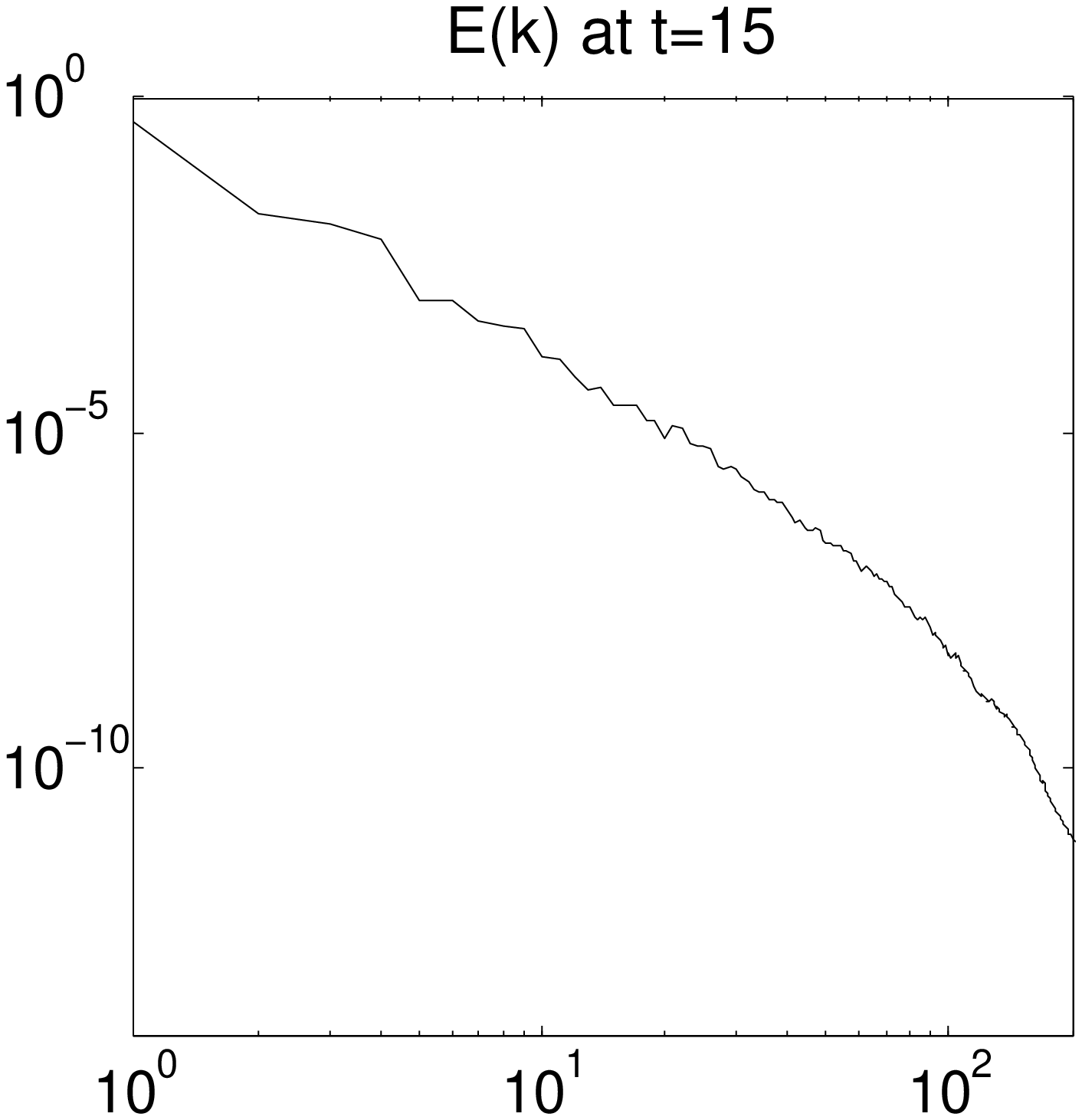}}
\end{minipage}
\begin{minipage}[c]{.323 \linewidth}
\scalebox{1}[1.]{\includegraphics[width=\linewidth]{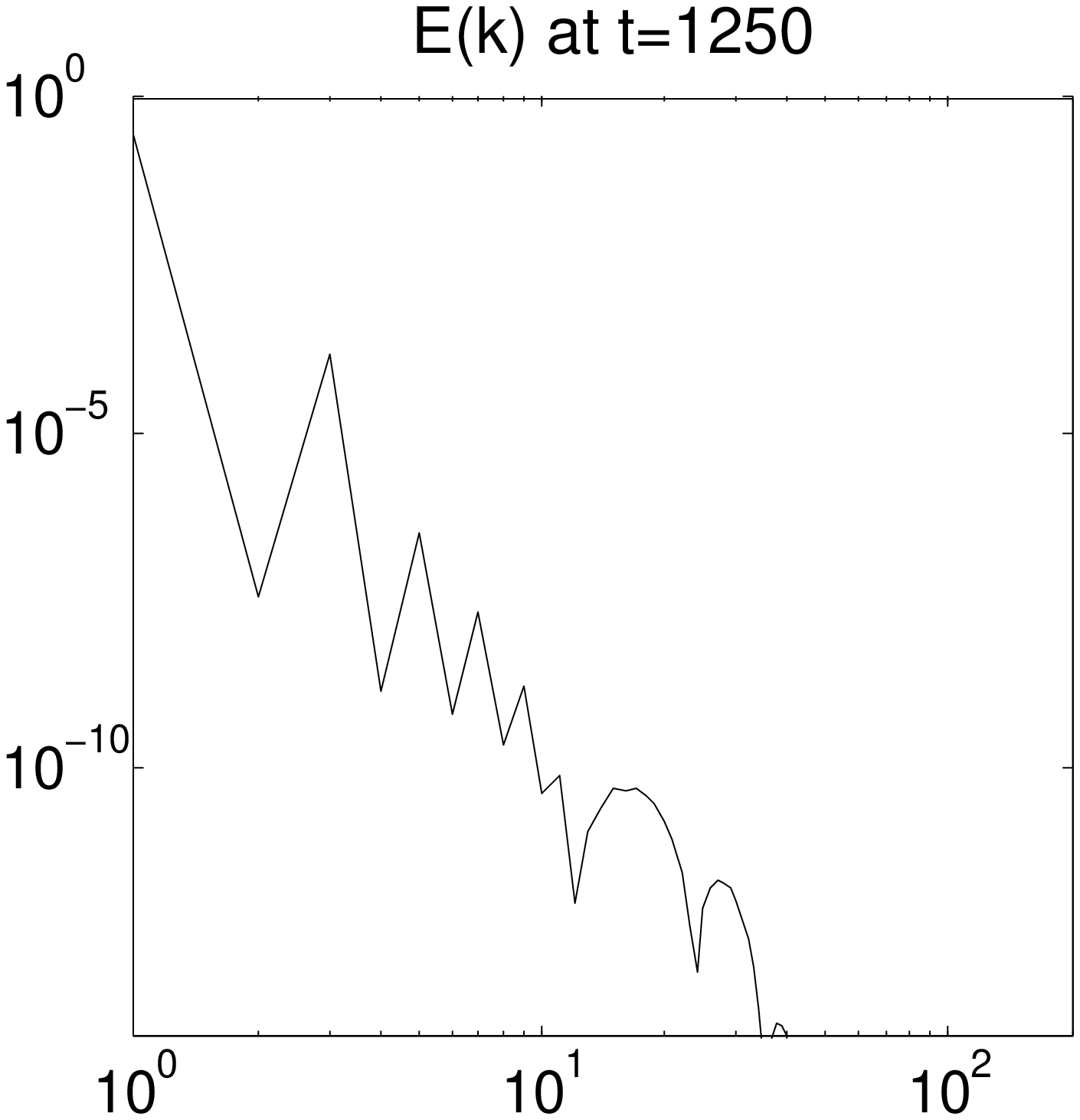}}
\end{minipage}
\begin{minipage}[c]{.323 \linewidth}
\scalebox{1}[1.]{\includegraphics[width=\linewidth]{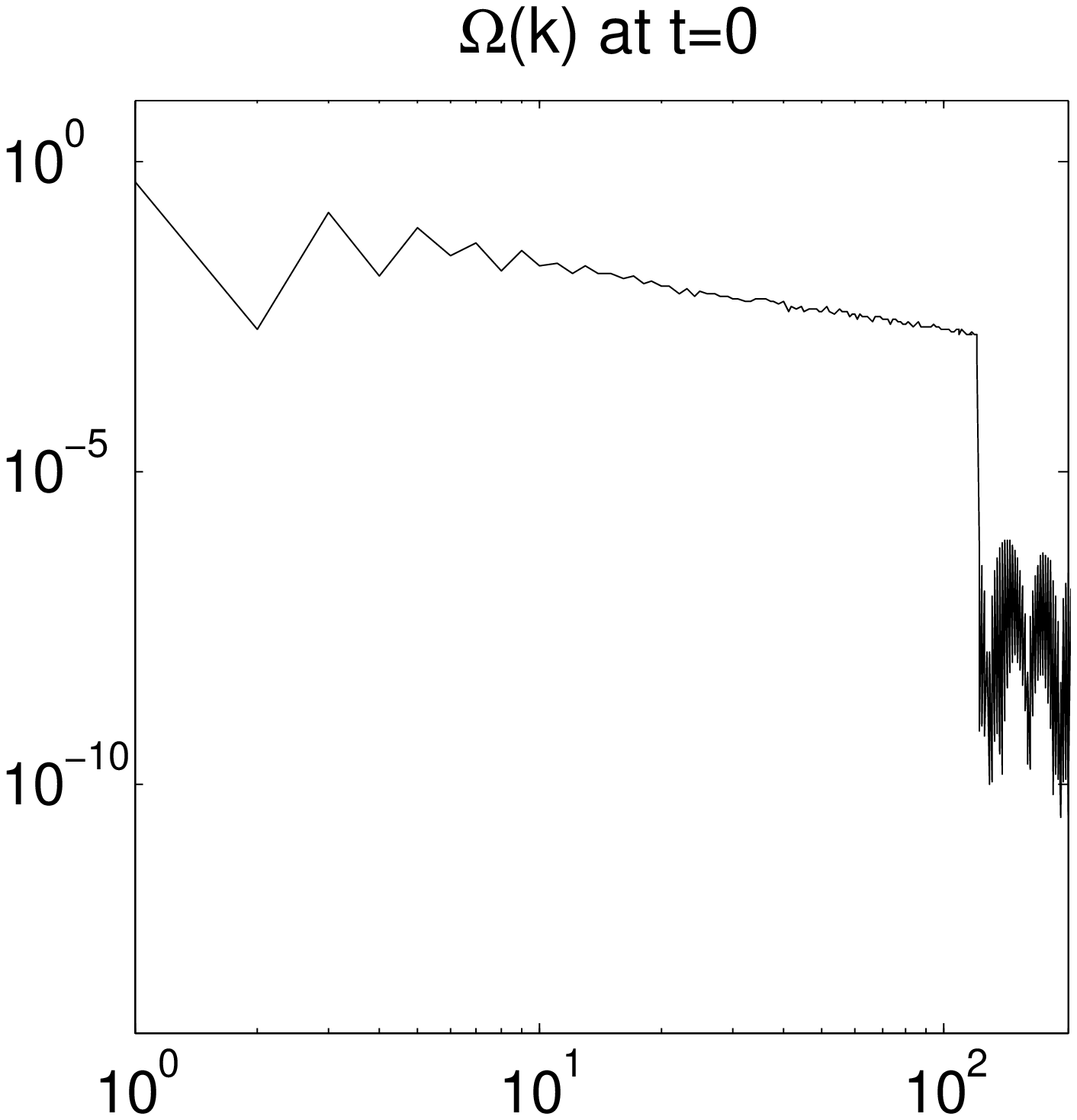}}
\end{minipage}
\begin{minipage}[c]{.323 \linewidth}
\scalebox{1}[1.]{\includegraphics[width=\linewidth]{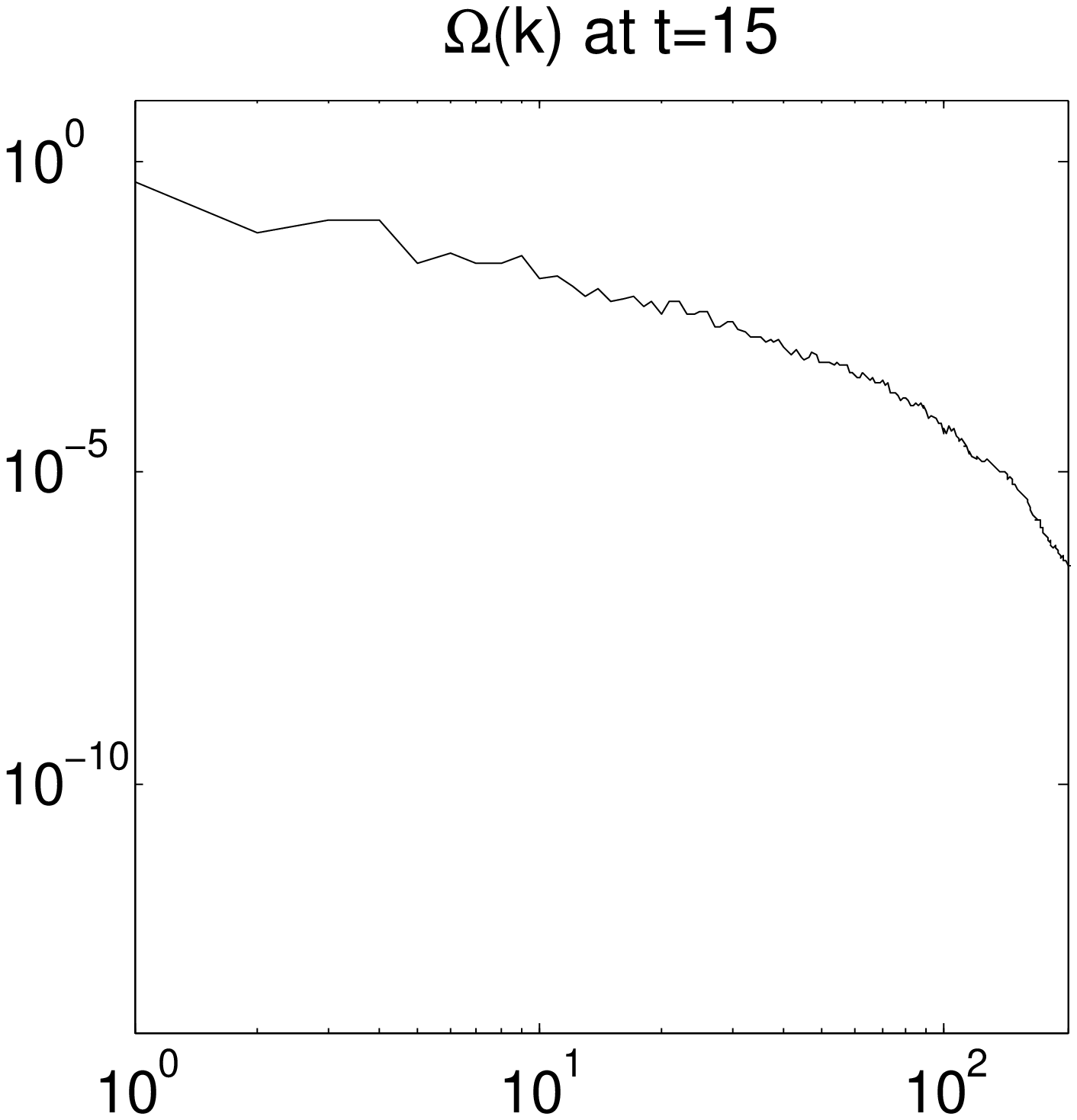}}
\end{minipage}
\begin{minipage}[c]{.323 \linewidth}
\scalebox{1}[1.]{\includegraphics[width=\linewidth]{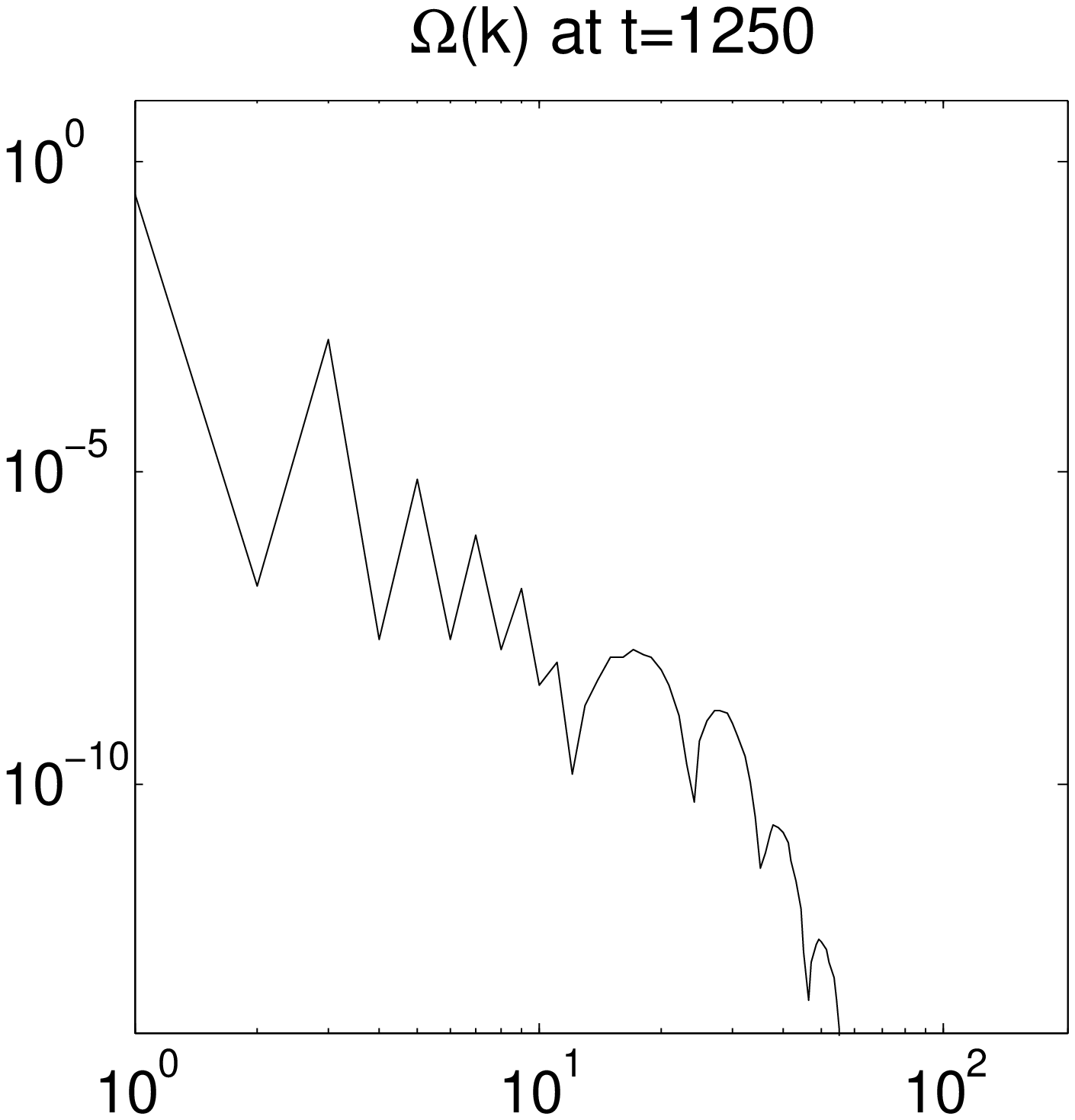}}
\end{minipage}

\caption{Angle-averaged modal energy and enstrophy  spectra at three different times for the run shown in Figs. \ref{fig:bar1}, \ref{fig:barst}, and \ref{fig:barmhhi}.  Note that these are not the usual omnidirectional spectra; since they do not contain the factor $2 \pi k$, their maxima occur at the minimum k.}
\label{fig:barvcd}
\end{figure}

\begin{figure}[!htbp]
\centering
\begin{minipage}[c]{.424 \linewidth}
\scalebox{1}[1.21]{\includegraphics[width=\linewidth]{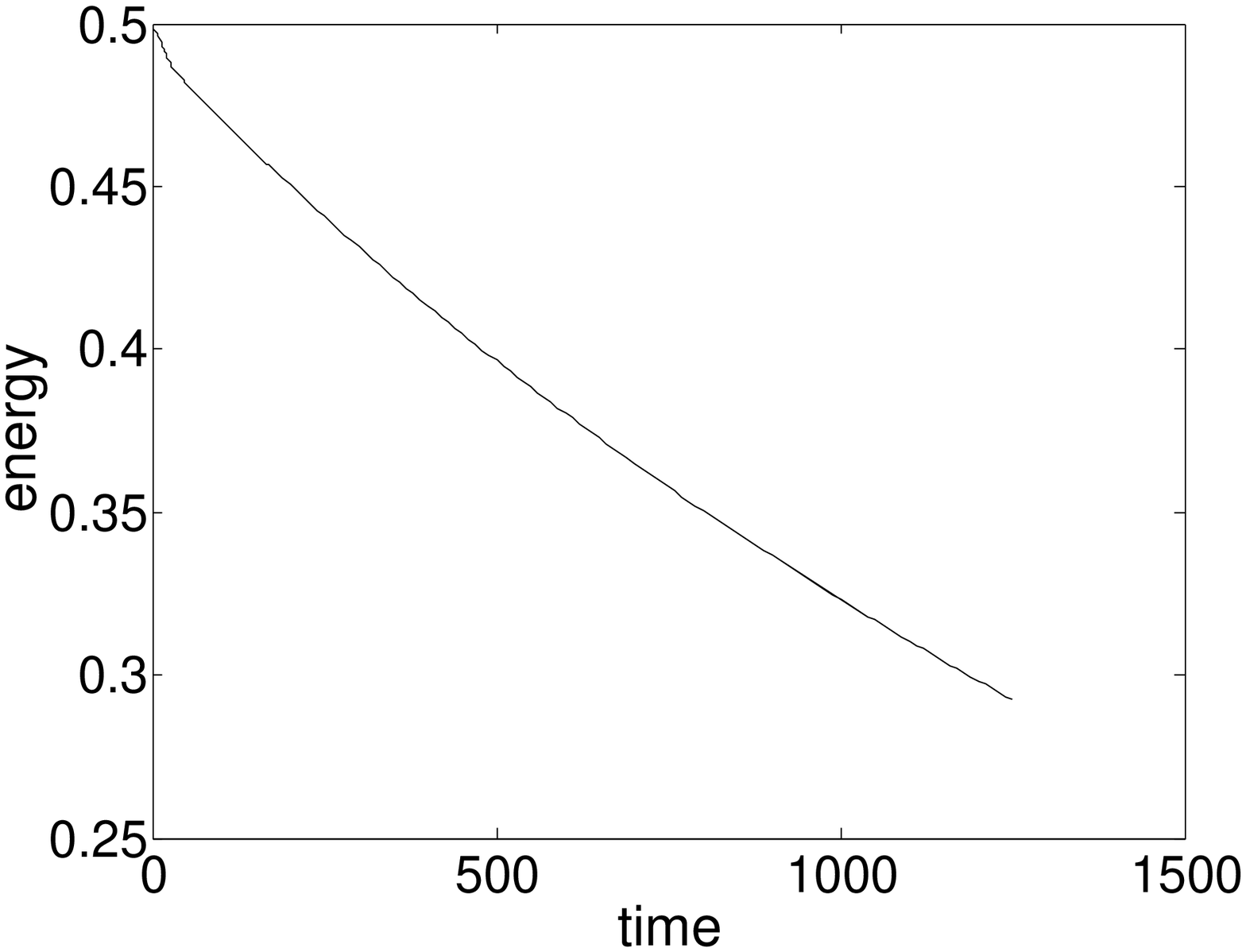}}
\end{minipage}
\begin{minipage}[c]{.424 \linewidth}
\scalebox{1}[1.21]{\includegraphics[width=\linewidth]{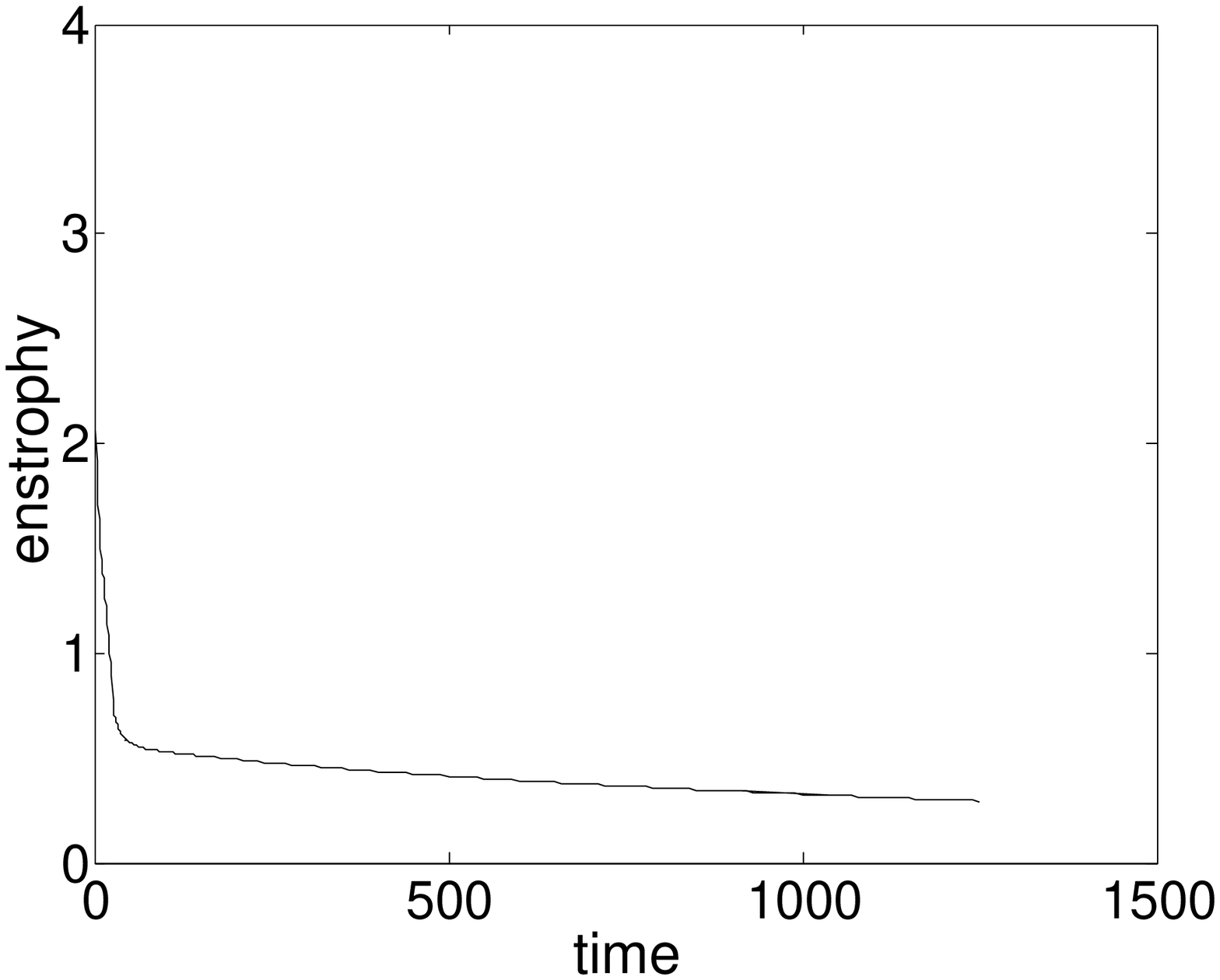}}
\end{minipage}
\begin{minipage}[c]{.424 \linewidth}
\scalebox{1}[1.21]{\includegraphics[width=\linewidth]{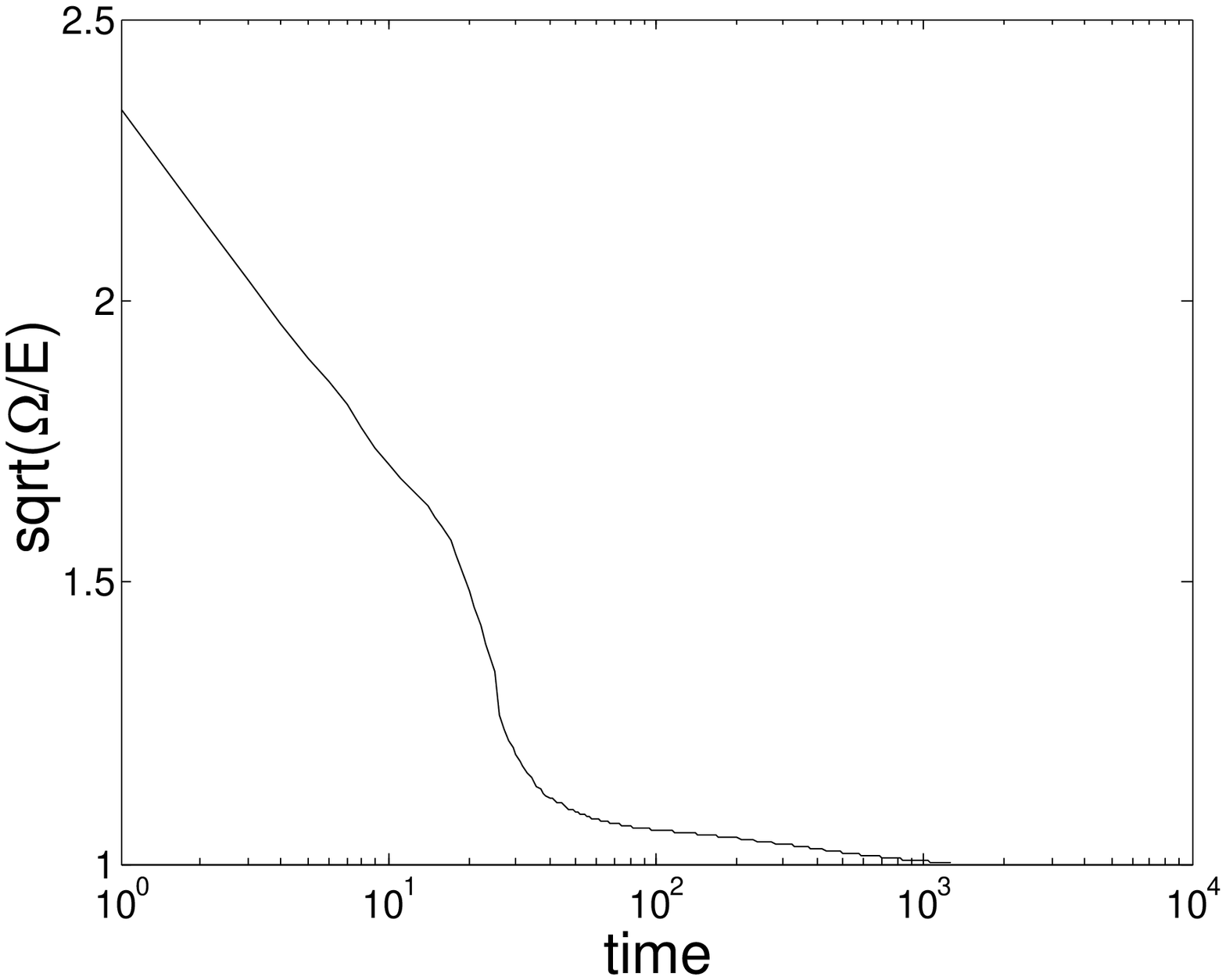}}
\end{minipage}
\begin{minipage}[c]{.424 \linewidth}
\scalebox{1}[1.21]{\includegraphics[width=\linewidth]{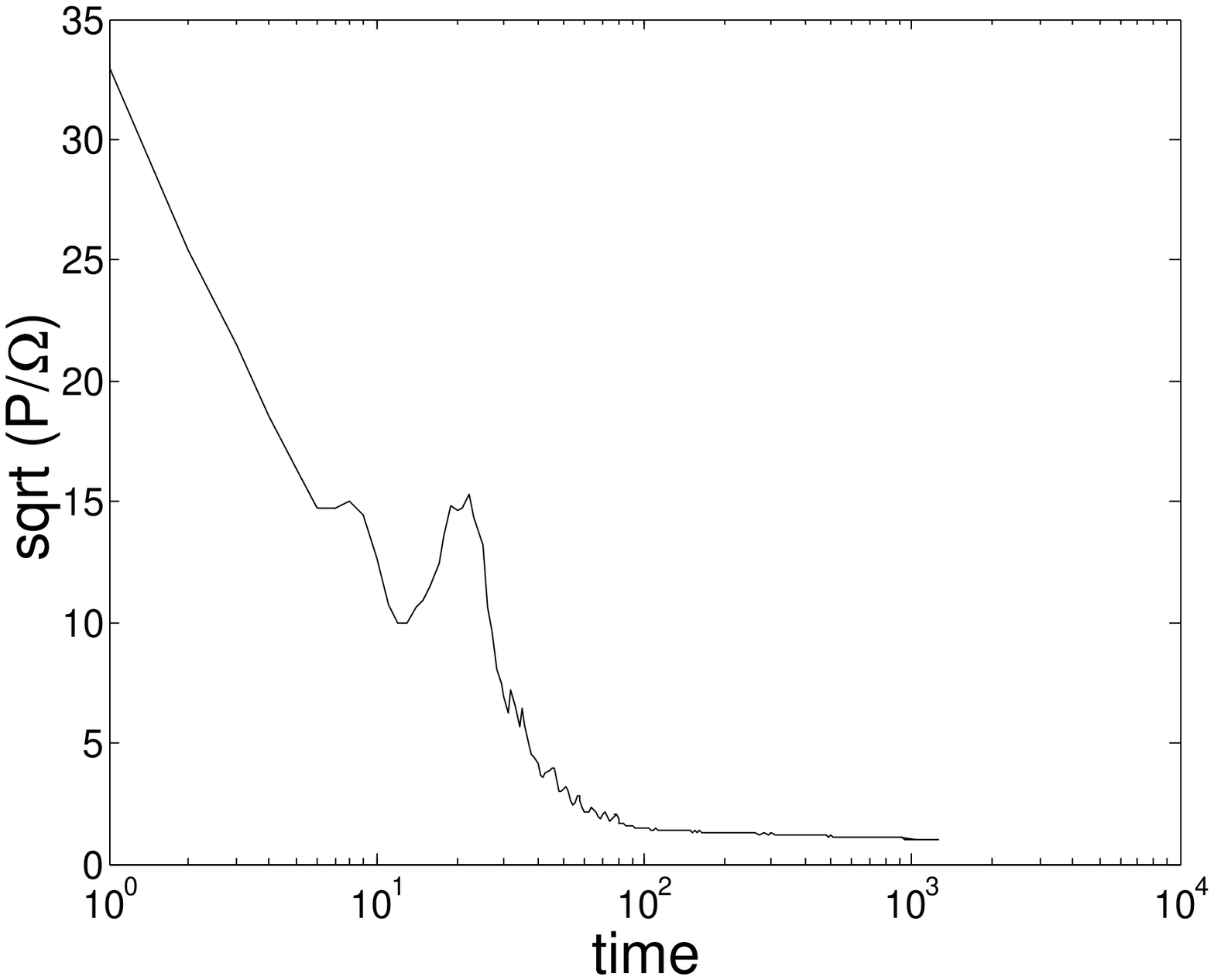}}
\end{minipage}
\caption{Time evolution, for the run shown in Figs. \ref{fig:banoise:b} through \ref{fig:barmhhi}, of the four global quantities energy, enstrophy, $\sqrt{\Omega /E}$ or mean wave number, and ratio of the square root of palinstrophy to enstrophy.  Note that the mean wave number (Taylor microscale) is never large, and so the strict definition of turbulence is not fulfilled at any time during the run. Nevertheless, the topology changes dramatically. This is effectively a selectively decayed state \cite{kn:q30}, and may not invite a probabilitistic explanation.}
\label{fig:barta}
\end{figure}

\begin{figure*}[!htbp]
\centering
\begin{minipage}[c]{.23489 \linewidth}
\scalebox{1}[1.1]{\includegraphics[width=\linewidth]{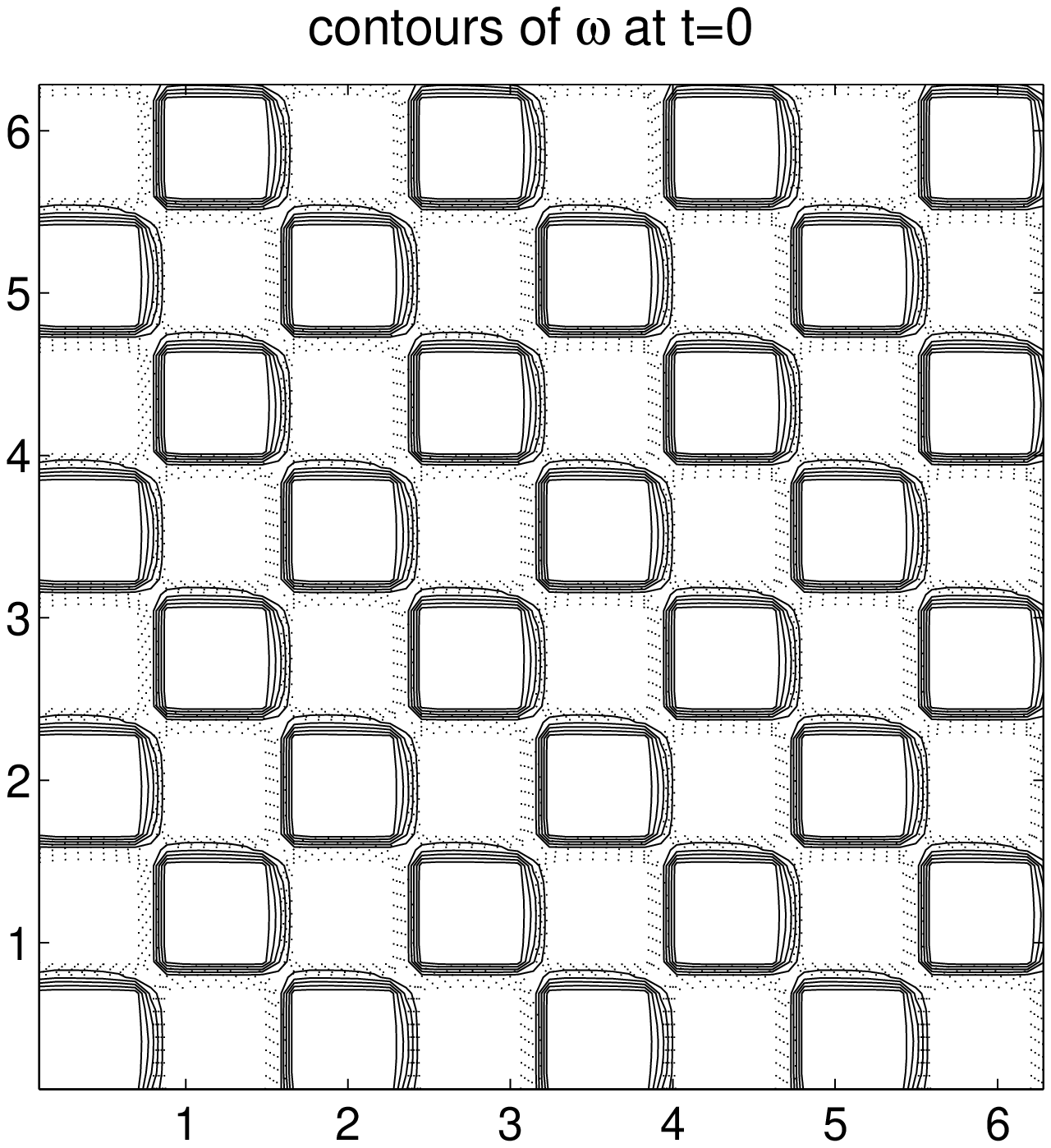}}
\end{minipage}
\begin{minipage}[c]{.23489 \linewidth}
\scalebox{1}[1.1]{\includegraphics[width=\linewidth]{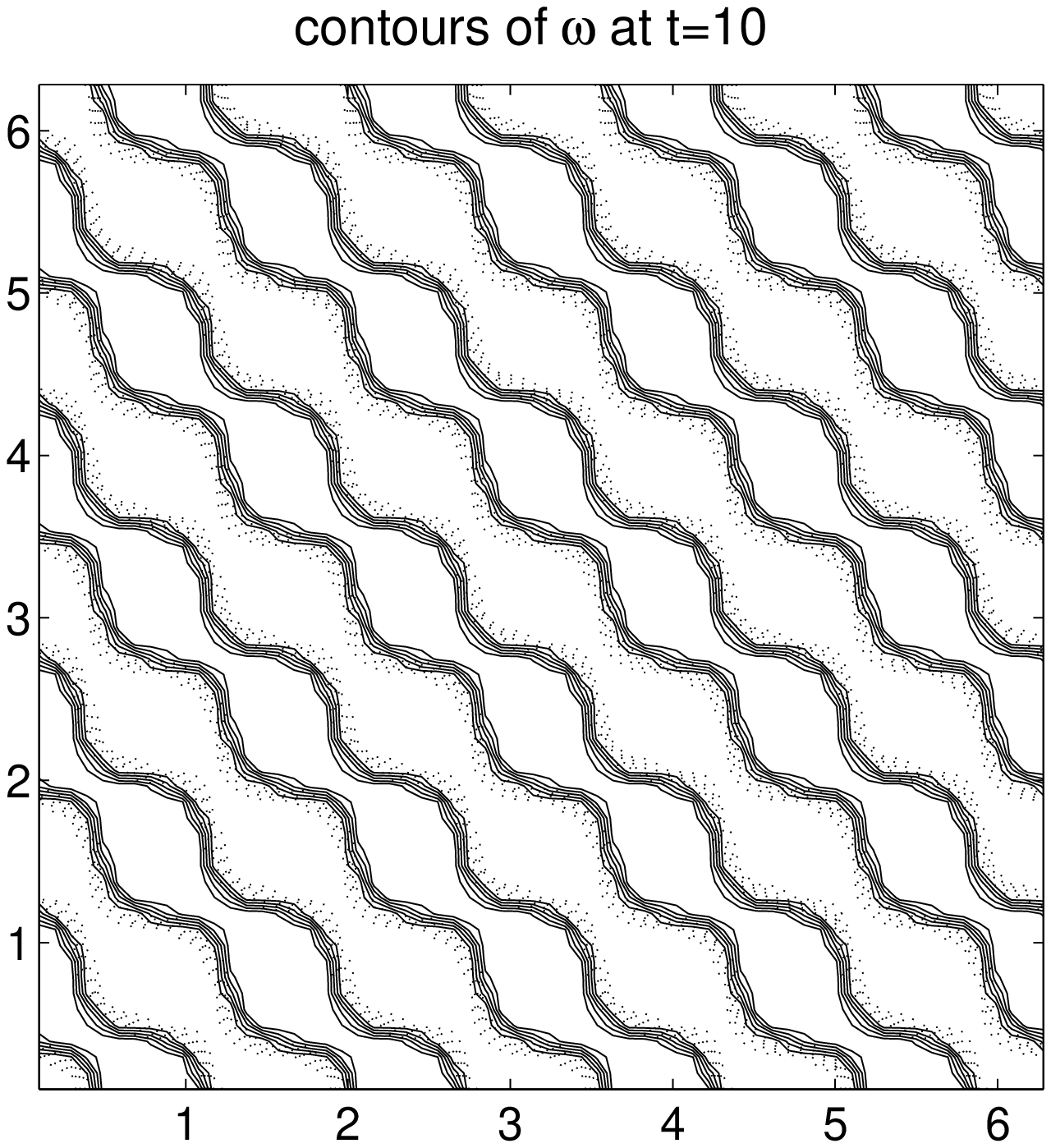}}
\end{minipage}
\begin{minipage}[c]{.23489 \linewidth}
\scalebox{1}[1.1]{\includegraphics[width=\linewidth]{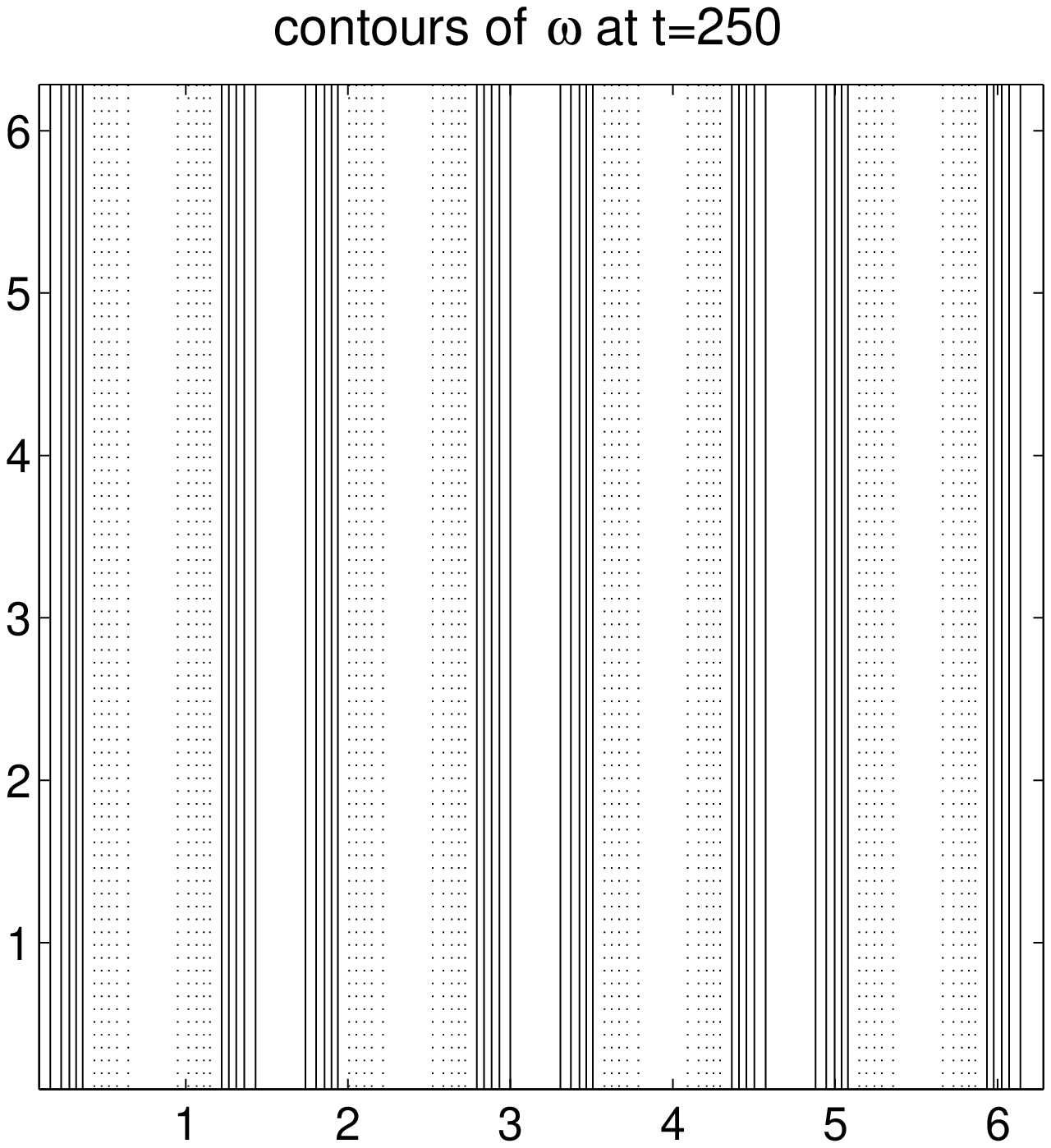}}
\end{minipage}
\begin{minipage}[c]{.23489 \linewidth}
\scalebox{1}[1.25]{\includegraphics[width=\linewidth]{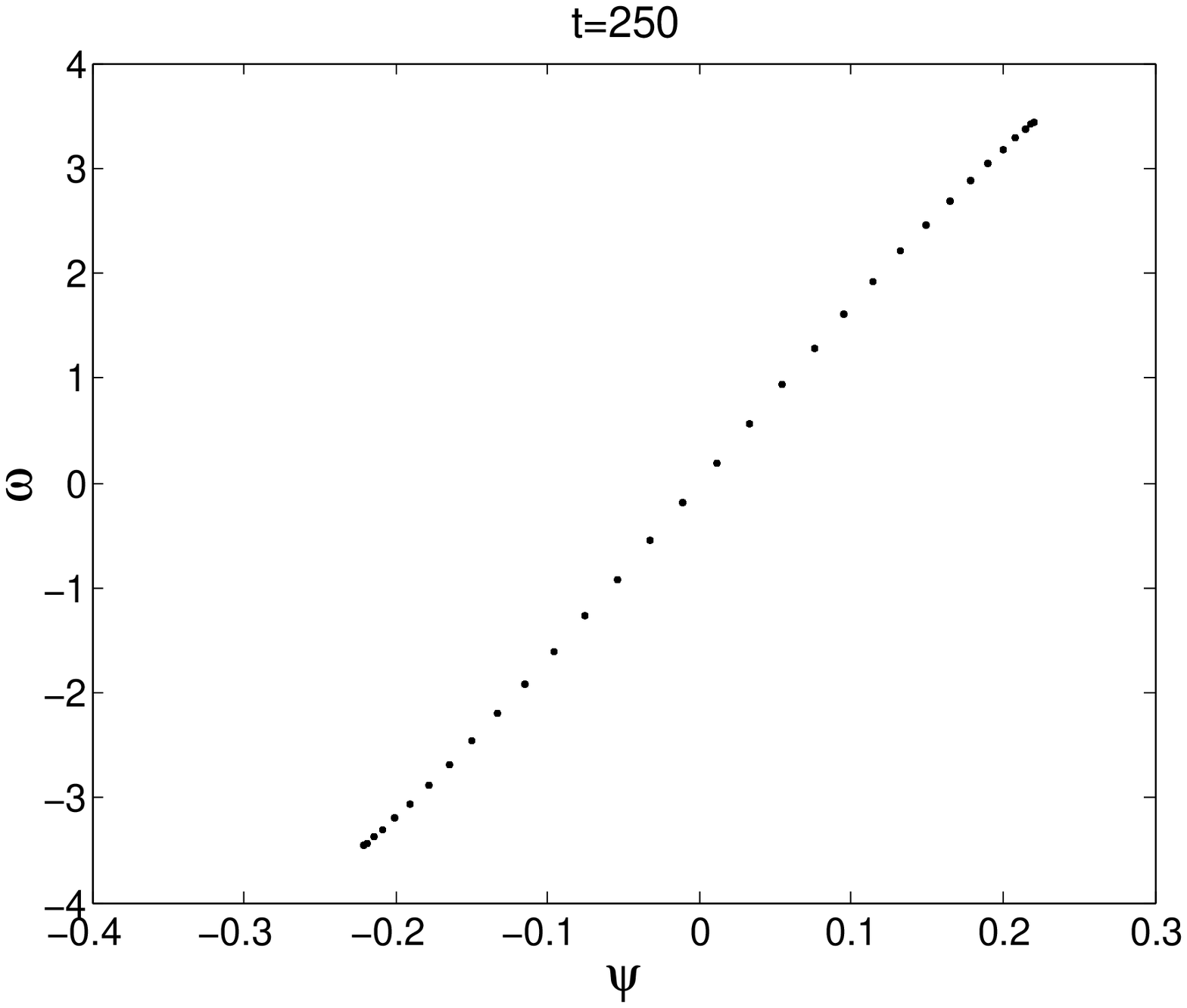}}
\end{minipage}
\caption{Time evolution of a 64-pole initial condition, triggered only by round-off error, into a bar state (the scatter plot of final state is close to Fig. \ref{fig:good:c}). No noise beyond round-off error has been added to the initial condition.}
\label{fig:64_0}
\end{figure*}

\begin{figure*}[!htbp]
\centering
\begin{minipage}[c]{.23489 \linewidth}
\scalebox{1}[1.1]{\includegraphics[width=\linewidth]{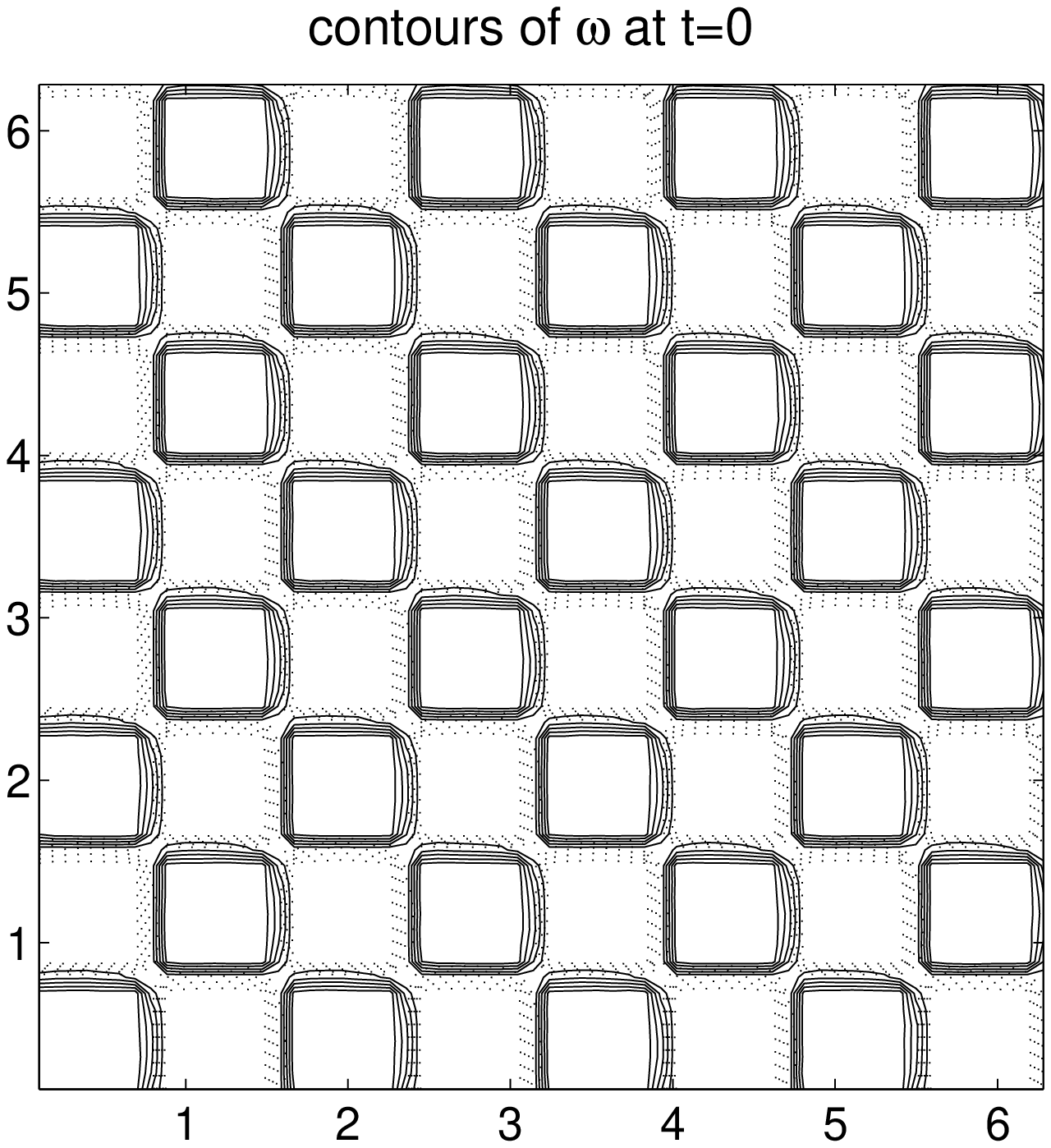}}
\end{minipage}
\begin{minipage}[c]{.23489 \linewidth}
\scalebox{1}[1.1]{\includegraphics[width=\linewidth]{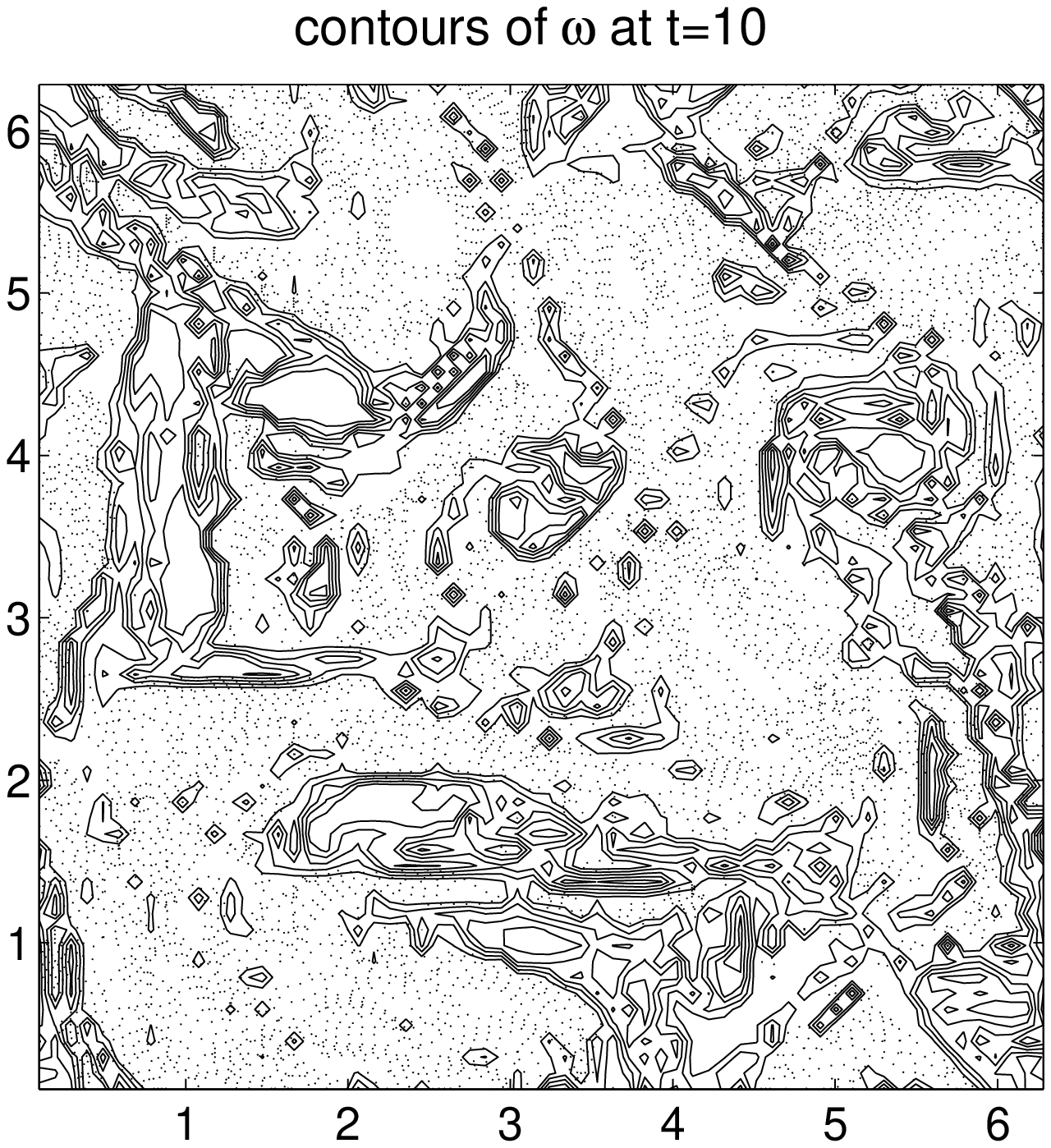}}
\end{minipage}
\begin{minipage}[c]{.23489 \linewidth}
\scalebox{1}[1.1]{\includegraphics[width=\linewidth]{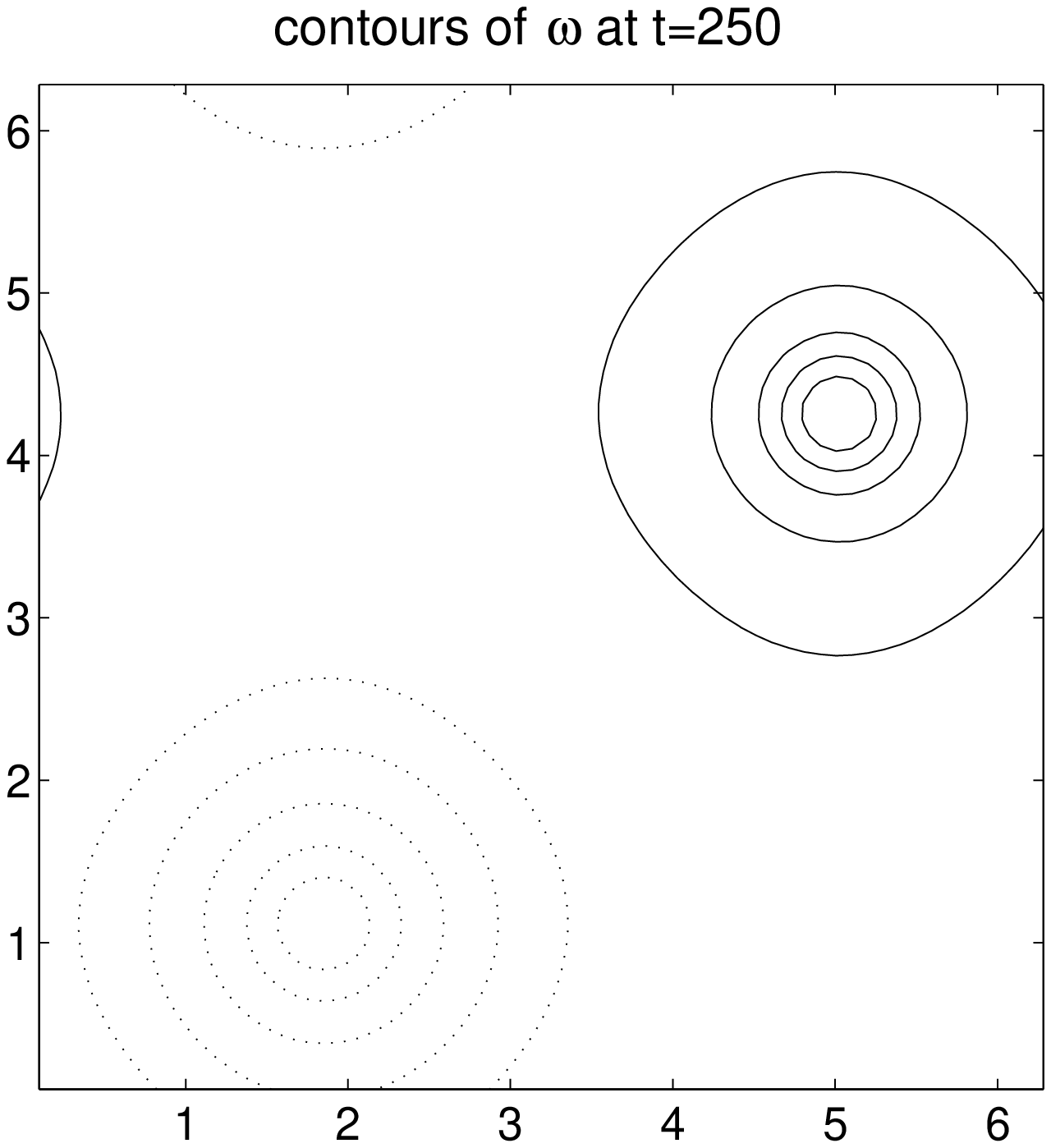}}
\end{minipage}
\begin{minipage}[c]{.23489 \linewidth}
\scalebox{1}[1.25]{\includegraphics[width=\linewidth]{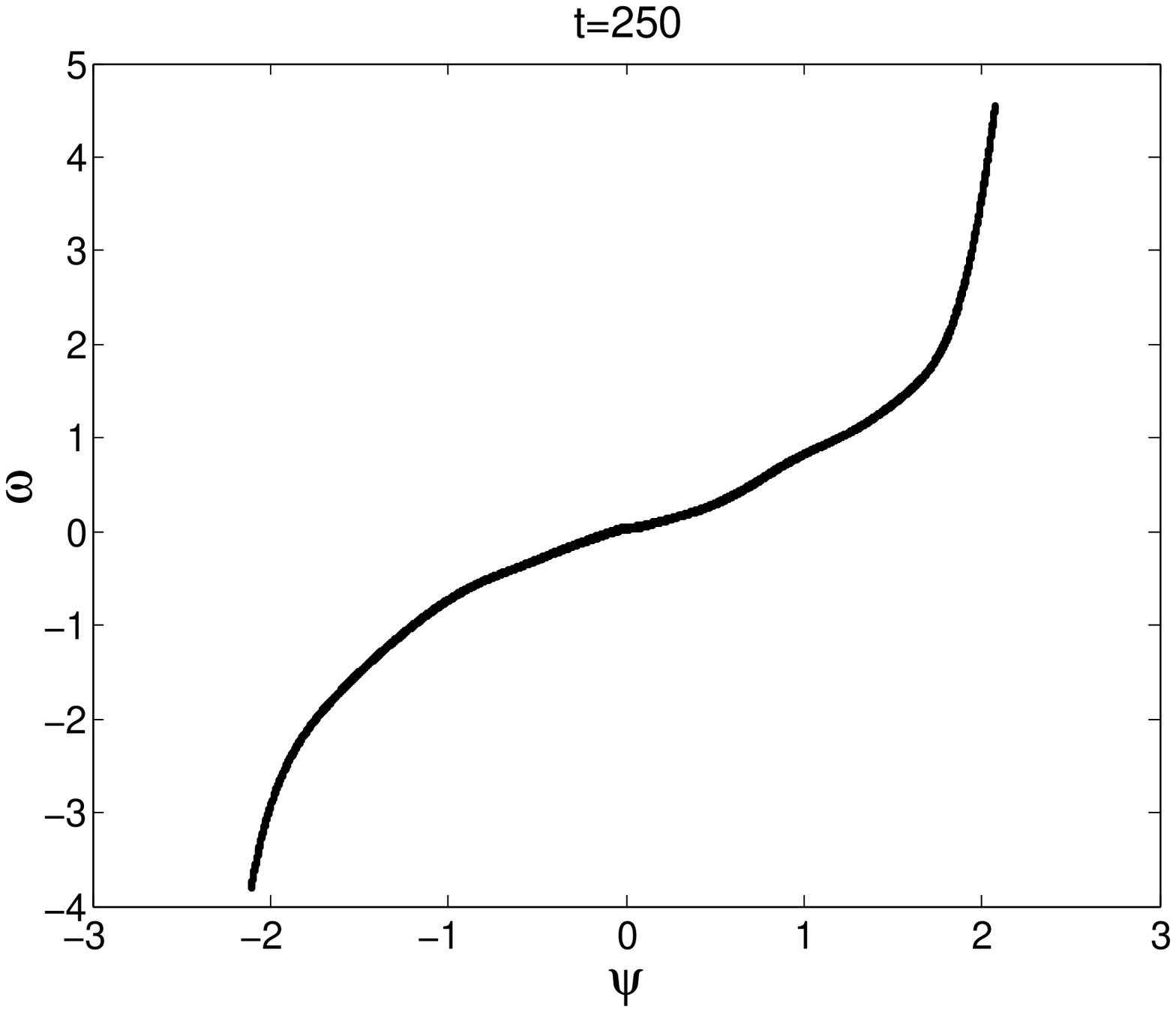}}
\end{minipage}
\caption{Time evolution of the vorticity contours, starting from the same initial condition as in Fig. \ref{fig:64_0}, but with a healthy addition of random noise (In this run, $R_{\lambda}$ increases from $2036$ initially to $11400$ at the end.). The last panel is the late-time $\omega - \psi$ scatter plot for the resulting dipole (which is close to Fig. \ref{fig:good:a}).}
\label{fig:64_noise}
\end{figure*}

\begin{figure*}[!htbp]
\centering
\begin{minipage}[c]{.23489 \linewidth}
\scalebox{1}[1.1]{\includegraphics[width=\linewidth]{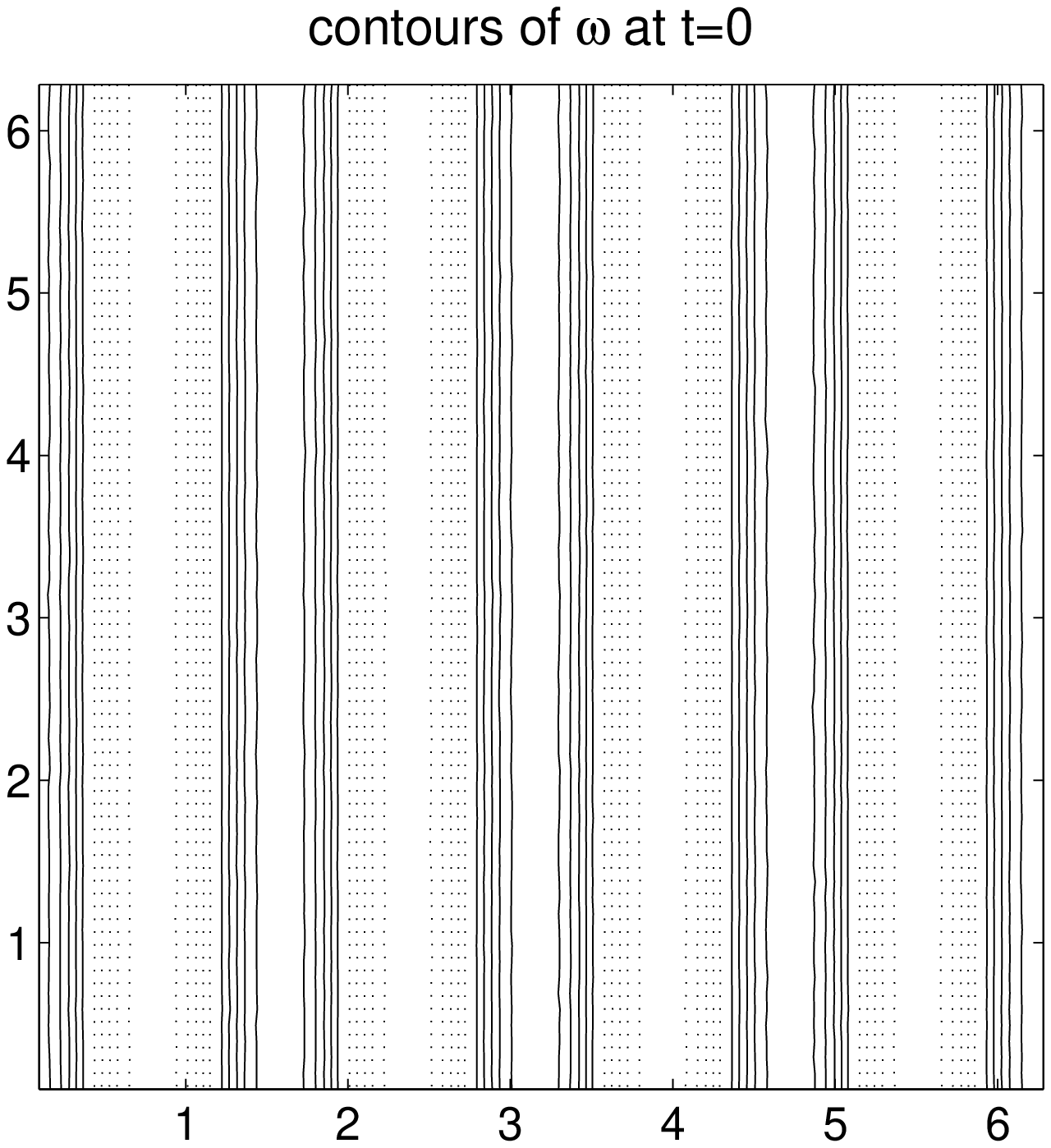}}
\end{minipage}
\begin{minipage}[c]{.23489 \linewidth}
\scalebox{1}[1.1]{\includegraphics[width=\linewidth]{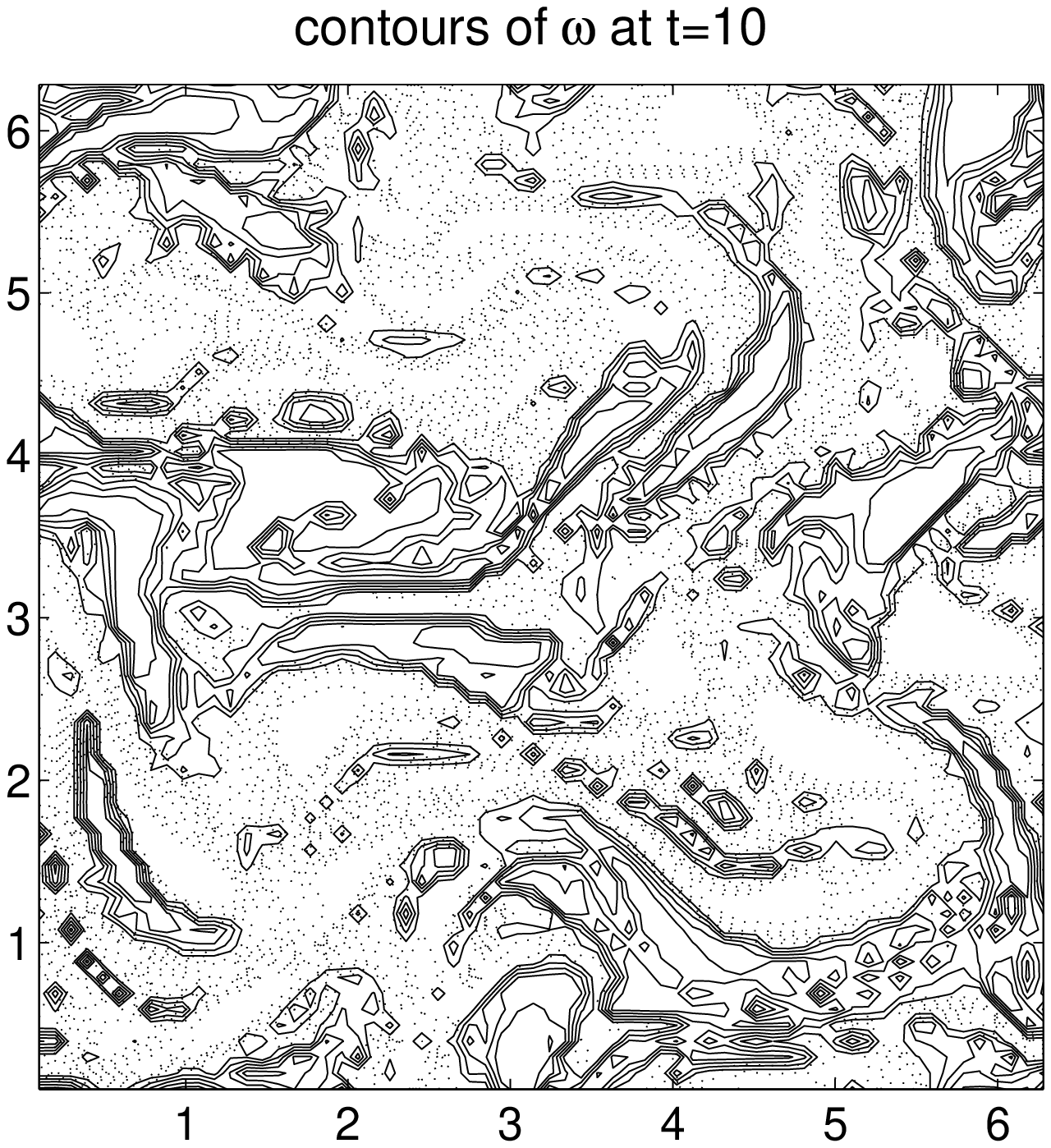}}
\end{minipage}
\begin{minipage}[c]{.23489 \linewidth}
\scalebox{1}[1.1]{\includegraphics[width=\linewidth]{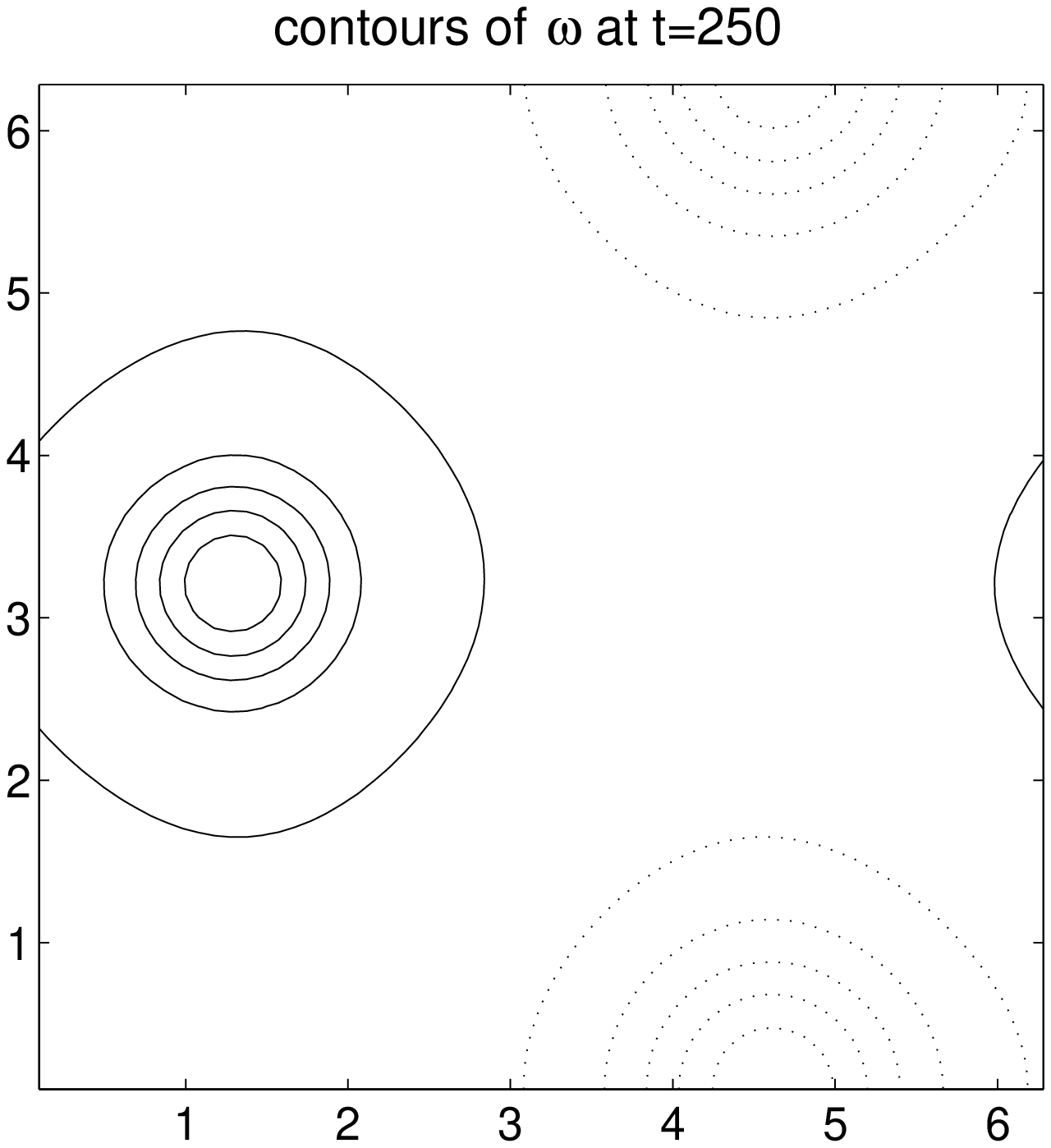}}
\end{minipage}
\begin{minipage}[c]{.23489 \linewidth}
\scalebox{1}[1.25]{\includegraphics[width=\linewidth]{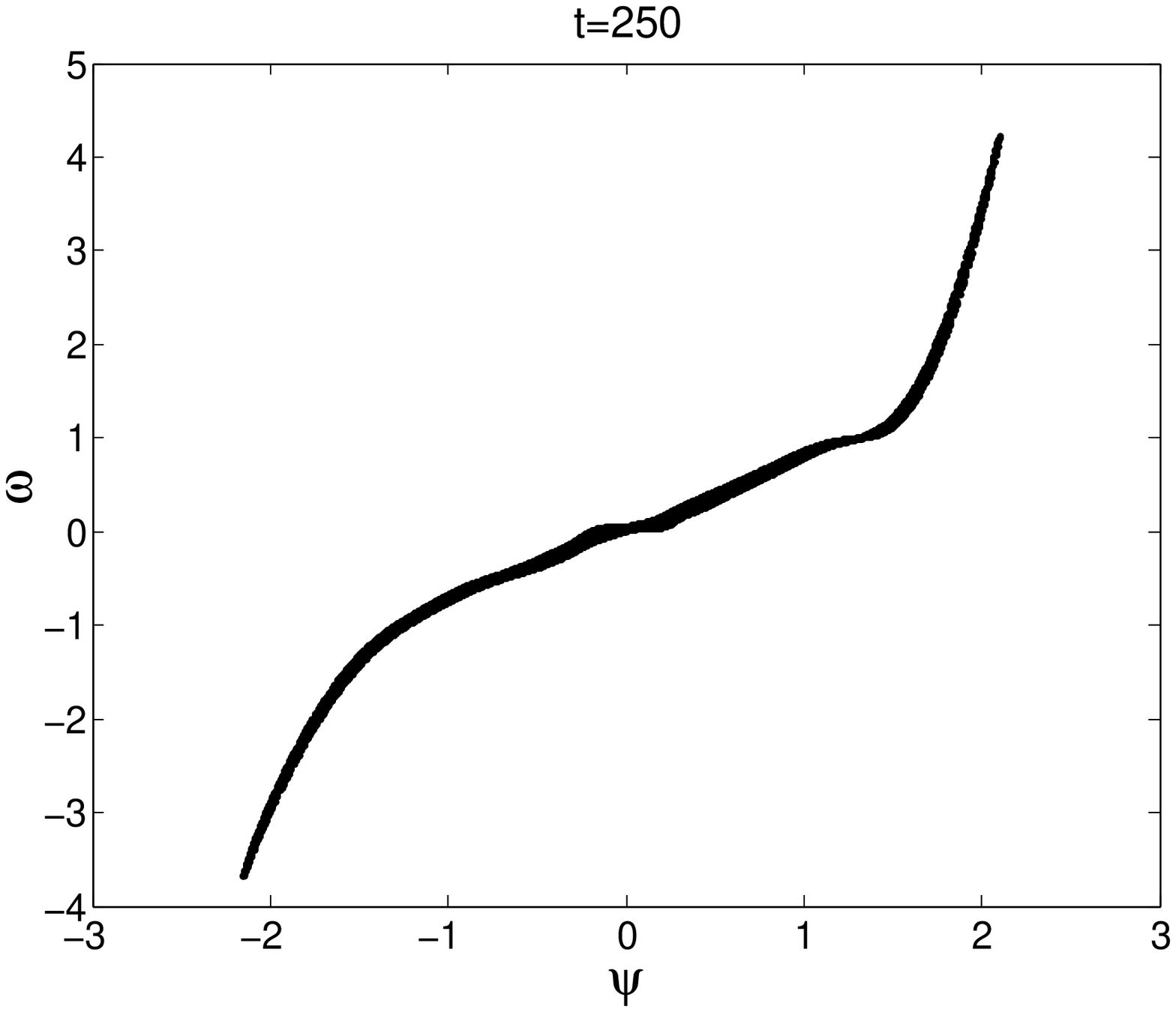}}
\end{minipage}
\caption{Evolution of the one-dimensional ``8-bar'' initial condition, with random noise added initially (In this run, $R_{\lambda}$ increases from $3228$ initially to $11500$ at the end.). The evolution is toward a dipole, now, as seen in the bottom panel (which is close to Fig. \ref{fig:good:a}).}
\label{fig:8_noise}
\end{figure*}

\begin{figure*}[!htbp]
\centering
\begin{minipage}[c]{.23 \linewidth}
\scalebox{1}[1.]{\includegraphics[width=\linewidth]{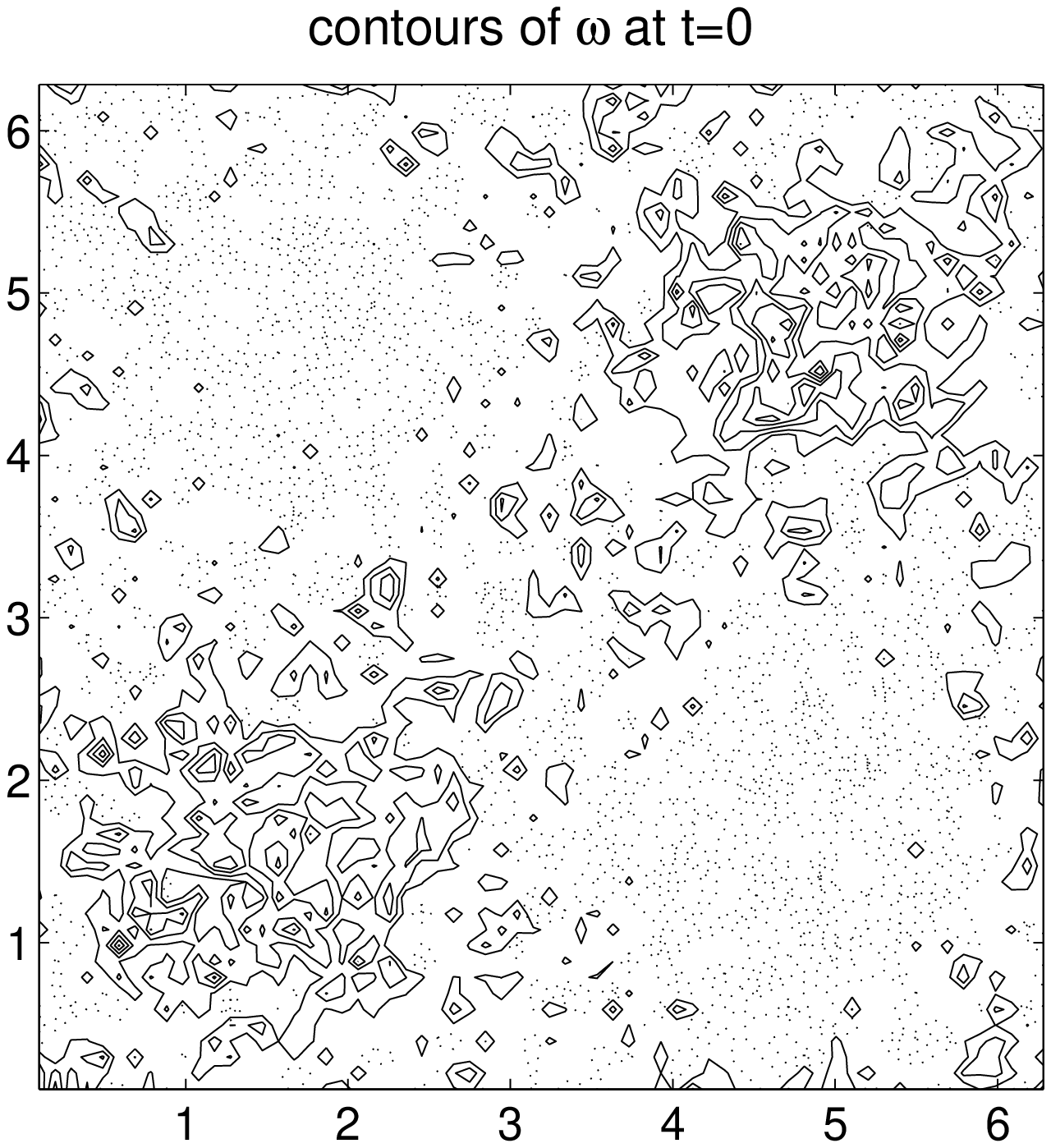}}
\end{minipage}
\begin{minipage}[c]{.23 \linewidth}
\scalebox{1}[1.]{\includegraphics[width=\linewidth]{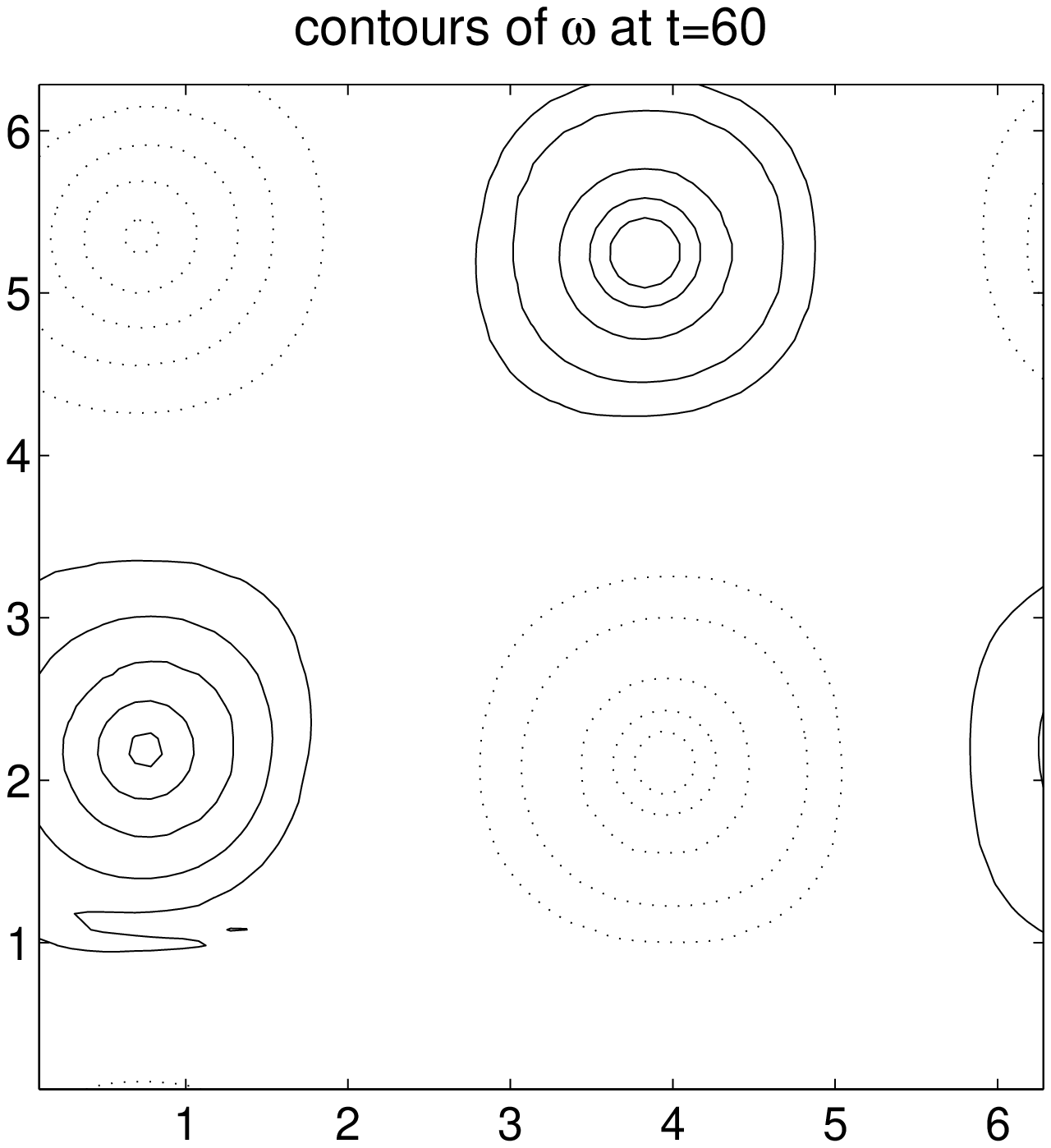}}
\end{minipage}
\begin{minipage}[c]{.23 \linewidth}
\scalebox{1}[1.]{\includegraphics[width=\linewidth]{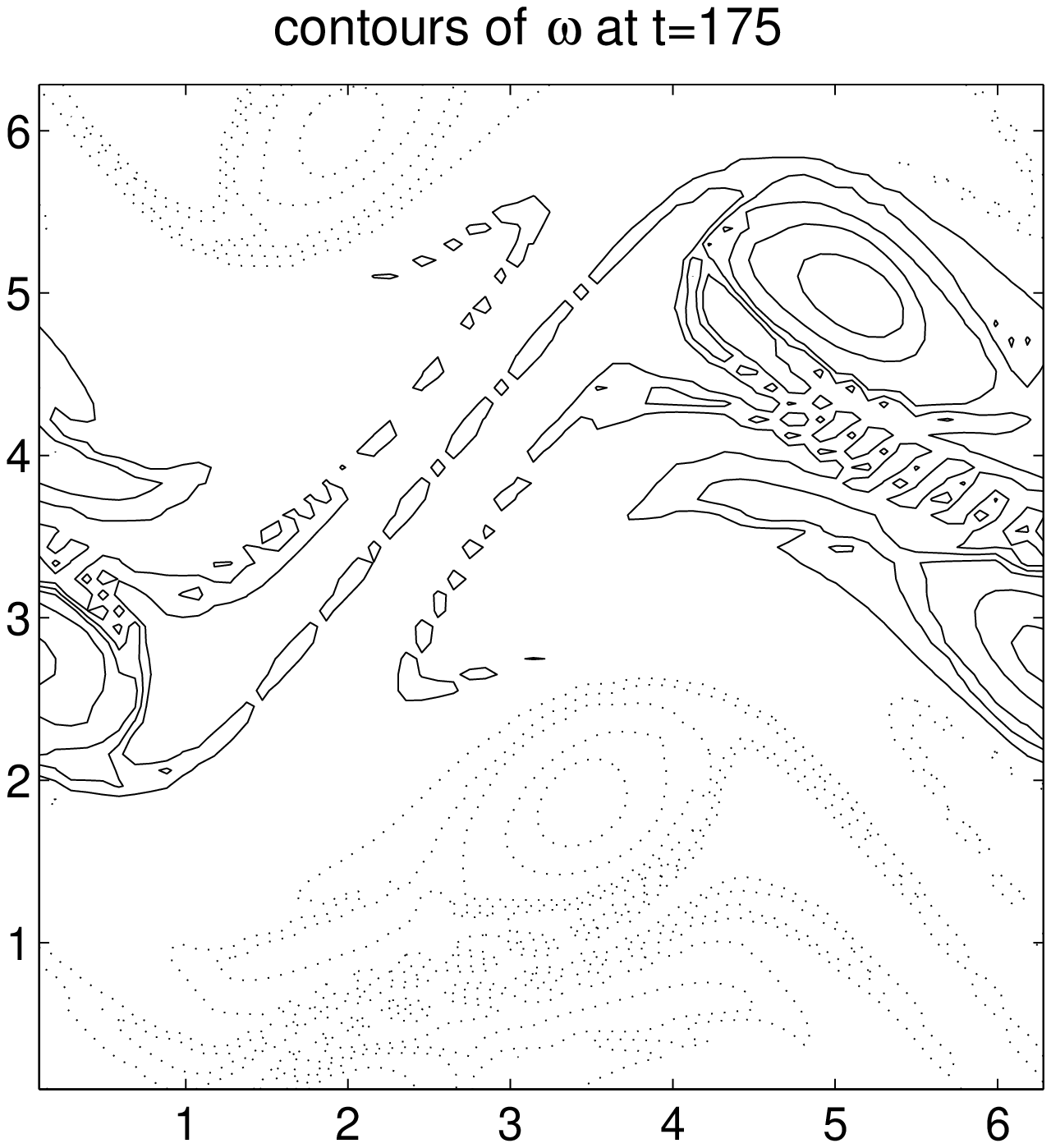}}
\end{minipage}
\begin{minipage}[c]{.23 \linewidth}
\scalebox{1}[1.]{\includegraphics[width=\linewidth]{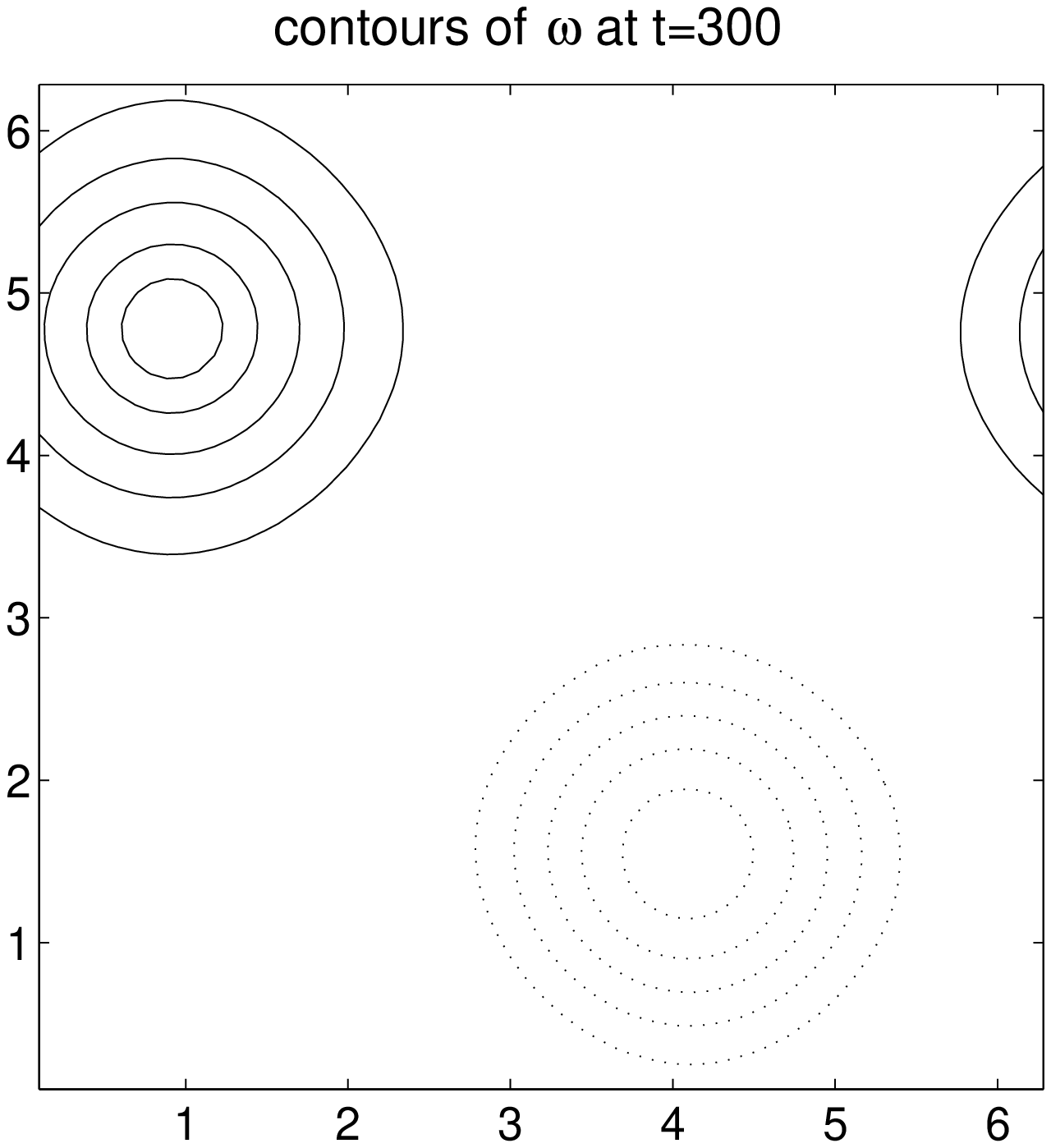}}
\end{minipage}
\begin{minipage}[c]{.23 \linewidth}
\scalebox{1}[1.]{\includegraphics[width=\linewidth]{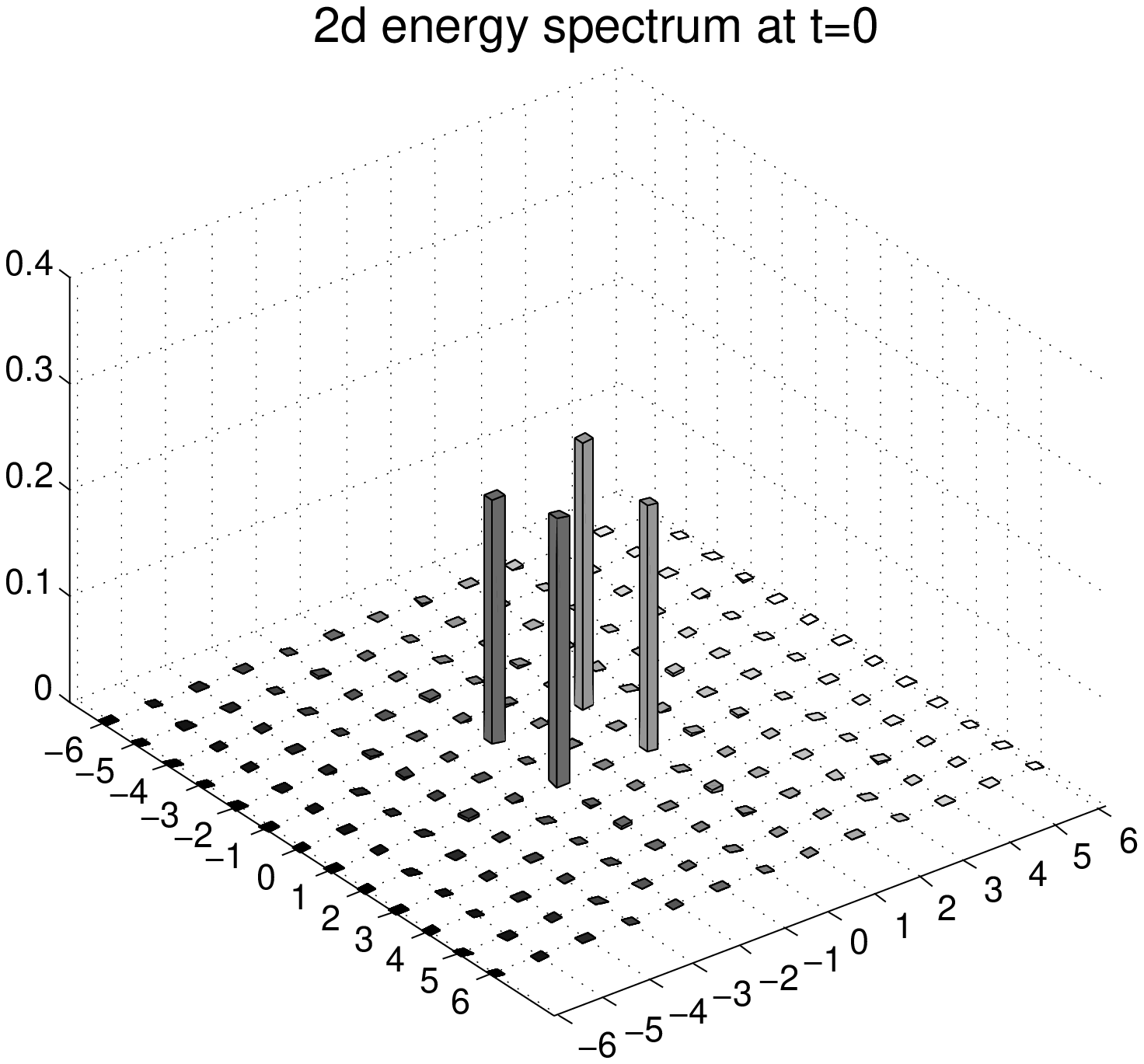}}
\end{minipage}
\begin{minipage}[c]{.23 \linewidth}
\scalebox{1}[1.]{\includegraphics[width=\linewidth]{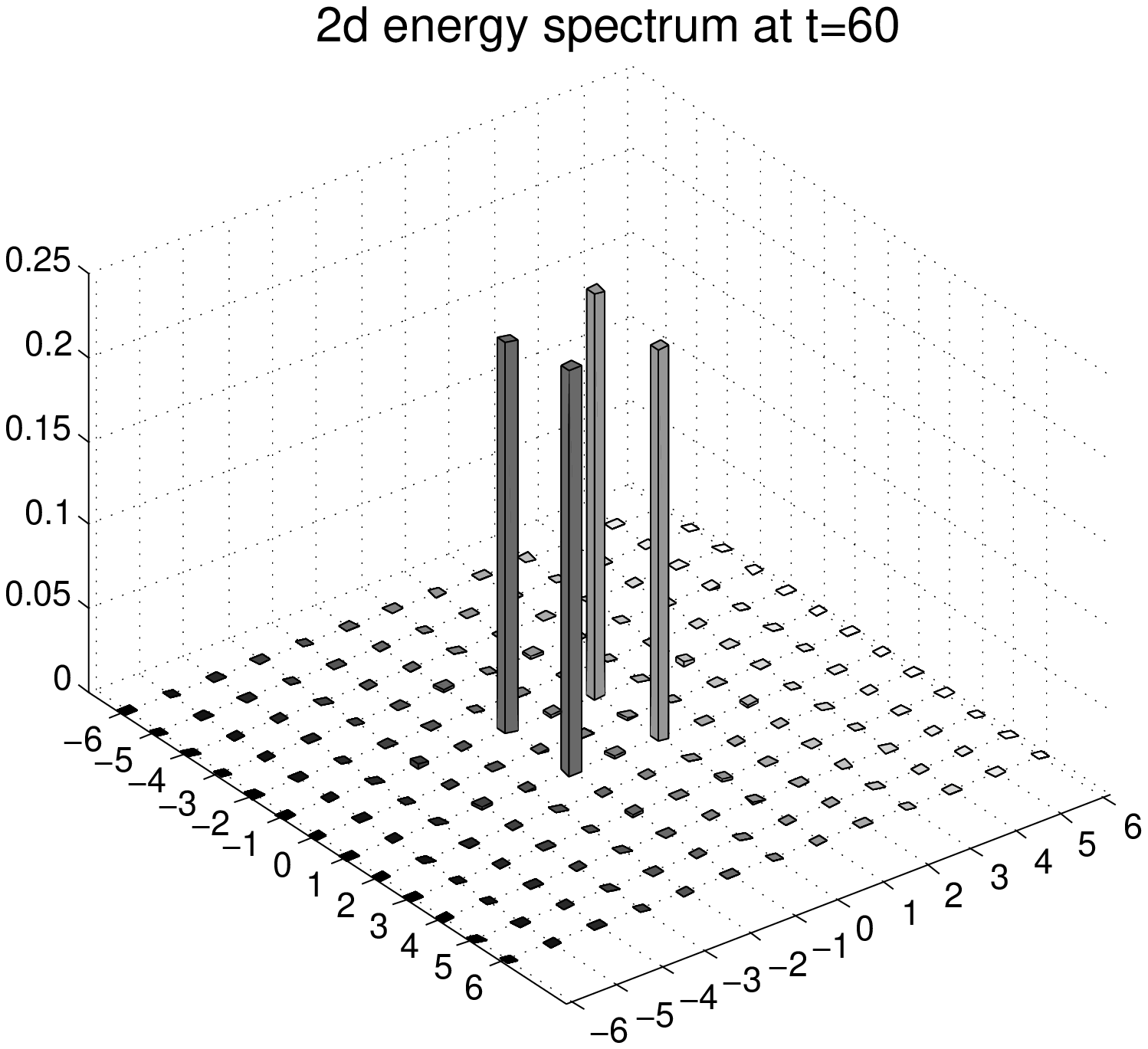}}
\end{minipage}
\begin{minipage}[c]{.23 \linewidth}
\scalebox{1}[1.]{\includegraphics[width=\linewidth]{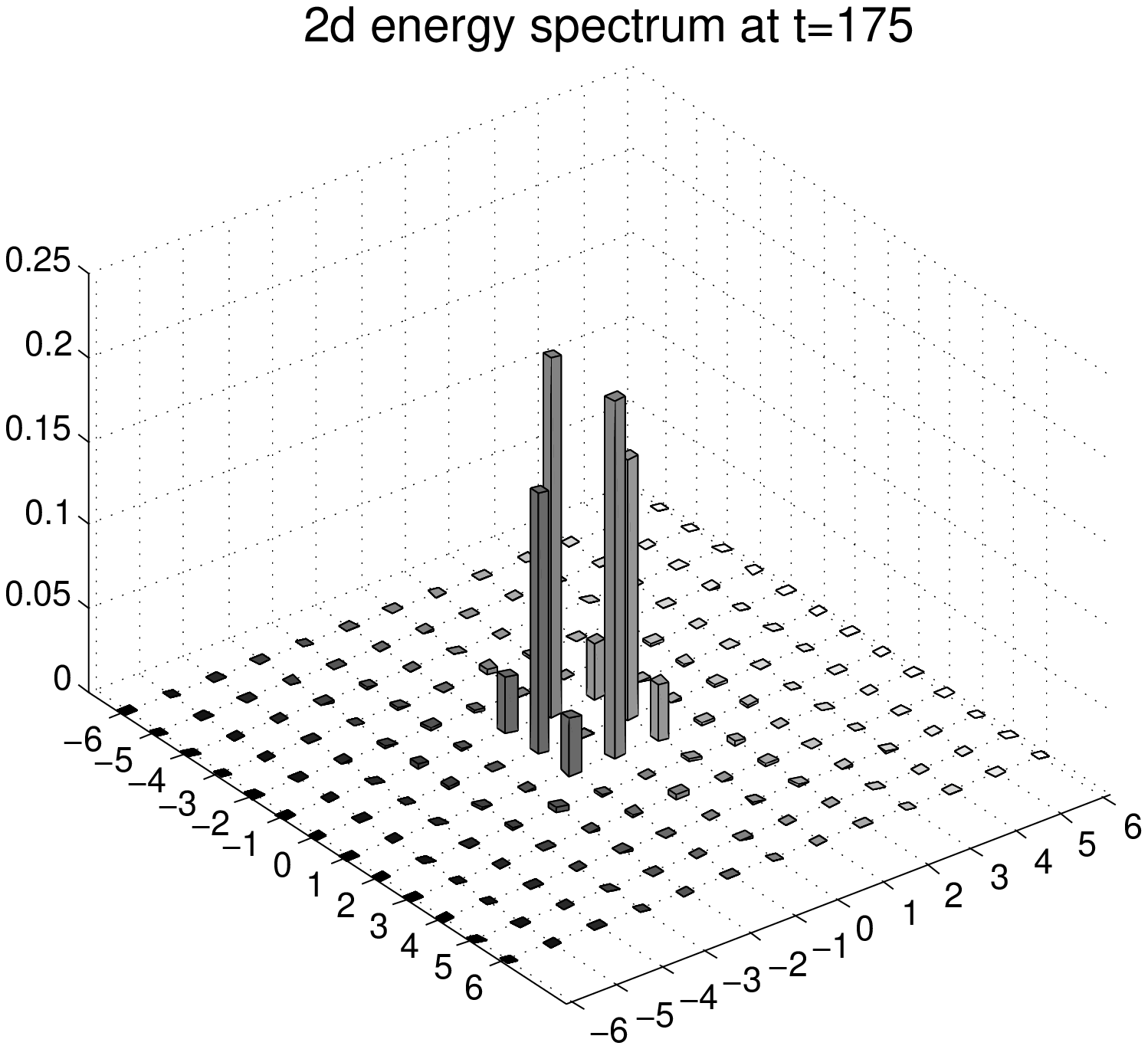}}
\end{minipage}
\begin{minipage}[c]{.23 \linewidth}
\scalebox{1}[1.]{\includegraphics[width=\linewidth]{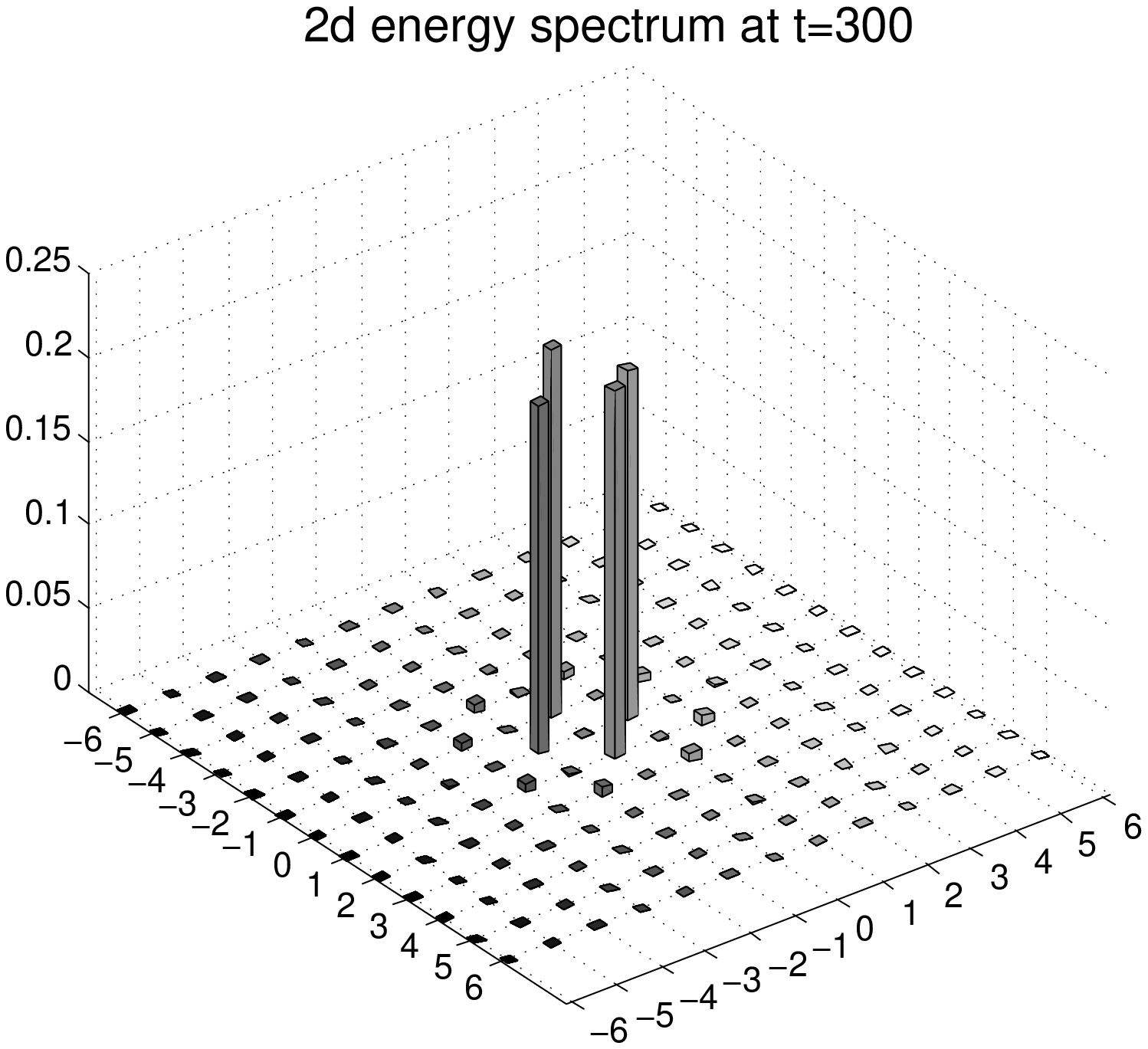}}
\end{minipage}
\caption{Evolution from initial conditions corresponding to a sinh-Poisson quadrupole, plus random noise. Note the differences in evolution from those shown in Figs. \ref{fig:banoise:b} - \ref{fig:barta}, whose initial conditions are superficially similar. The low-k part of the Fourier space contains most of the energy throughout, so the evolution again is not classically ``turbulent.''}
\label{fig:4_noise1}
\end{figure*}

\begin{figure*}[!htbp]
\centering
\begin{minipage}[c]{.26 \linewidth}
\scalebox{1}[0.95]{\includegraphics[width=\linewidth]{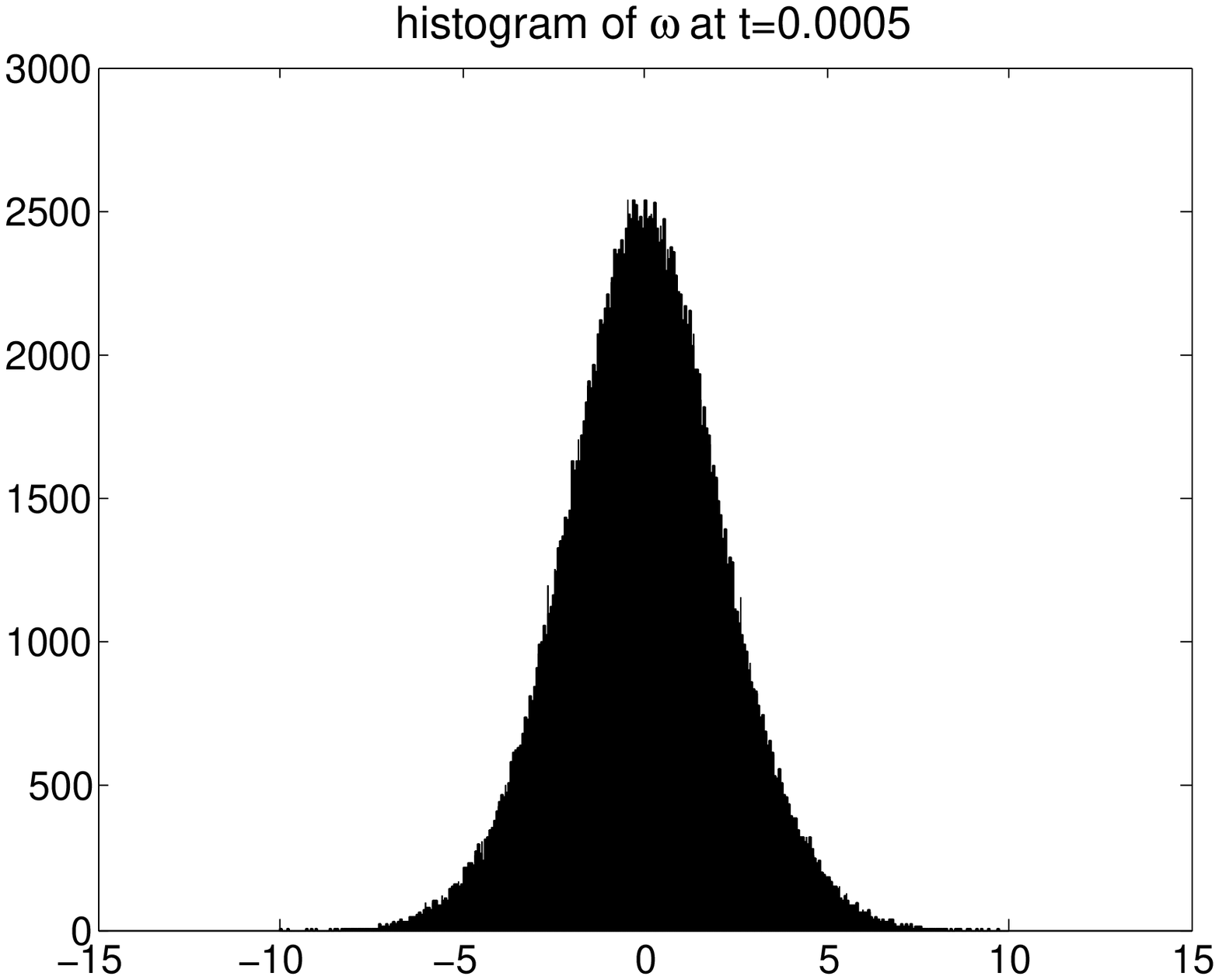}}
\end{minipage}
\begin{minipage}[c]{.26 \linewidth}
\scalebox{1}[0.95]{\includegraphics[width=\linewidth]{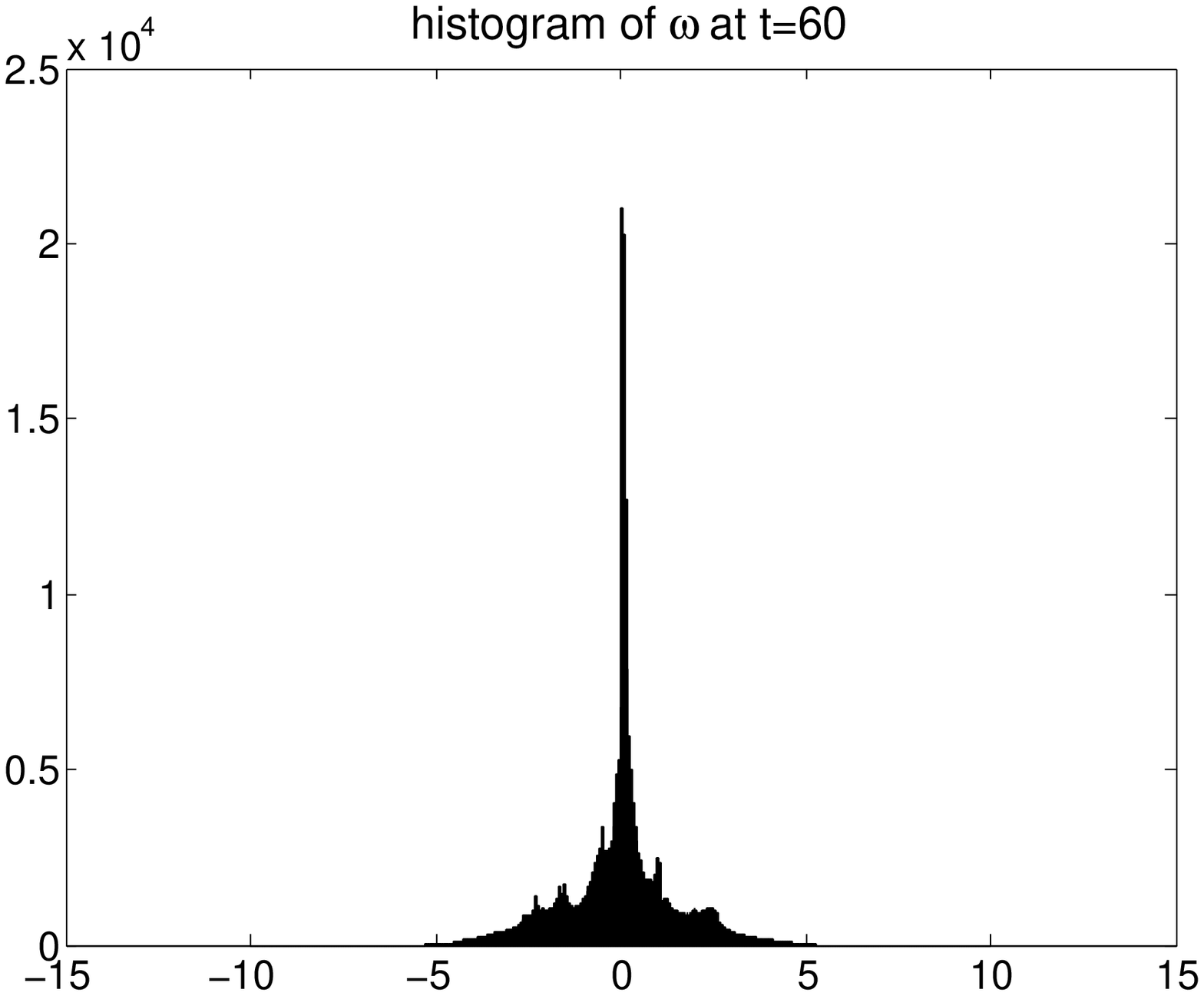}}
\end{minipage}
\begin{minipage}[c]{.26 \linewidth}
\scalebox{1}[0.95]{\includegraphics[width=\linewidth]{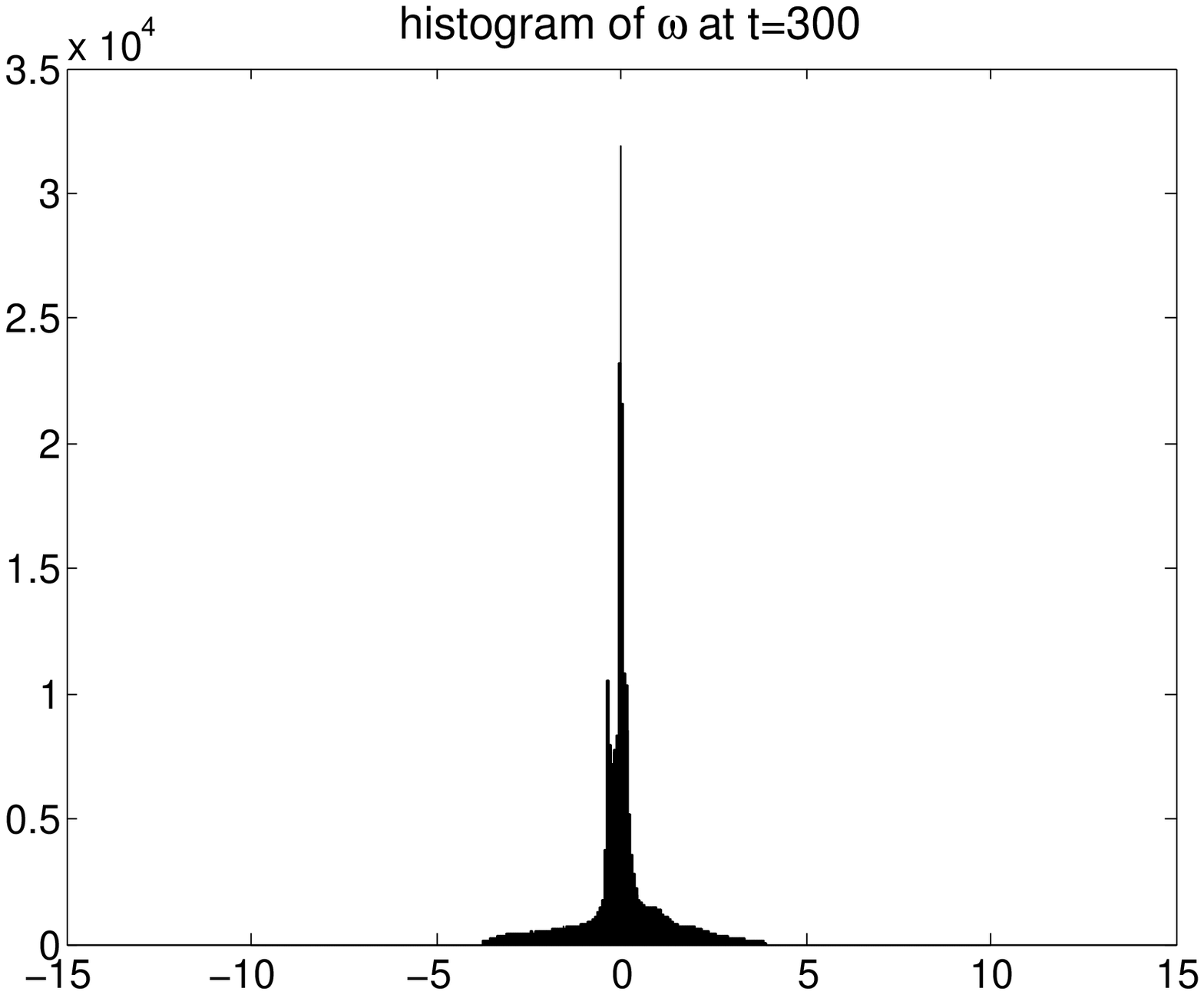}}
\end{minipage}
\caption{Vorticity histogram at three different times for the evolution shown in Figs. \ref{fig:4_noise1}.}
\label{fig:4_noise2}
\end{figure*}

\begin{figure}[!htbp]
\centering
\begin{minipage}[c]{.63 \linewidth}
\scalebox{1}[0.7]{\includegraphics[width=\linewidth]{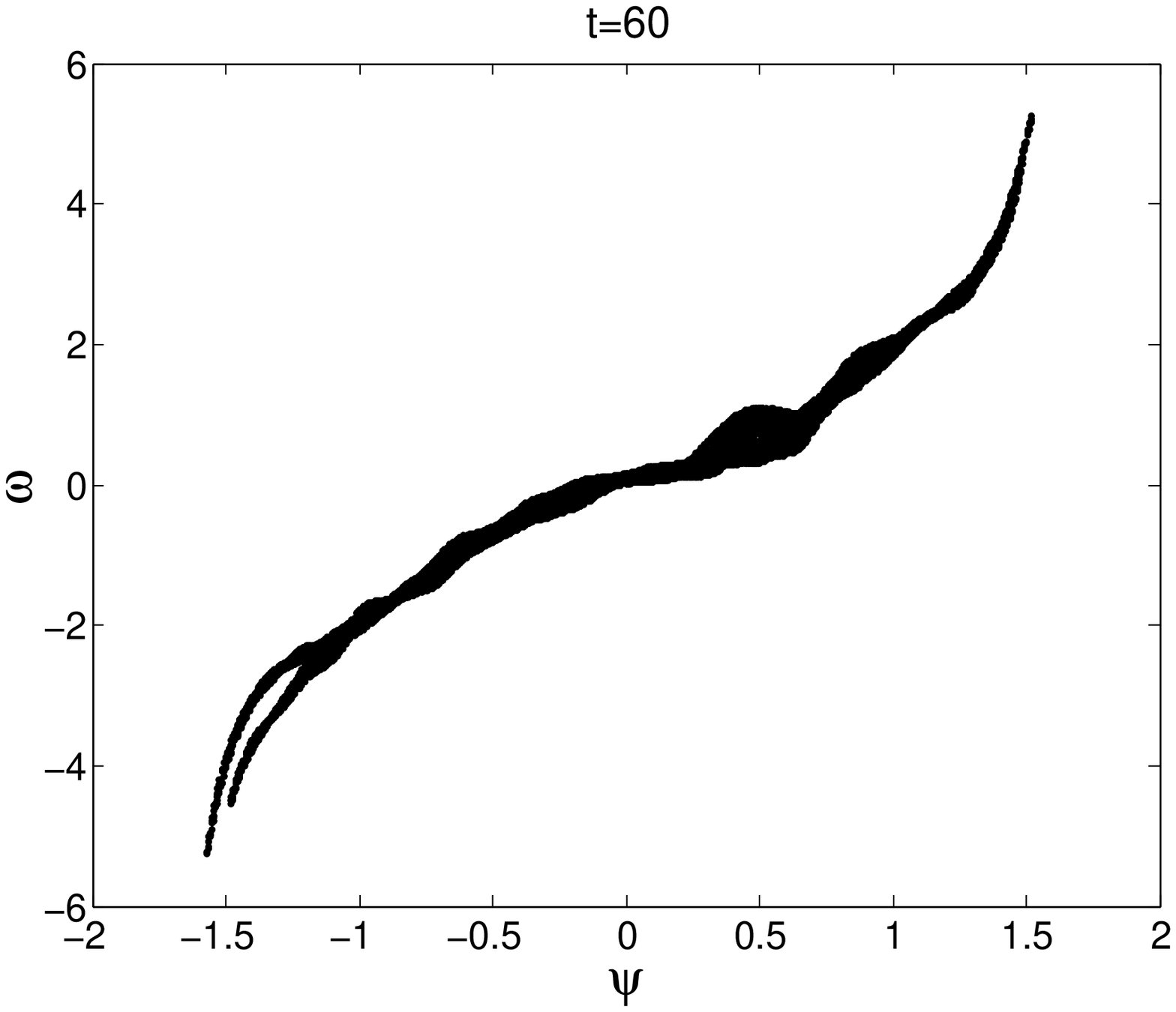}}
\end{minipage}
\begin{minipage}[c]{.63 \linewidth}
\scalebox{1}[0.7]{\includegraphics[width=\linewidth]{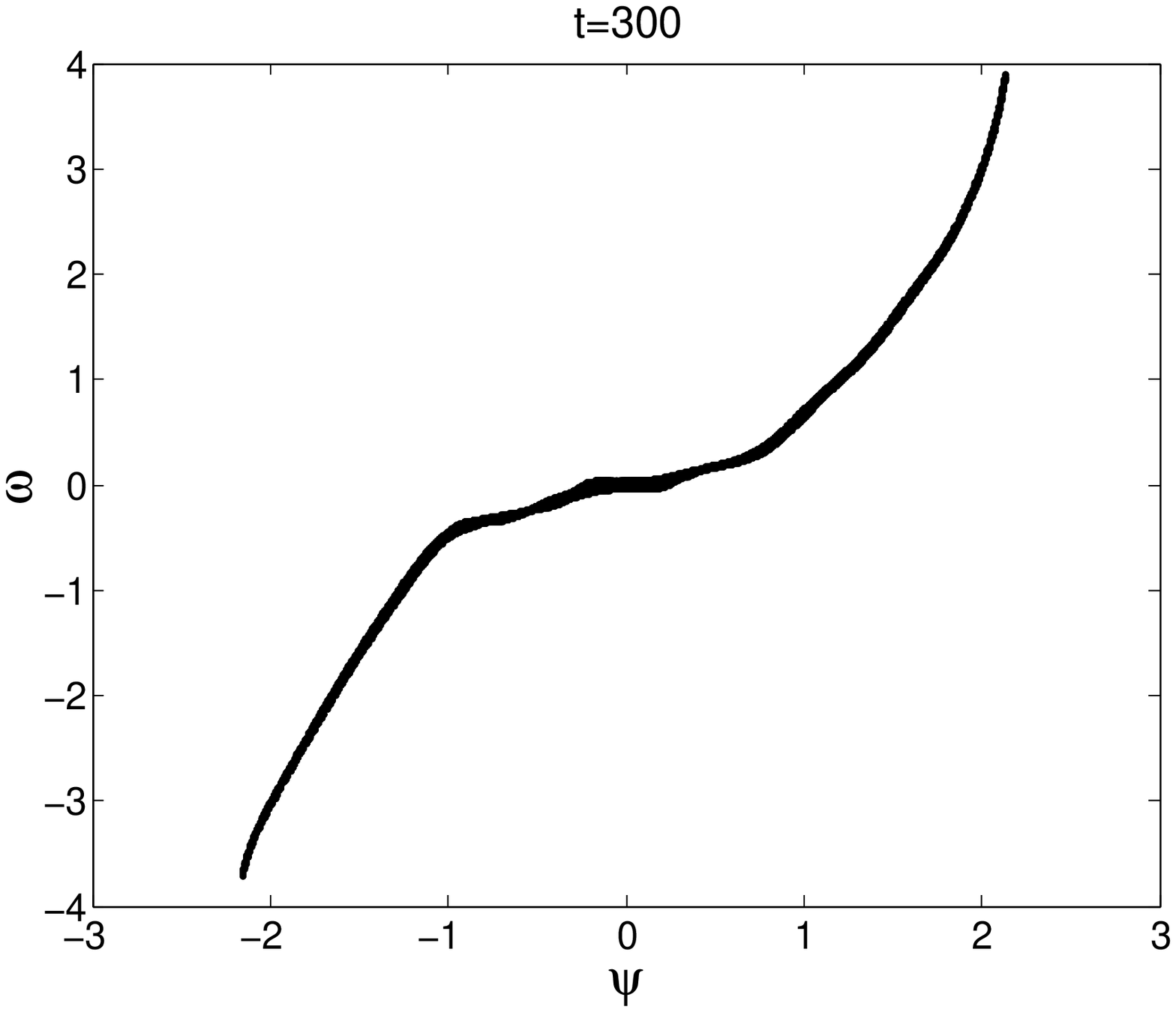}}
\end{minipage}
\caption{Scatter plot of the pointwise dependence of $\omega$ upon $\psi$ at two different times, for the evolution shown in Figs. \ref{fig:4_noise1} and \ref{fig:4_noise2} (both of them are close to Fig. \ref{fig:good:a}).}
\label{fig:4_noise3}
\end{figure}

\begin{figure}[!htbp]
\centering
\begin{minipage}[c]{.487 \linewidth}
\scalebox{1}[1]{\includegraphics[width=\linewidth]{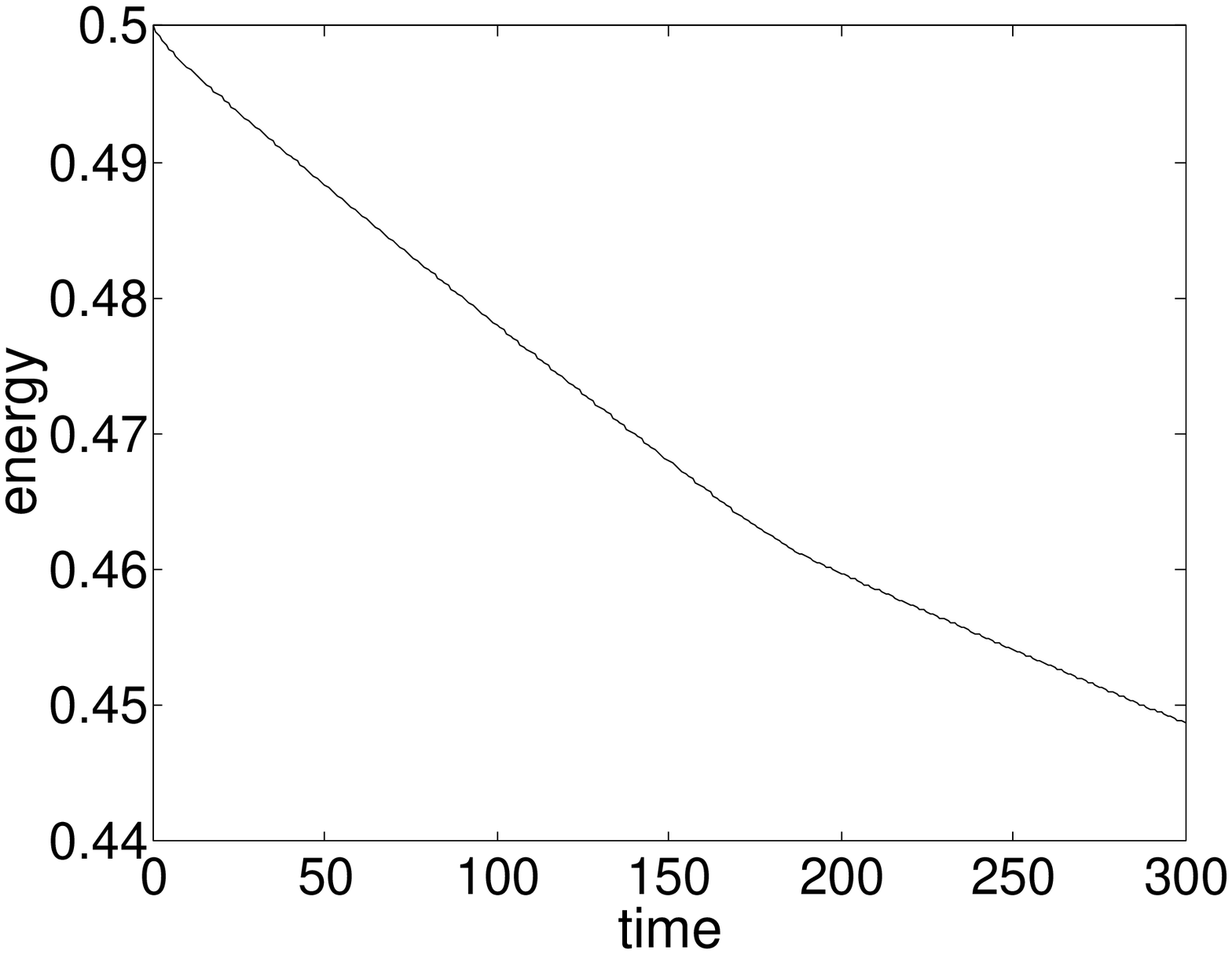}}
\end{minipage}
\begin{minipage}[c]{.487 \linewidth}
\scalebox{1}[1]{\includegraphics[width=\linewidth]{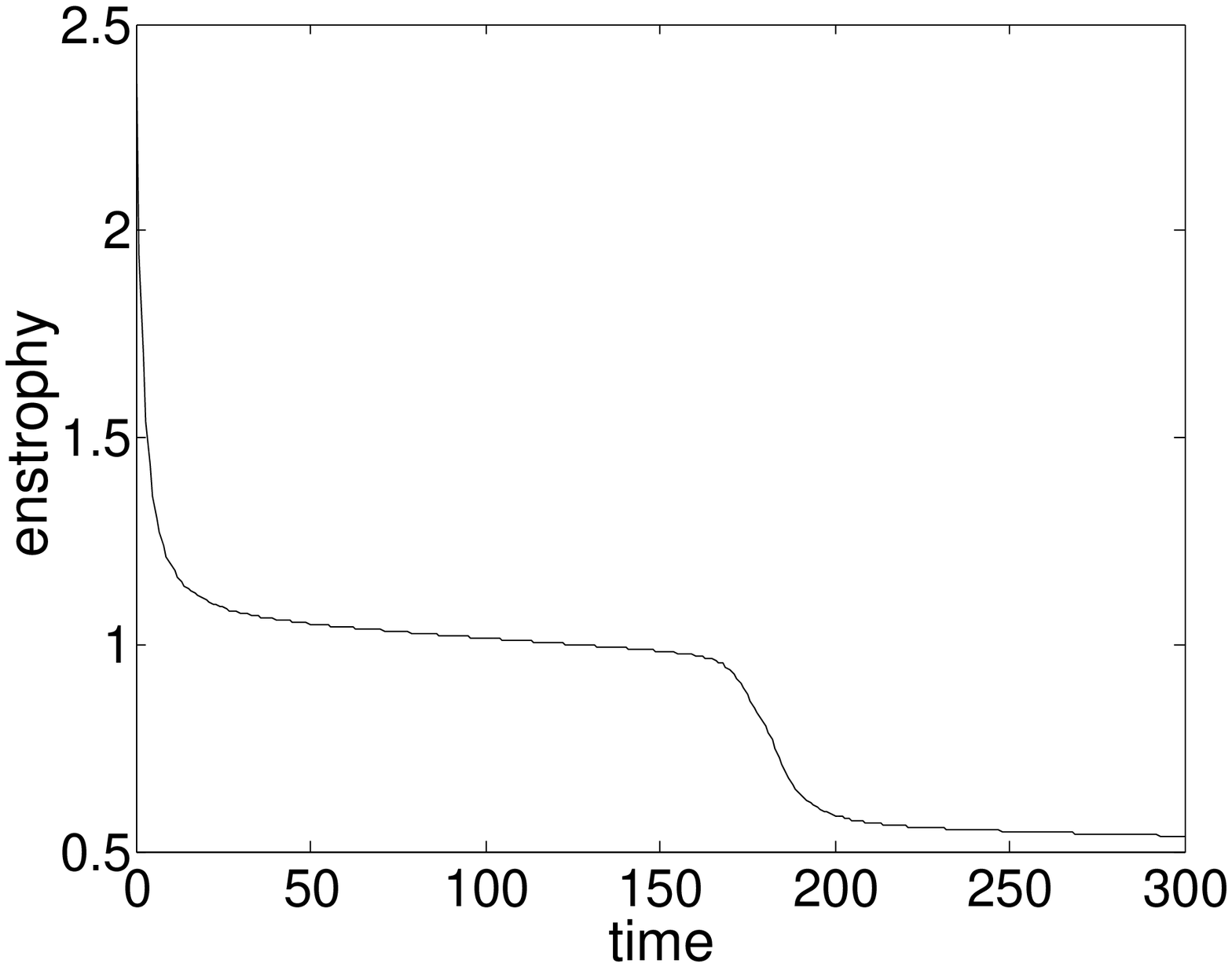}}
\end{minipage}
\begin{minipage}[c]{.487 \linewidth}
\scalebox{1}[1]{\includegraphics[width=\linewidth]{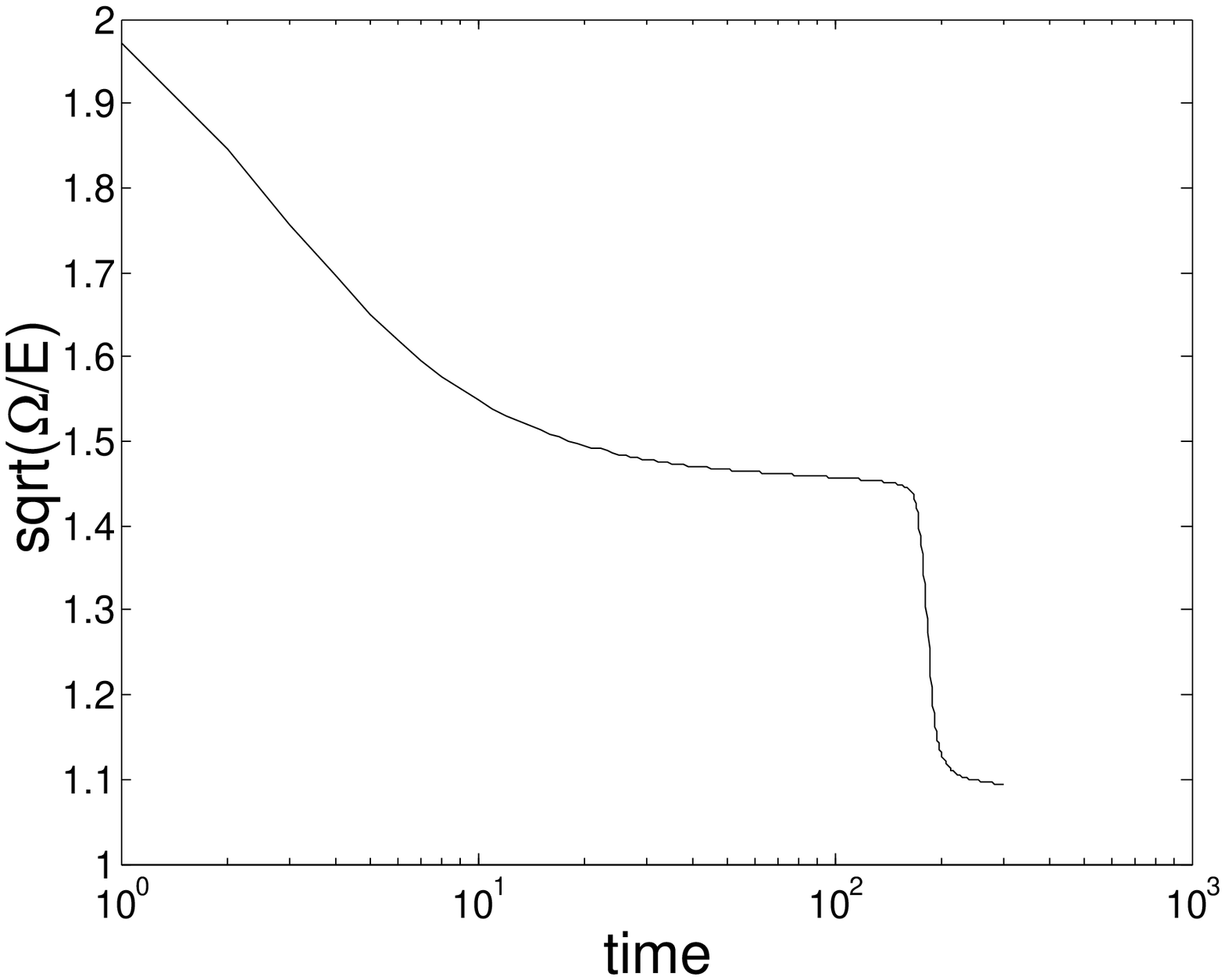}}
\end{minipage}
\begin{minipage}[c]{.487 \linewidth}
\scalebox{1}[1]{\includegraphics[width=\linewidth]{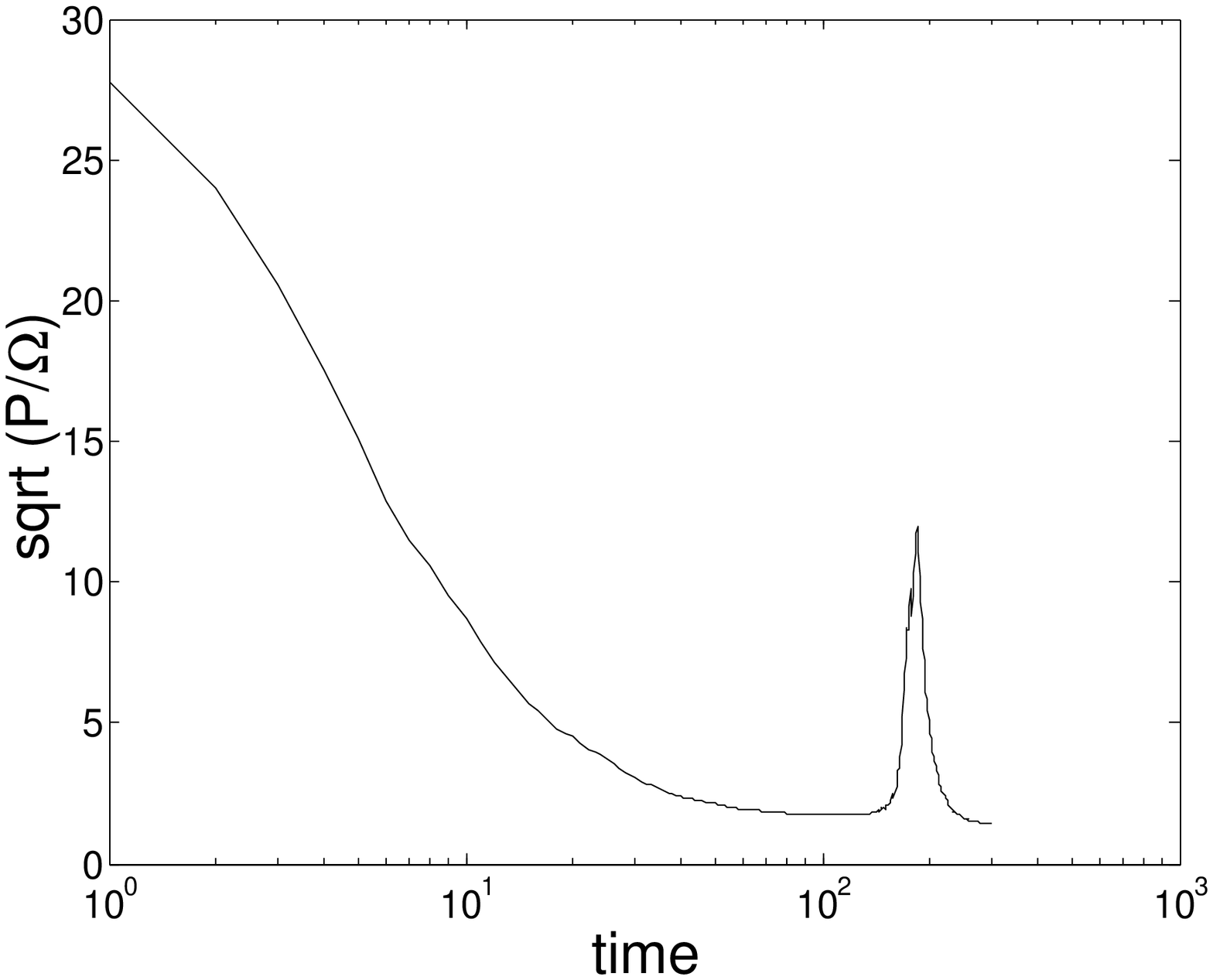}}
\end{minipage}
\caption{Evolution of the global quantities as functions of time for the run shown in Figs. \ref{fig:4_noise1} through \ref{fig:4_noise3}.}
\label{fig:4_noise4}
\end{figure}

\begin{figure*}[!htbp]
\centering
\begin{minipage}[c]{.245 \linewidth}
\scalebox{1}[1.15]{\includegraphics[width=\linewidth]{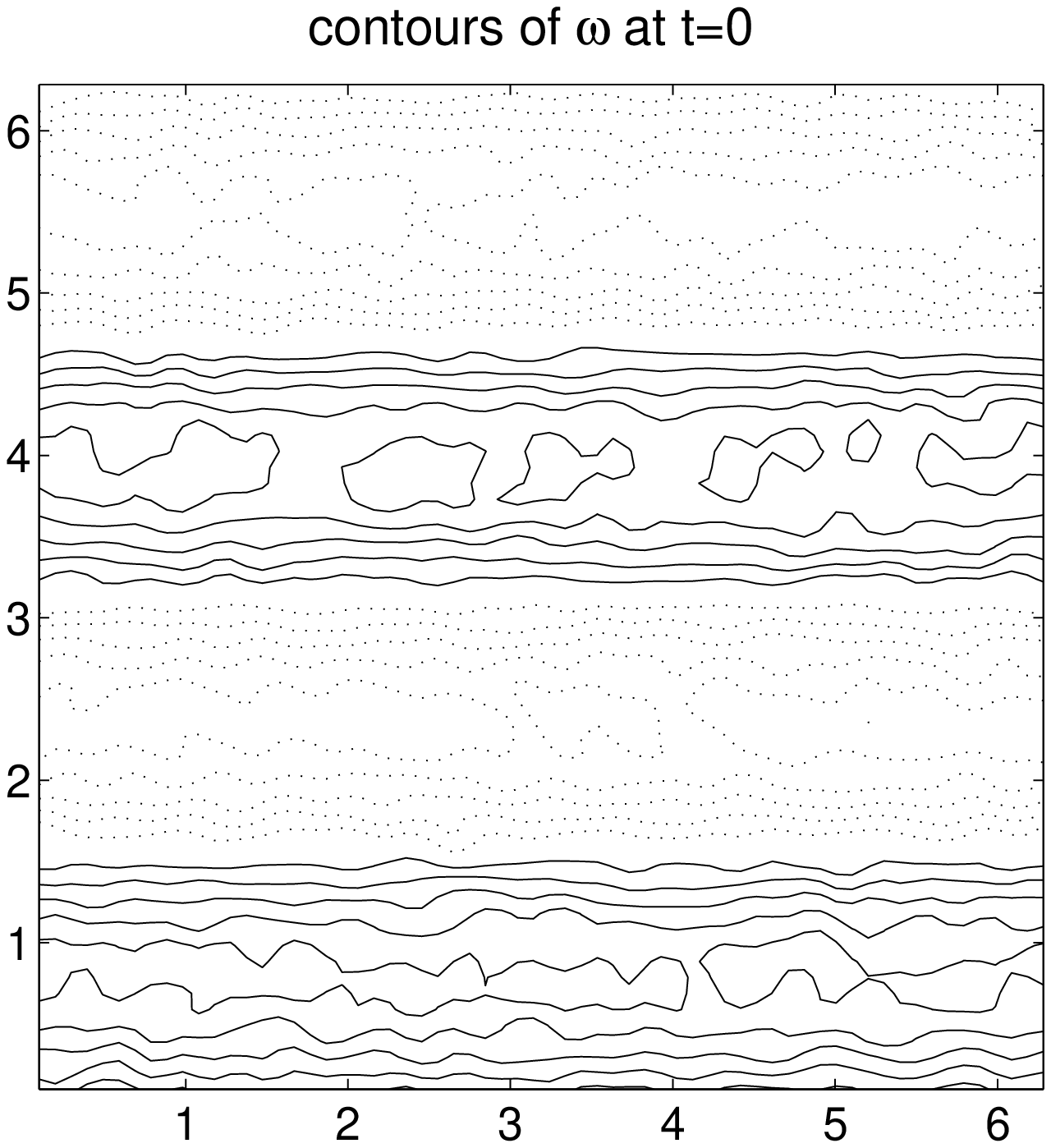}}
\end{minipage}
\begin{minipage}[c]{.245 \linewidth}
\scalebox{1}[1.15]{\includegraphics[width=\linewidth]{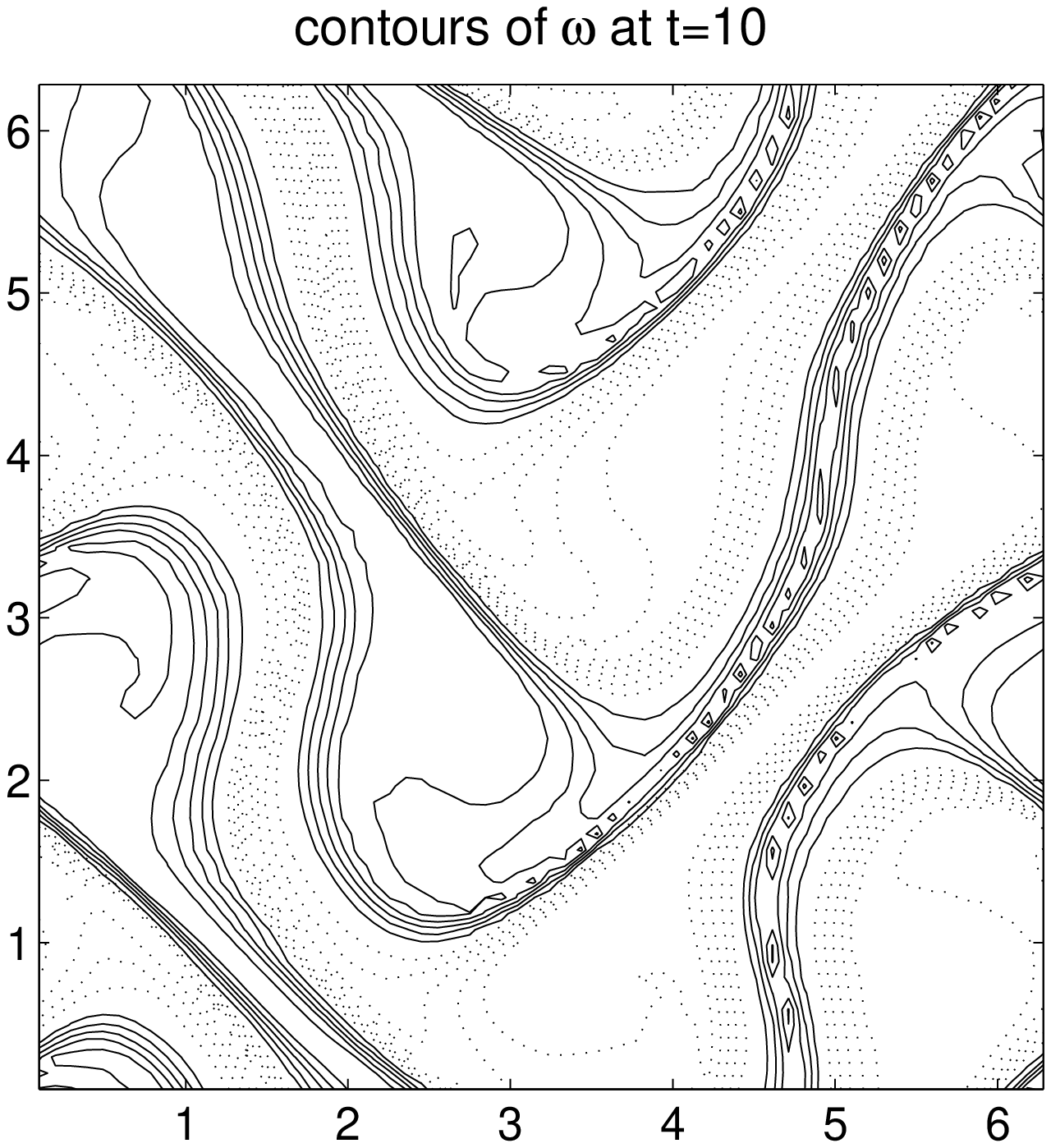}}
\end{minipage}
\begin{minipage}[c]{.245 \linewidth}
\scalebox{1}[1.15]{\includegraphics[width=\linewidth]{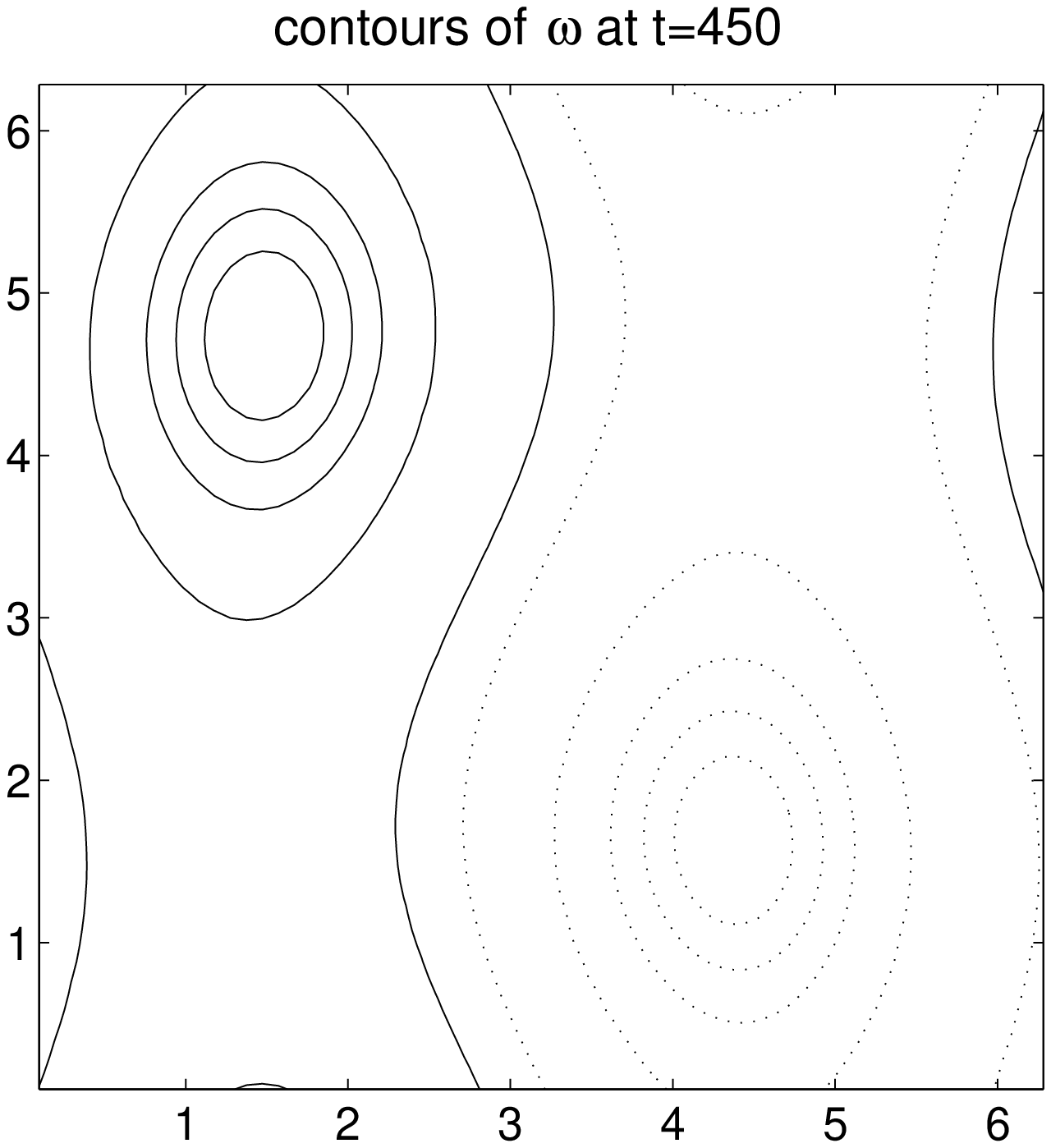}}
\end{minipage}
\begin{minipage}[c]{.245 \linewidth}
\scalebox{1}[1.4]{\includegraphics[width=\linewidth]{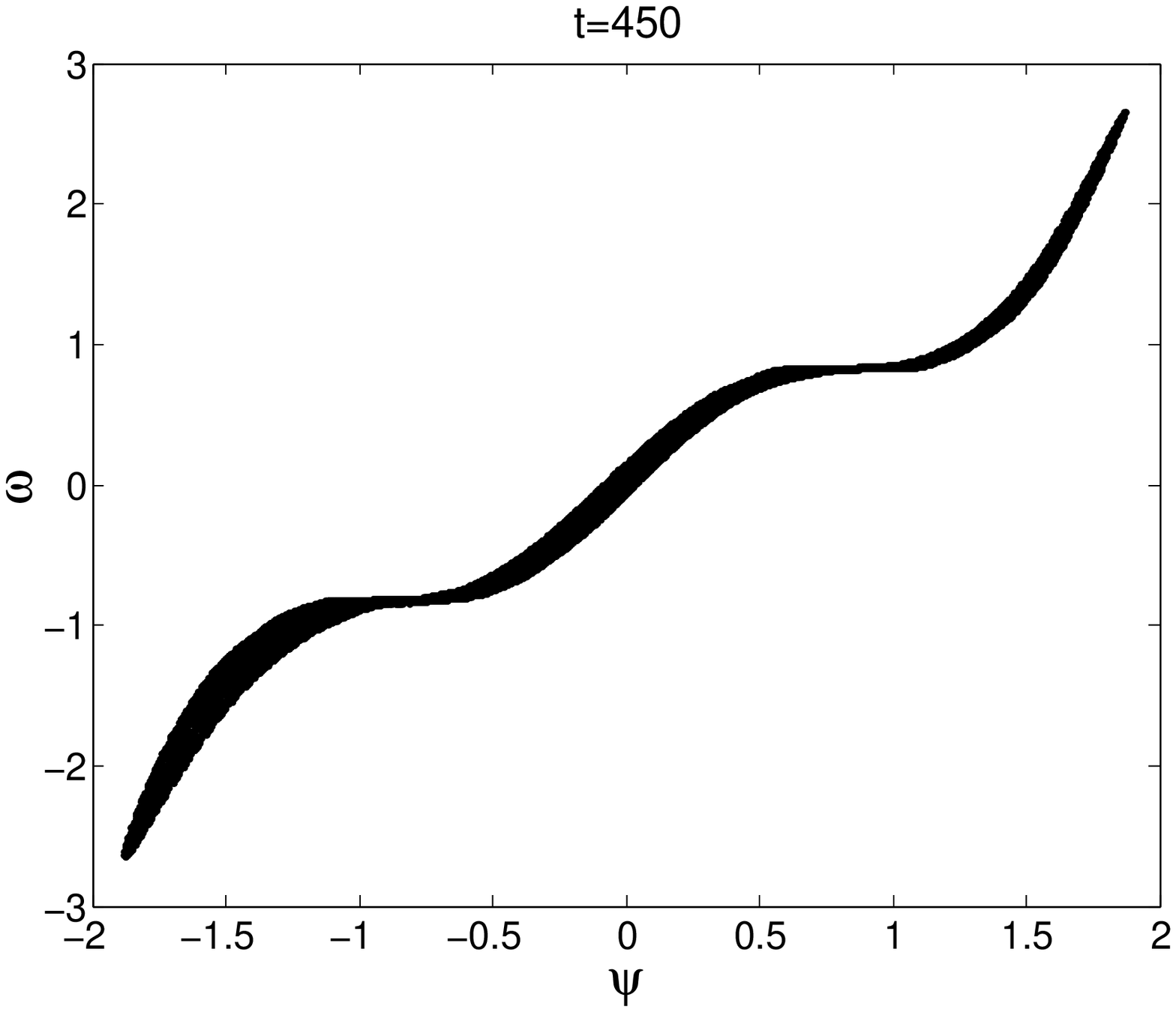}}
\end{minipage}
\caption{Evolution of a 4-bar initial condition, plus random noise, that ceases to evolve (perhaps because of loss of nonlinearity due to decay) before either a dipole or a bar final state is achieved.}
\label{fig:together}
\end{figure*}

	When we consider the modal energy spectra in Fig. \ref{fig:barmhhi}, for this evolution, we see that both the initial and final states are dominated by four and two Fourier modes, respectively. In fact, the evolution after t=40 has this character. The energy spectrum at t=15 is somewhat more broad-band than either, but whether it should properly be called ``turbulent'' may be debated; it never achieves the fully broad-band character of the evolution shown in Fig. \ref{fig:noise1}, for example. The final state achieved is consistent with the maximum-entropy prediction of the ``patch'' formulation for suitably large-sized patches (Fig. \ref{fig:bbrag:3entropy4}), but not with the same formulation for smaller-sized ones (Fig. \ref{fig:bbrag:3entropys2}). The right column of Figs. \ref{fig:barmhhi} shows a histogram of the vorticity at three different times in this evolution; it is in effect a frequency distribution for the appearance of various values of the vorticity over the plane, one per computational cell in the $512^2$ array. It will be seen that this distribution changes enormously over the course of the run, which it cannot do, of course, in any ideal Euler picture.  Figs. \ref{fig:barvcd} are binned plots of the angle-averaged modal energy spectrum ($E(k)$) and enstrophy ($\Omega(k)$) spectrum at the three different times.  Fig. \ref{fig:barta} shows the time history of several global quantities for this run: energy, enstrophy, mean wave number (square root of the ratio of enstrophy to energy), and square root of the ratio of palinstrophy ($P \equiv \Sigma_\mathbf{k} (1/2) k^2 \left| \omega(\mathbf{k}) \right|^2$) to enstrophy, showing spikes that are characteristic of vortex merger events at high but finite Reynolds numbers. 
	
	Figs. \ref{fig:banoise:b}  through \ref{fig:barta}, show, then, a reproducible evolution of an initial condition consisting of a patch quadrupole plus large amounts of random noise into a one-dimensional ``bar'' most-probable state.  This state is more probable than the dipole, according to a ``patch'' analysis if the patch size is big enough, though it is not for a smaller but finite patch size.
	
        A similiar initial condition is used by Segre and Kida \cite{kn:q33}. However, their computation made use of hyperviscosity and appears not to have run long enough for the proper late-time state to evolve.

\subsubsection{Local maximum states}

\paragraph{Evolution from a 64-pole initial condition.}

	The contrasting evolutions of the two initial states in the previous sub-section naturally arouse curiosity about whether there is a limit in which one behavior goes over into the other. In this sub-section, we attempt an answer to this by considering initial conditions which originate in high-order multipole solutions of the patch formulation, not maximum entropy states in either formulation, but intuitively closer to the random initial conditions that led to the dipole solution before, in the first run reported.

	There are four numerical solutions in this group, with $1/ \nu = 10,000$ for all four runs. They all originate in the 64-pole solution of the patch formulation, using the 3-level equation, Eq. (\ref{eqn:s3_level}).  The same conclusions that we reach can also be reached using 16-pole initial states, but we will not display those results here.

	The motivation was to see if the high-order multipole initializations would lead to ``bar'' final states, the way the quadrupole does. Intuitively, they would seem to be closer to our picture of what true turbulence might look like. The bar states are no longer found, but there are ``local maximum'' entropy states that can be achieved in the limit of low noise in the initial conditions. 

	First consider what happens to the 64-pole initial conditions without any noise, as shown in Fig.~\ref{fig:64_0}. A straightforward laminar evolution occurs, with the end product ($t=250$) being a one-dimensional bar state with a total of eight maxima and minima. This state is essentially dominated by one Fourier mode, as indicated by the essentially linear pointwise dependence of vorticity on stream function. In Figs.~\ref{fig:64_noise}, we show the evolution of the same initial conditions with only a low level of noise: $1/2000$ of the amplitude of the 64-pole solution, not visible on the $t=0$ contour plot. In this case, fully-developed turbulence does seem to result, with a dipole final state as the end result. 
	
	The third run in the group concerns the result of taking the final bar state (we raise the initial energy: $E(t = 0) = 0.5$), without putting any noise into it, from Figs. ~\ref{fig:64_0}, and allowing it to run.  This appears to be a time-independent state, stable in the presence of round-off error for the duration of the run, but not a maximum-entropy state and not stable in the presence of noise of greater amplitude. This is clear from Figs. ~\ref{fig:8_noise}, which show the development of the unstable evolution induced by putting on the same level of random noise as in Figs. \ref{fig:64_noise}, with a fully-developed turbulence and a dipole final state as the result. 

	There are thus some subtleties revealed by these four runs. Bar final states, predicted by the patch theory with sufficiently large patch size, do result from the evolution of a patch quadrupole plus sufficient noise. However, higher order multipoles from the patch formulation seem to be unstable at low levels of random noise, and evolve into dipoles, consistently. Still less easy to fit into the picture is the laminar evolution of the ``local maximum'' 64-pole state without noise into the bar state shown in Figs. ~\ref{fig:64_0}. Neither state is in this case an absolute maximum entropy one, though both are in some sense local maxima.

\paragraph{Evolution starting from a sinh-Poisson quadrupole.}

	We have started several runs from initial conditions that originate with a sinh-Poisson quadrupole state, with $1/ \nu = 10,000$. The first is the quadrupole state without any added random noise, and the others are the same state plus smaller or larger amounts of random noise. 
	
	The zero-random noise initial conditions seems to be stable and to remain in the same shape in the presence of only round-off error for the duration  of the run, as reported previously ~\cite{kn:sinh4}. Even after putting in enough random noise to manifestly break the symmetry (Figs. ~\ref{fig:4_noise1}), the quadrupolar shape is maintained for quite a long time before breaking down into a dipole by $t=300$. Figs. ~\ref{fig:4_noise2} are vorticity histograms computed at three different times, showing more concentration of $\omega$ near zero values, characteristic of sinh-Poisson states. The $\omega - \psi$ scatter plots shown in Figs. ~\ref{fig:4_noise3} also illustrate this feature. Evolution of the global quantities for this run is shown in Figs. ~\ref{fig:4_noise4}, showing abrupt changes in enstrophy and palinstrophy when vortex merger occurs, but otherwise not displaying characteristics of turbulent behavior. Note that energy is well conserved for this case. In this run, the quadrupole persists for a long time because the energy is concentrated in the four well-separated vortices, before breaking down into a dipole. This is noticeably different behavior than when we started with the patch quadrupole.

\subsubsection{Oddities}

\paragraph{} A run ($1/ \nu = 20,000$) that leads to an ``unclassifiable'' final state is shown in Figs. \ref{fig:together}. It begins with a four-fold bar state, plus a considerable amount of random noise, and ends at a state that shows features of both dipoles and bars. A tentative interpretation of this evolution is that the nonlinearity is simply exhausted by the viscous decay before either evolution can be completed. Once the amplitudes fall below amounts or in Fourier configurations at which the activity is effectively nonlinear, the pattern in place is ``frozen'' in its topology, and can only slowly decay. The scatter plot has features of both the point and patch predictions.

\begin{figure*}[!htbp]
\centering
\begin{minipage}[c]{.27 \linewidth}
\scalebox{1}[1.1]{\includegraphics[width=\linewidth]{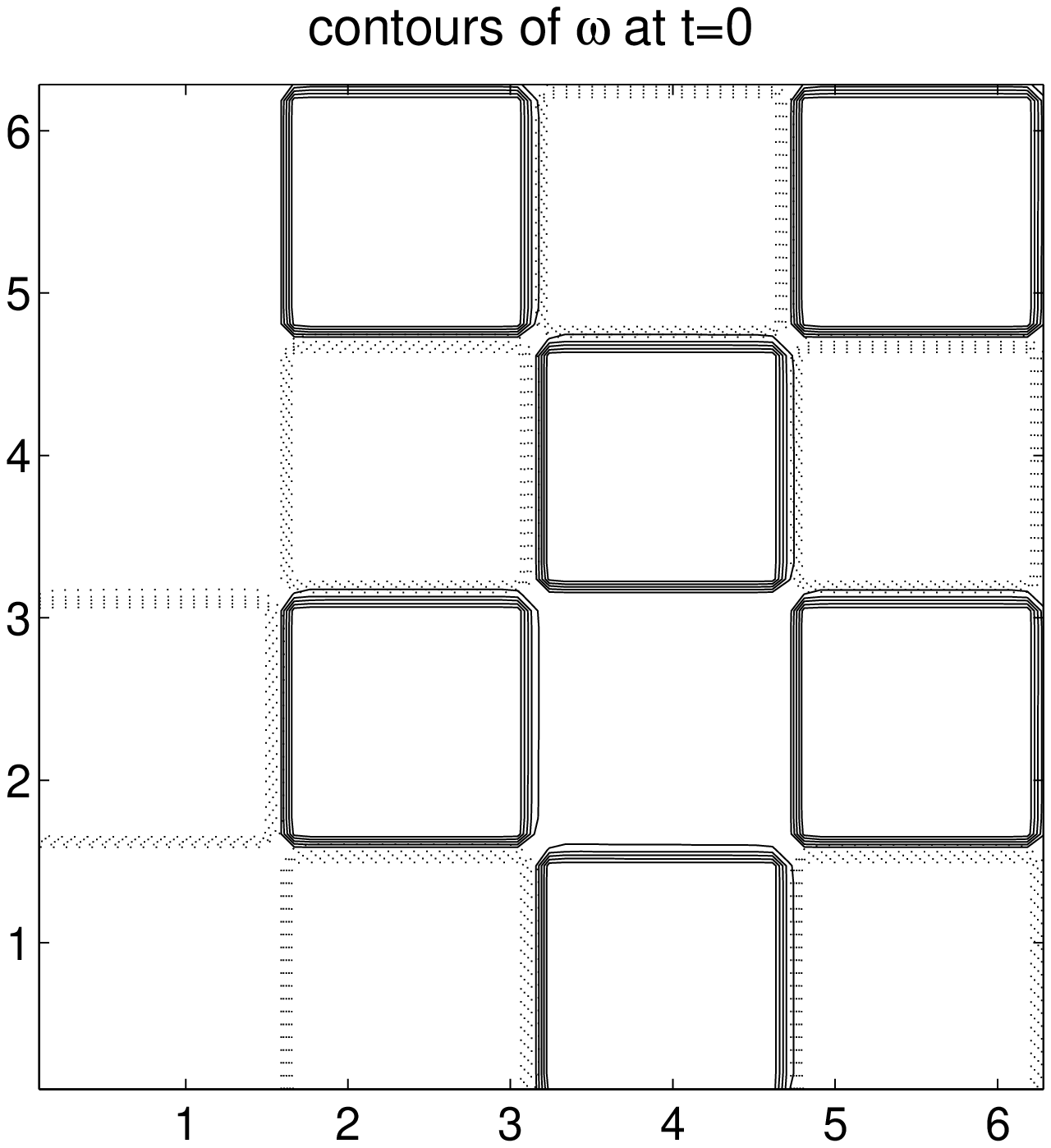}}
\end{minipage}
\begin{minipage}[c]{.27 \linewidth}
\scalebox{1}[1.1]{\includegraphics[width=\linewidth]{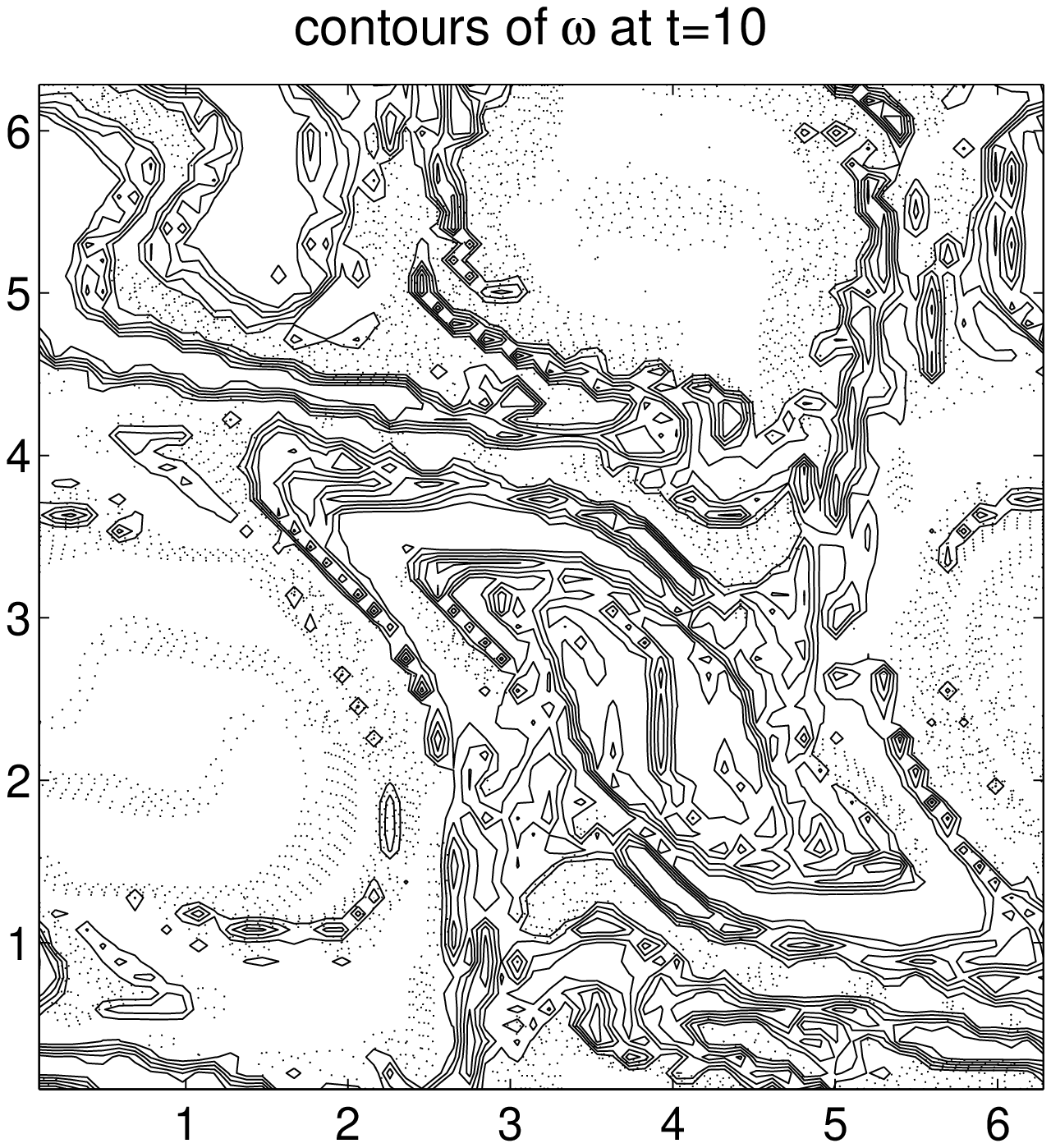}}
\end{minipage}
\begin{minipage}[c]{.27 \linewidth}
\scalebox{1}[1.1]{\includegraphics[width=\linewidth]{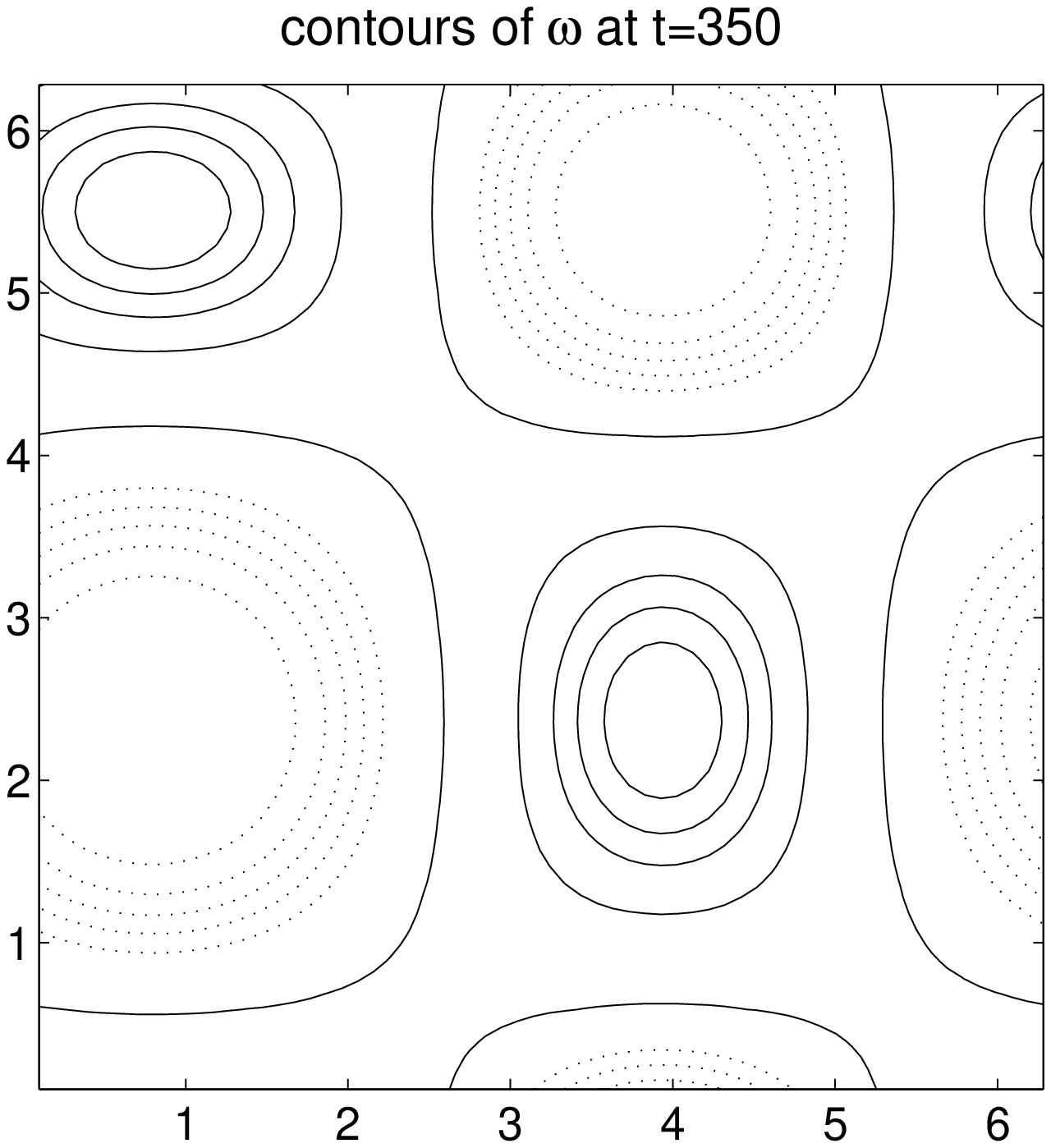}}
\end{minipage}
\caption{A second run, whose late-time state is outside the patch/point classifications, that seems to have reached some metastable late-time state. Here, a large area of zero vorticity was created in the initial conditions by removing, asymmetrically, four pieces of a 16-pole patch solution.}
\label{fig:argue1}
\end{figure*}

\begin{figure}[!htbp]
\centering
\begin{minipage}[c]{.9 \linewidth}
\includegraphics[width=\linewidth]{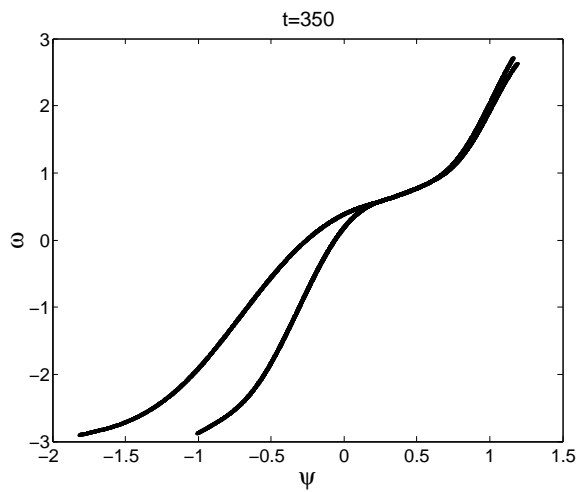}
\end{minipage}
\caption{The $\omega - \psi$ scatter plot of the late-time state achieved in Figs. \ref{fig:argue1}, suggesting two independent co-existing sinh-Poisson states which have developed asymmetrically. This is for the evolution shown in Figs. \ref{fig:argue1}.}
\label{fig:argue3}
\end{figure}

\begin{figure}[!htbp]
\centering
\begin{minipage}[c]{.48 \linewidth}
\scalebox{1}[1.21]{\includegraphics[width=\linewidth]{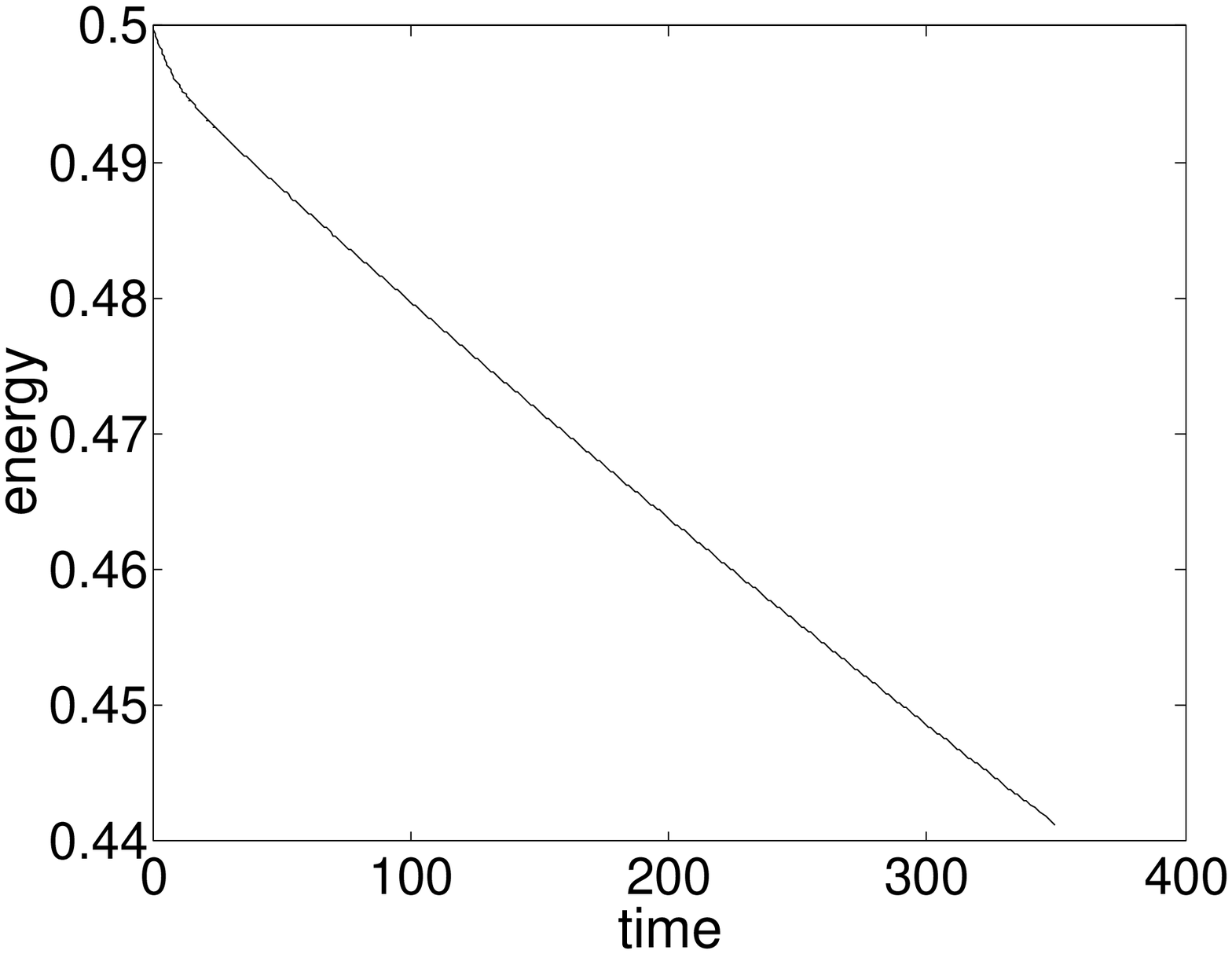}}
\end{minipage}
\begin{minipage}[c]{.48 \linewidth}
\scalebox{1}[1.21]{\includegraphics[width=\linewidth]{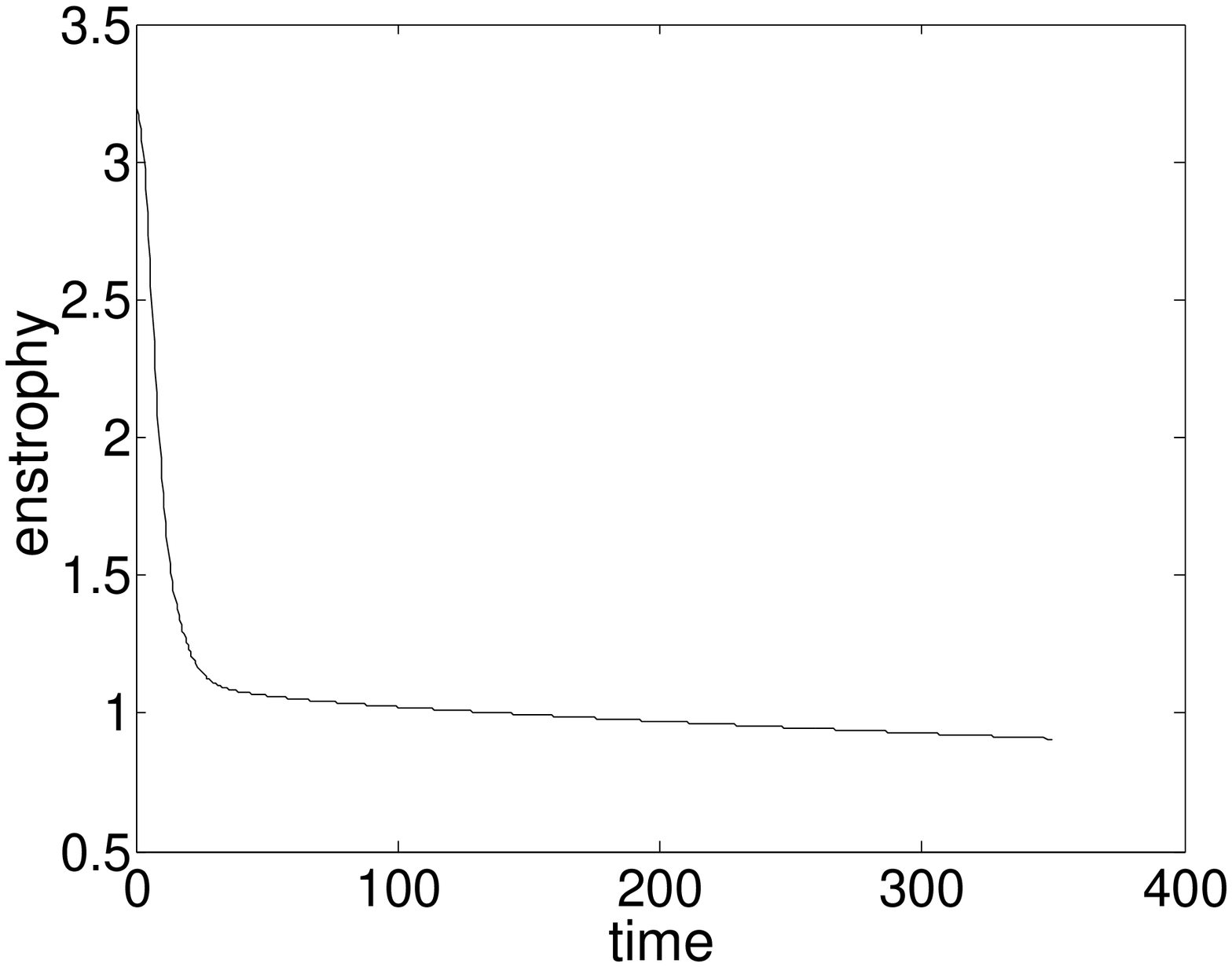}}
\end{minipage}
\begin{minipage}[c]{.48 \linewidth}
\scalebox{1}[1.21]{\includegraphics[width=\linewidth]{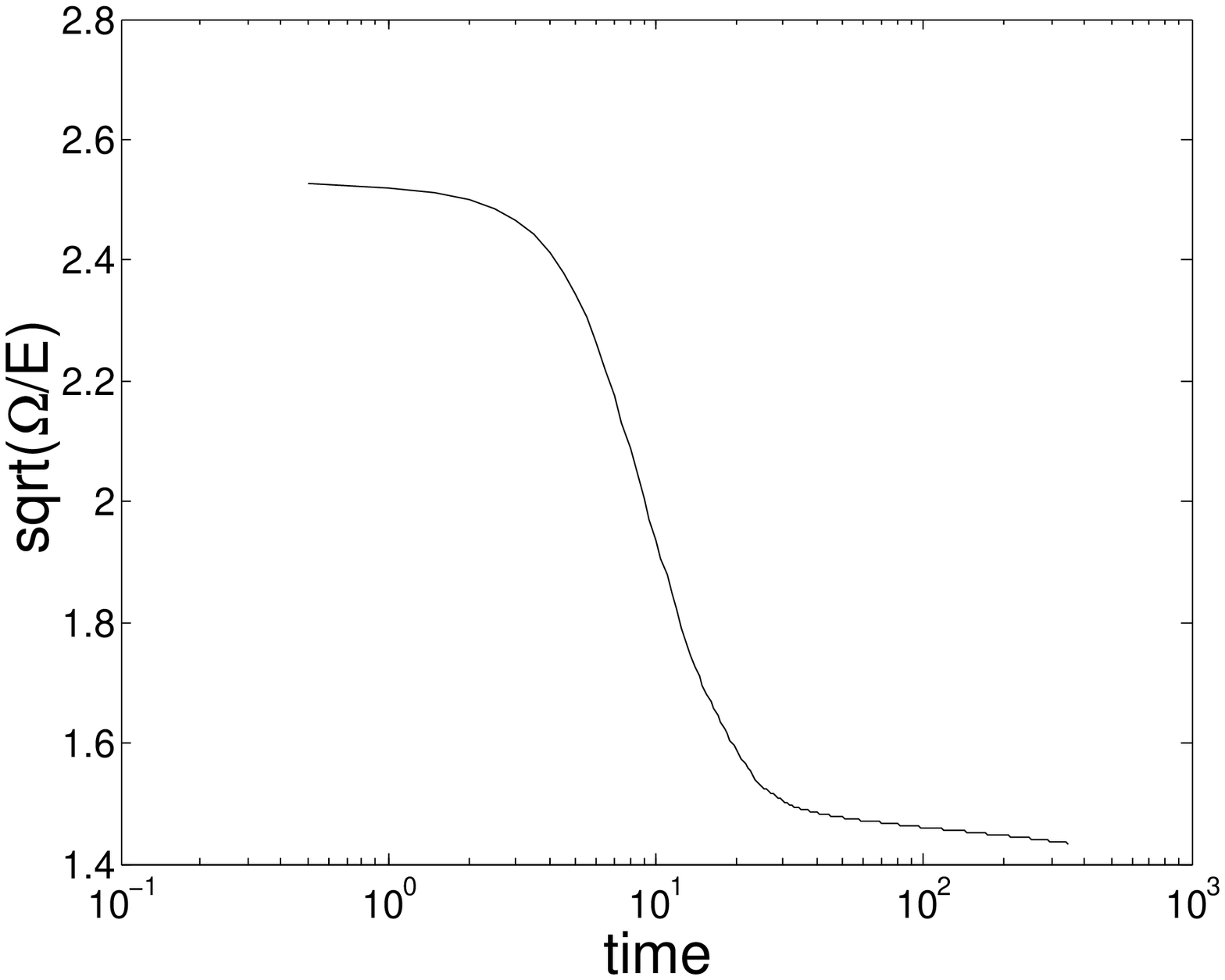}}
\end{minipage}
\begin{minipage}[c]{.48 \linewidth}
\scalebox{1}[1.21]{\includegraphics[width=\linewidth]{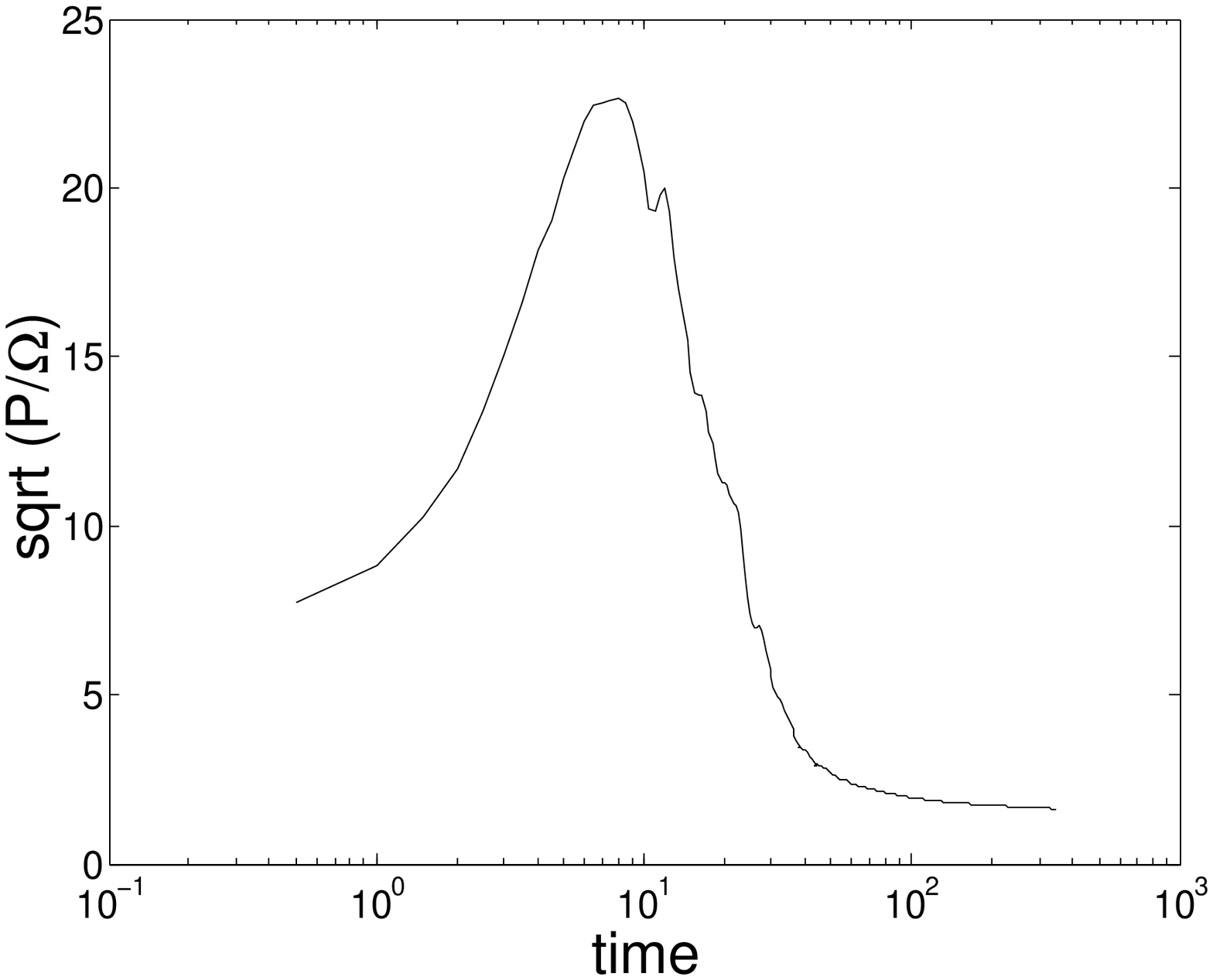}}
\end{minipage}
\caption{Time evolution of the global quanties for the run shown in Figs. \ref{fig:argue1} and \ref{fig:argue3}. Note that the mean wave number never reaches above 2.5.}
\label{fig:argue4}
\end{figure}

\paragraph{} Several other runs demonstrated odd features. Metastable states were found that would persist for a long time as a consequence of evolution that, though disordered, might be thought to be less than totally turbulent. An example appears in Figs. ~\ref{fig:argue1}. In this run ($1/ \nu = 12,500$), we began with a 16-pole solution of the 3-level patch equation, removed some of the squares of vorticity, and used the rest as an initial condition. Most of these went to the dipolar states, but one of the untypical ones reached a vaguely quadrupolar one, in the last panel of Figs. ~\ref{fig:argue1}. Fig. ~\ref{fig:argue3} is a $\omega - \psi$ scatter plot, showing a bilinear form, as if two locally ``most probable'' states had formed into a quasi-equilibrium that was slow to break up and decay toward anything globally ``most probable.'' Figs. ~\ref{fig:argue4}  show the evolution of the global quantities for this run. 
	
	Several runs ($1/ \nu = 10,000$) were carried out using this metastable quasi-quadrupole as the basis for initial conditions, plus a significant amount of added random noise. The evolution then was typically that the system, after having lingered awhile, evolved into the dipole configuration.

\section{CONCLUSIONS AND DISCUSSION}
\label{sec:s4}
	We have set out to test the relevance of the predictions of Eqs. (\ref{eqn:point1}) and (\ref{eqn:general1}) and their respective vorticity discretizations to the numerically-determined, long-time states of a 2D NS fluid with Reynolds numbers of several thousand subject to doubly-periodic boundary conditions. Eq. (\ref{eqn:point1}) has been derived by modeling the vorticity distribution as a mean-field limit of equal delta-function point vortices, while Eq. (\ref{eqn:general1}) models the vorticity distribution as made up of flat mutually exclusive patches of an area whose choice is arbitrary. Both equations have an infinite number of solutions, characterized by different topologies, but for fixed vorticity fluxes and total energy, all but one are only local maxima and there seems to be always one uniquely defined maximum-entropy prediction. As seen in Figs. \ref{fig:pentropy} through \ref{fig:bbrag}, it is sometimes a matter of painstaking analysis to determine which state this is, however. 
	We did computational runs of two basic types: runs with broad-band, totally disordered initial conditions of the type previously investigated, and runs originating from vorticity distributions that consisted of large areas of nearly flat vorticity patches plus random noise to break possible unwanted symmetries. In the former case, there were no surprises: the previously found dipolar late-time states inevitably resulted. This was not the case for the second kind of initial conditions, however. They did not generate broad-band turbulence in which energy was shared widely among many Fourier modes, but did exhibit considerable nonlinear activity in which the energy remained primarily in the lower parts of Fourier space. It is perhaps a semantic quibble as to whether the evolution should be called fully turbulent. 
	This second class of initial conditions, in any case, exhibited a more diverse range of possible behavior than the broad-band initializations did, and often came close to a late-time state that could be identified as one of the solutions of Eq. (\ref{eqn:general1}). Most interesting of those was the one-dimensional ``bar'' state exemplified by Figs. \ref{fig:bar1} and \ref{fig:64_0}; in the former case, it is attained with the help of added initial random noise, and in the latter, without it. As Figs. \ref{fig:pentropy} through \ref{fig:bbrag} illustrate, it may  or may not be considered to be the most-probable patch state, depending upon the choice of the patch size $\Delta /M$, which seems to be arbitrary. In any case, it is not the most probable point state predicted from Eq. (\ref{eqn:point1}).
	In the finite-size patch theory leading to Eq. (\ref{eqn:general1}), the arguments for it proceed from ``coarse graining'' the Euler equation behavior and presuppose a minimum observational scale below which fine spatial structure cannot be resolved. In a well-resolved NS computation (presumably including all of the ones reported here), there is no such scale, and observation of all scales participating is feasible. Thus that the patch formulation of Eq. (\ref{eqn:general1}), with a large enough value of $\Delta$, can predict a radically different topology than it does with a smaller $\Delta$, and then substantiate that prediction in a dissipative NS computation, is another of the accumulating puzzles which remain to be deciphered.
	In our opinion, there can be said to be an interesting 2D NS regime to be explored that has opened up in these investigations in which nonlinear evolution, perhaps not fully turbulent in the conventional sense, nevertheless leads from one state that can be identified to another laminar, late-time state that is quantitatively more probable than its initial conditions: an essentially thermodynamic behavior. The conditions for assigning this probability remain incompletely defined, and seem to us worthy of further numerical investigation.

\begin{acknowledgments}

    We are grateful to Prof. W.H. Matthaeus who supplied the original Fortran 77 pseudospectral dynamic code upon which our MPI Fortran 90 code is based. We also thank Dr. B.N. Kuvshinov and Dr. B.T. Kress for useful discussions. We also thank Prof. G.J.F. van Heijst, Dr. R.R. Trieling and Dr. X. Ke, whose help has been constant and invaluable. 
    
    This work is sponsored by the Stichting Nationale Computerfaciliteiten (National Computing Facilities Foundation, NCF) for the use of supercomputer facilities, with financial support from the Nederlandse Organisatie voor Wetenschappelijk Onderzoek (Netherlands Organization for Scientific Research, NWO).

\end{acknowledgments}
   
\appendix
\section{Statistical Mechanics In 2d Turbulence}

	In this Appendix, we attempt to review the most elementary calculations of the ``most probable'' vorticity distribution, inside a periodic square, that is compatible with a fixed value of the total kinetic energy and fixed (equal) fluxes of positive and negative vorticity. We do this for the case of Lynden-Bell statistics, which can then be specialized to Boltzmann statistics as a special case. It should be emphasized that in both pictures, vorticity is regarded as a pointwise conserved quantity, so that viscosity is not involved.

Consider the problem of rearranging a continuous periodic vorticity distribution $\omega(x,y;t=0)$ to constitute a ``more probable'' arrangement of the vorticity elements. This can only be approached through some kind of discretization. We consider coarse graining the vorticity distribution in the following sense.

Cut the periodic box into many cells of equal volume; $i$ will label the cells.The cells are of area $\Delta$. The coarse-grained vorticity distribution will be defined by the numbers
\begin{equation}
 \omega_i=\frac{1}{\Delta}\int \int_{cell \, i}\omega(x,y)\,dxdy.
 \label{eqn:numbers}
\end{equation}
We now represent the vorticity $\omega_i$ inside the $i$th cell in terms of smaller sub-units. These can clearly either have finite areas (``patches'') or zero areas (``points''). We should be able to take the ``patch'' formulation and achieve the ``point'' formulation by suitable limiting processes, so we start with the patches. The ultimate question will be which, if either, satisfactorily represents the long-time state of the Navier-stokes evolution of the $\omega(x,y;t=0)$ at high Reynolds number. It is an academic and not-very-interesting question as to which one better predicts the long-time Euler equation evolution of $\omega(x,y;t=0)$, since we can never compute that anyway.

Note that if $\omega(x,y;t=0)$ is an analytic function and we look at the set of points $x$, $y$ at which $\omega$ takes on any value $\omega_0$, say, that set is always a set of measure zero$-$$-$i.e., zero area. So arriving at any finite number of ``levels'' will always implicity assume that some kind of ``coarse-graining,'' as in Eq. (\ref{eqn:numbers}), has already been done.

We now cut each cell up into equal areas (they do not have to have any particular shape), each one the size of a ``patch.'' We say $M_i$ is the maximum number of patches allowed in the $i$th cell. Without apparent loss of generality, we can take all the patches as of equal area, but we will carry along the index $i$ on $M_i$. There is no obvious theoretical basis for choosing the size of a ``patch,'' $\Delta/M_i$, except that it is far smaller than $\Delta$, so $M_i \gg 1$.

We assume there are different ``strengths'' of the patches, though this will later come to seem perhaps unnecessary. Let the superscript $j$ represent the ``species,'' or strength of vortices of type $j$, which will be called $K^j$. $K^j$ is the integral of the (uniform) vorticity of the patch over its area. Let $j=0,1,2,3...q$, where $K^0=0$ (an ``empty'' site, with no patch in it), and $q$ different strengths represented,which can be positive or negative, but not zero. For convenience in notation, we will let $K^j=K_j$, hereafter, to avoid confusion with exponents.  

Giving a set of occupation numbers $N_i^j$ will give a discretized microscopic representation of all the $\omega_i$:
\[
\tilde\omega_i=\sum_j N_i^j K_j.
\]
 Obviously, there is no unique way to do this.

Note that
\begin{equation}
\sum_{j=1}^q N_i^j  \leq M_i \mbox{ and } \sum_{j=0}^q N_i^j = M_i.
\label{eqn:consrtains}
\end{equation}
Also $\sum_i N_i^j = N^j = \mbox{const.}$ for any rearrangement that does not create or destroy patches. ($\sum_i$ will always run over all the cells.)

For the ``probability'' of any total set of occupation numbers $N_i^j$, we adopt the expression
\begin{equation}
W=Const. \prod_i \prod_{j=1}^q \frac{N^j!}{(N_i^j!)(M_i-\sum\limits_{l=1}^q N_i^l)!}.
\label{eqn:total}
\end{equation}
Notice that the second of the constraints (\ref{eqn:consrtains}) has been built in and the $N_i^j$ may be varied independently for $j=1,2,3,4...q,$ so long as we demand that in the variation $ N_i^j  \rightarrow N_i^j + \delta N_i^j $
\[
 \sum_i N_i^j =Const. =N^j, j=1,2,3,...q,
\]
or that
\begin{equation}
 \sum_i \delta N_i^j = 0.
 \label{eqn:another}
\end{equation}

The last term in the denominator of (\ref{eqn:total}) comes from the empty sites or ``holes'' in which no finite-vorticity patch resides.

Assuming that all the $N_i^j \gg 1 $ and $M_i \gg 1$, we have for the entropy, using Stirling's approximation,
\begin{eqnarray}
S &=& \ln W \simeq Const. - \sum_i \sum_{j=1}^q (N_i^j \ln N_i^j -N_i^j) -     \nonumber \\ 
  & & \sum_i \{(M_i - \sum_{l=1}^q N_i^l) \ln (M_i - \sum_{l=1}^q N_i^l)       \nonumber \\
  & &  -(M_i - \sum_{l=1}^q N_i^l)\} + \mbox{(negligible terms).}
\end{eqnarray}

The discretely represented energy is 
\begin{equation}
E  \equiv  \frac{1}{2} \sum_i \sum_k \sum_{j,l} K_j N_i^j \psi_{ik} K_l N_k^l.
\label{eqn:c101}
\end{equation}
The $j = 0$ and $l = 0$ terms will give nothing, since $K_0 = 0$. Note that $\psi_{ik} = \psi_{ki}$ is just the interaction energy of two unit-strength vortices in cells $i$ and $k$, and that $\psi_{ik}$ is species-independent.

We may also write 
\begin{equation}
  E = \sum_{i,j} \frac{1}{2} K_j N_i^j \psi_i,
  \label{eqn:c102}
\end{equation}
where 
\begin{equation}
\psi_i \equiv \sum_{l,k} K_l N_k^l \psi_{ik},
\label{eqn:c103}
\end{equation}
$\psi_i$ is the discretized stream function. 

When $N_i^j \rightarrow N_i^j + \delta N_i^j$,
\begin{equation}
\delta E = \sum_i \sum_j K_j \psi_i \delta N_i^j.
\label{eqn:c104}
\end{equation}

For $\delta S$, we have
\begin{equation}
\delta S = - \sum_i \sum_{j=1}^q \delta N_i^j \ln N_i^j 
+ \sum_i \sum_{j=1}^q \ln(M_i - \sum_{l=1}^q N_i^l) \delta N_i^j.
\label{eqn:c105}
\end{equation}

To maximize $S$ under the constraints(\ref{eqn:another}) and $\delta E = 0$, using lagrange
multipliers, we set 
\[
\delta S + \tilde{\beta} \delta E + \sum_i \sum_{j=1}^q \tilde{\alpha}_j \delta  N_i^j =0,
\]
where the Lagrange multipliers are $\tilde{\beta}$ and $\tilde{\alpha}_j$, and the 
$\delta  N_i^j$ are now regarded as independent, so their coefficients may be set equal to $0$
for all $i$ and $j=1,2,3,...q$. We get
\begin{equation}
- \ln N_i^j + \tilde{\alpha}_j + \tilde{\beta} K_i \psi_i + \ln(M_i - \sum_{l=1}^q N_i^l) = 0,
\label{eqn:c106}
\end{equation} 
we let $\tilde{\beta}= - \beta$, $\tilde{\alpha}_j = \alpha_j$ to conform with earlier
notation (they are still undetermined, anyway).

Exponentiating,
\begin{equation}
\frac{N_i^j}{M_i -\sum\limits_{l=1}^q N_i^l}
= e^{\alpha_j - \beta K_j \psi_i},  \, j=1,2,3 \dots q.
\label{eqn:Exponentiating}
\end{equation}
But $ M_i - \sum\limits_{l=1}^q N_i^l =N_i^0 =$ number of empty sites in cell $i$, always.

So 
\begin{equation}
N_i^j=N_i^0 e^{\alpha_j - \beta K_j \psi_i},
\label{eqn:c107}
\end{equation}
and
\begin{equation}
\sum_{j=0}^q N_i^j = M_i =N_i^0 \sum_{j=0}^q e^{\alpha_j - \beta K_j \psi_i},
\label{eqn:c108}
\end{equation}
and so
\begin{equation}
 N_i^j = M_i \frac
{e^{\alpha_j - \beta K_j \psi_i}}{\sum\limits_{l=0}^q e^{\alpha_l - \beta K_l \psi_i}}, \, j=1,2, \dots q.
\label{eqn:expression}
\end{equation}
The same expression will work with $j=0$ if we  take $\alpha_0 = 0$. So 
\begin{equation}
N_i^0 = M_i \frac{1}{\sum\limits_{l=0}^q e^{\alpha_l - \beta K_l \psi_i}},
\label{eqn:expression2}
\end{equation}
and Eq. (\ref{eqn:expression}) can be extended to the case $j=0$, also.
 
The equation (\ref{eqn:expression})-(\ref{eqn:expression2}) give the most probable set of
occupation numbers $N_i^j$ for fixed $E$ and $N^j$. If we take the $N_i^j$ from 
Eqs. (\ref{eqn:expression}) and (\ref{eqn:expression2}), we have
\begin{equation}
N^j = \sum_i N_i^j, \,  j=1,2,3 \dots q,
\label{eqn:c109}
\end{equation}
and
\begin{equation}
E = \frac{1}{2} \sum_{i,k} \sum_{j,l} K_j N_i^j \psi_{ik} K_l N_k^l,
\label{eqn:c110}
\end{equation}
as $q+1$ equations which determine the $q+1$ unknowns $\alpha_j$ and $\beta$, $j = 1,2,3 \dots
q$, as functions of $E$ and the $N^j$.

If we want associate a ``partial vorticity'' with each species, we may write
\begin{equation}
\omega_i^j = \frac{M_i}{\Delta} K_j \frac{e^{{\alpha_l - \beta K_l \psi_i}}}{Z_i}, \, \omega_i = \sum_j \omega_i^j,
\label{eqn:c111}
\end{equation}
where
\begin{equation}
Z_i \equiv \sum_{l=0}^ q e^{\alpha_l - \beta K_l \psi_i},  \,  (\alpha_0 = 0).
\label{eqn:c112}
\end{equation}

Now the physical vorticity associated with the most probable state is 
\begin{equation}
\omega(x_i,y_i) = \sum_j \int \int_{cell i} \omega_i^j dxdy,
\label{eqn:c113}
\end{equation}
where ($x_i,y_i$) lie inside the $i$th cell. In this formulation, the constraints are
\begin{eqnarray}
\frac{1}{2} \sum_i \omega_i \psi_i =E,
\\
\sum_i \frac{\omega_i^j}{K^j} = N^j.
\end{eqnarray}

We need a tractable expression for $S$, in order to compare different entropies for different
sets of $N_i^j$  for the same $E$ and $N^j$. We may rewrite Eq. (\ref{eqn:total}) as
\begin{equation}
W = Const. \prod_i \prod_{j=0}^q \frac{N^j!}{N_i^j!},
\label{eqn:c114}
\end{equation} 
so that
\begin{equation}
 S = Const. - \sum_i \sum_{j=0}^q (N_i^j \ln N_i^j -N_i^j) + \mbox{(negligible terms)},
\label{eqn:c115}
\end{equation}
or
\begin{eqnarray}
 S &=& Const. - \sum_i \sum_{j=1}^q N_i^j ( \alpha_j - \beta K_j \psi_i +\ln N_i^0)
                                                       \nonumber \\
   & &  - \sum_i N_i^0 \ln N_i^0 + \sum_i N_i^0,                  \\
 S &=& Const. - \sum_j \alpha_j N^j + 2 \beta E        \nonumber \\
   & & - \sum_i (\ln N_i^0 )(\sum_{j=1}^q N_i^j + N_i^0) + \sum_i N_i^0.
\end{eqnarray}
Notes that  $\sum_{j=1}^q N_i^j + N_i^0 = M_i$, and $\sum_i N_i^0 =$total unoccupied sites, and
both of these are just fixed constants. So up to unimportant additive constants, the entropy of
the stationary states is
\begin{equation}
  S_{eq.} = - \sum_j \alpha_j N^j + 2 \beta E - \sum_i M_i \ln N_i^0. 
  \label{eqn:stationarystates}
\end{equation}
   
The first two terms of Eq. (\ref{eqn:stationarystates}) are familiar from the ``point'' theory. The
last one is new, and is a consequence of the finite area of the patches. If the ``patch'' area
shrinks to zero, the last term becomes an infinite additive constant,apparently, and the
``point''  entropy is recovered.

If we assume all the $M_i$ are equal (there seems to be no obvious reason for not doing so), the last term simplifies:
\[
 - \sum_i M_i \ln N_i^0 = - \sum_i (M \ln M) + M \sum_i \ln Z_i.
\]
The first term is constant, and the second term has the integral representation
\begin{equation}
      M \sum_i \ln Z_i =\frac{M}{\Delta} \int\int_{whole \, box} dxdy 
                     \ln \sum_{j=0}^q e^{\alpha_j - \beta K_j \psi(x,y)}.
\label{eqn:c116}
\end{equation}
Note that $M / \Delta$ is still entirely up to us to choose.

There is still far more freedom than we need to represent any initial $\omega_i$. We can still
do it, for example, by using only three value of $K$: namely, $K_j = -1, \, 0, \, \mbox{or}, \, +1$. Any
coarse-grained ``level'' can be achieved, that is, with only one kind of positive patch, one
kind of negative one, and a zero or ``empty'' site. Then we have 
\begin{equation}
S = - \alpha_+ N_+ - \alpha_- N_- + 2 \beta E + M \sum_i \ln Z_i,
\label{eqn:c117}
\end{equation}
or in the continuous representation,
\begin{eqnarray}
S &=& \frac{M}{\Delta} {\int\!\!\int}_{(box)} dxdy \ln [ 1 + e^{\alpha_+ - \beta \psi} + 
                                                  e^{\alpha_- + \beta \psi} ]  \nonumber\\
  & & - \alpha_+ N_+ - \alpha_- N_- + 2 \beta E .
\end{eqnarray}

If we now demand that $\psi$ obey Possion's equation, passing to the limit of arbitrarily many
patches of arbitrarily small area,
\begin{equation}
\nabla^2 \psi = - \omega = - \frac{M}{\Delta} \left[ \frac
{e^{\alpha_+ - \beta \psi} -e^{\alpha_- + \beta \psi}}
{1 + e^{\alpha_+ - \beta \psi} + e^{\alpha_- + \beta \psi}} \right].
\label{eqn:c118}
\end{equation}

This would be the analogue of the ``sinh-Poisson'' or ``tanh-Poisson'' equation discussed
 later. 
 
It is simple, computationally, to consider initial states for which there is, for every
$\omega_i$, another level with the value $- \omega_i$. This will in fact guarantee 
$\int \int_{(box)} \omega dxdy =0$. It also makes it plausible to assume $+$, $-$ symmetry, so that
$\alpha_+ = \alpha_-$. Then we would have 
\begin{equation}
\nabla^2 \psi = - \omega = \frac{M}{\Delta} \left[ \frac
  {2 \sinh(\beta \psi)}
  {e^{- \alpha} + 2 \cosh(\beta \psi) } \right].
\label{eqn:threeequation}  
\end{equation}
Suppose we want to let the patch area $\rightarrow 0$, now. This would mean
$M/ \Delta \rightarrow \infty$ . The only way the right hand side could 
stay finite and have a finite flux, $\int \int |\omega| dxdy < \infty$, would be for 
$e^{- \alpha} \rightarrow \infty$, and do it in such a way that $(M/ \Delta) / e^{- \alpha}$
stayed finite:
\[
\frac{M}{\Delta e^{-\alpha}} \rightarrow \frac{\lambda^2}{2} = Const..
\] 
So
\begin{equation}
 \nabla^2 \psi = \lambda^2 \sinh(\beta \psi),
 \label{eqn:point}
\end{equation}
and the sinh-Poisson equation is recovered.

It appears that for the symmetric periodic case, we can also get by with no unoccupied sites,
that is $N_i^0 \equiv 0$, $\forall i$. Every $\omega_i$ could be made up of just the right
number of $K = +1$ and $K = -1$ patches to give all the desired $\omega_i$, including zero. To
treat this case, we can re-start the whole procedure, with only one species independent($K=+1$,
say), so that $N_i^- = M_i - N_i^+$, always.

Then
\begin{eqnarray}
S &\cong& \ln W \cong -\sum_i(N_i^+ \ln N_i^+ - N_i^+)    \nonumber \\
  &     &   - \sum_i[(M_i - N_i^+) \ln (M_i - N_i^+)- (M_i - N_i^+)], \nonumber 
\end{eqnarray} 
and the analogue of Eq. (\ref{eqn:Exponentiating}) is
\[
N_i^+ = e^{\alpha - \beta \psi_i}(M_i - N_i^+),
\]
or
\[
N_i^+ = \frac{M_i e^{\alpha - \beta \psi_i }}{1 + e^{\alpha - \beta \psi_i}}=
\frac{M_i e^{\frac{1}{2}(\alpha - \beta \psi_i) }}
{e^{\frac{1}{2}(\alpha - \beta \psi_i)}+ e^{-\frac{1}{2}(\alpha - \beta \psi_i) }},
\]
with
\[
N_i^- = \frac{M_i e^{-\alpha + \beta \psi_i }}{1 + e^{-\alpha + \beta \psi_i}}=
\frac{M_i e^{-\frac{1}{2}(\alpha - \beta \psi_i) }}
{e^{\frac{1}{2}(\alpha - \beta \psi_i)}+ e^{-\frac{1}{2}(\alpha - \beta \psi_i) }}.
\]

This gives an overall vorticity of 
\begin{equation}
\omega_i = - \frac{M_i}{\Delta} 
\frac{\sinh(\frac{1}{2}(\alpha - \beta \psi_i))}{\cosh(\frac{1}{2}(\alpha - \beta \psi_i))}.
\label{eqn:c119}
\end{equation}

In this case, $E$ is invariant to letting $\psi_i \rightarrow \psi_i +const.$, so we can always
choose the constant so that $- \beta \times Const. + \alpha = 0$, leaving
\[
\omega_i = - \frac{M_i}{\Delta} \tanh \frac{1}{2} \beta \psi_i.
\]
Since $\beta$ is as yet undetermined, we might as well let $\beta \rightarrow 2 \beta$,
$M_i/ \Delta = \lambda^2$, and we have 
\begin{equation}
\nabla^2 \psi = - \omega = \lambda^2 \tanh(\beta \psi).
\label{eqn:patch}
\end{equation}

Eq. (\ref{eqn:point}) corresponds to the limiting case in which the ``patches'' have no area, and
Eq. (\ref{eqn:patch}) to the limiting case in which they have all the area, and there is no empty
space. Both apply to the symmetric case in which for every intial $\omega_i$, there is somewhere
another with $-\omega_i$ as its value.

The most general, completely unrestricted, case has, as its analogue of sinh-Poisson,
\begin{equation}
\nabla^2 \psi = - \omega = - \sum_{j=1}^q \frac{M}{\Delta} K_j 
\frac{e^{\alpha_j - \beta \psi K_j}}{\sum\limits_{l=0}^q e^{\alpha_l - \beta \psi K_l}}.
\label{eqn:generalequation}
\end{equation}

Comparing Eqs. (\ref{eqn:generalequation}), (\ref{eqn:patch}), (\ref{eqn:point}), and
(\ref{eqn:threeequation}), it seems clear that every choice of the $M/ \Delta$ and the possible
$K_j$ will lend to a different equation and a different $\psi$. 
There seem to be no compelling arguments as to why one might be superior to the other in 
predicting long-time Navier-Stokes behavoir, and
no point in speculating which will do a better job of predicting long-term Euler equation
evolution, since we can never know that. It is perhaps worth remarking that if we want to
shrink $\Delta$ so much that there is no approximation involved in letting the $\omega_i$ of
Eq. (\ref{eqn:numbers}) represent an analytic function $\omega(x,y,0)$, then the ``points'' are
the only possibility. Every ``level'' approximation of the type Eq. (\ref{eqn:numbers}), with $i$
bounded from above is already a function significantly different from $\omega(x,y,0)$. For
example, it has an infinite palinstrophy, and thus an infinite intial enstrophy decay rate,
because of the sharp discontinuities between the cells.

A possible stategy is to search for intial conditions where the presence of finite-area
conserved ``patches'' would seem to have the most important consequences: 
Eq. (\ref{eqn:threeequation}) with positive $\alpha$. For example, in 
Eq. (\ref{eqn:threeequation}), which presumably can be solved numerically, choose an
``$\alpha$'' and a ``$\beta$'' which will give a relatively flat scatter plot (like Fig. \ref{fig:good:b}).

Pick an $M/ \Delta$ that is large enough that $ \psi $ seems slowly varying on the 
scale $\sim \sqrt{\Delta}$, but $M \gg 1$. This will give a function $\omega(x,y)$ which 
is presumably well-represented by patches, if anything can be. Do not pick the maximum entropy
state but something like a quadrupole or higher multipole which can be expected to break up and
generate turbulence when used as an initial condition for Navier-Stokes code at small $\nu$.

There may have to be a small random perturbation on those initial conditions to get it to go
turbulent as soon as possible. Notice that choosing $\alpha$, $\beta$ and $M/ \Delta$ in advance
provides computable expressions for energy and vorticity fluxes. The first question to ask is, what
``final'' state does the system relax to for large times -- what does its scatter plot look
like?
   
The second question is, then, using those expressions for energy and vorticity flux in the
sinh-Poisson code, what does the final maximum-entropy state look like -- its scatter-plot in
particular? Is there any sense in which the initially flat areas of vorticity flux are still
there? The same program could be carried out starting from the more restrictive 
Eq. (\ref{eqn:patch}). 

A really convincing, but perhaps over-ambitious plan would be to take the $\alpha$, $\beta$ and
vorticity flux from the solution to Eq. (\ref{eqn:threeequation}), and use them as input for
generating a sinh-Poisson solution, comparing its scatter plot with the one resulting
dynamically.

\section{Solving The Poisson Equation}

The sinh-Poisson equation (Eq. (\ref{eqn:s_point})) has been solved by a lot of people numerically and analytically (see Ref.~\cite{kn:sinh1} and references therein). Probably the most direct way to do it is to put the equation into spectral space and do the iteration there:

\begin{equation} 
 \widehat{(\Psi_{n+1})}_{\bf k}= \frac{\lambda^2}{{\left| {\bf k} \right|}^2}
              \widehat{(\sinh \Psi_{n})}_{\bf k}. 
 \label{eqn:iteration}
\end{equation}

We show three solutions in Fig. \ref{fig:Dipole}. In addition to the solutions exhibited, there are an infinite number more, characterized by more and more maxima and minima. They are obtained by starting with a trial function on the right hand side of Eq. (\ref{eqn:iteration}) that has the desired number of final maxima and minima, as noted by several authors: {McDonald ~\cite{kn:sinh1}, Book \textit{et al.} ~\cite{kn:q6}, Lundgren and Pointin ~\cite{kn:q7}}. Typically, the solutions with many maxima and minima are lower entropy solutions, for fixed energy and vorticity fluxes, and so we do not discuss them in any detail.

The spectral method loses it advantage when we try to use it to solving Eq. (\ref{eqn:s3_level}), because we might have some step function solutions. Here we adopt the numerical method used by McDonald~\cite{kn:sinh2}. 

The first step is to get the nontrivial solution for Eq. (\ref{eqn:s3_level}) with $ \Psi = 0 $ on a square boundary of $[0,\pi] \times [0, \pi]$. Starting from a trial solution $W(x,y)$, we get the solution of $v(x,y)$ from 

\begin{equation}
  \nabla^2 v + v f^{\prime}(W) = R(W,W),
 \label{eqn:vget}
\end{equation}
subject to $v = 0$ on the boundary. Here

\begin{equation}
   (A,B) = \int_0^{\pi}\int_0^{\pi} A(x,y)B(x,y) dxdy,
 \label{eqn:vget1}
\end{equation}

\begin{equation}
   f(\Psi) = \frac{\lambda^2 \sinh \Psi}{g + \cosh \Psi},
 \label{eqn:vget2}
\end{equation}

\begin{equation}
   R(x,y) = \frac{\nabla^2 W + f(W)}{(W,W)}.
 \label{eqn:vget3}
\end{equation}
Then we can correct the trial solution:
\begin{equation}
W \rightarrow W + \frac{v(x,y)(W,W)}{2(v,W) - (W,W)},
 \label{eqn:vget4}
\end{equation}
until we get a sufficient accurate solution.

\begin{figure}[!htbp]
\centering
\begin{minipage}[c]{.98 \linewidth}
\scalebox{1}[1]{\includegraphics[width=\linewidth]{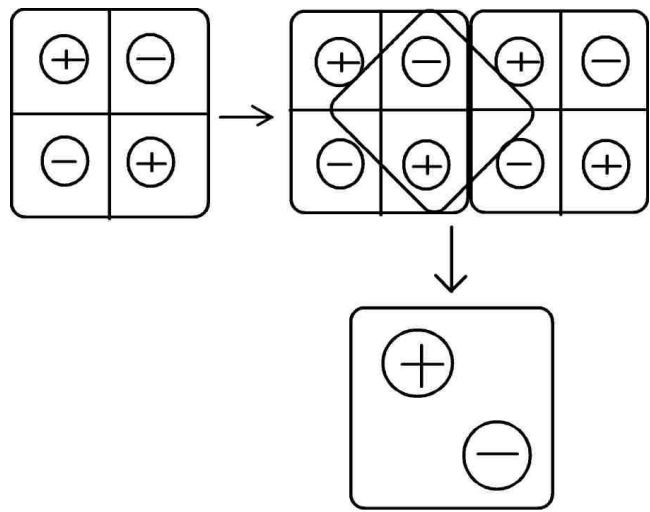}}
\end{minipage}
\caption{A periodic quadrupolar solution may be used to generate a dipolar one, depending upon the spatial sub-volume chosen, as shown. The value of $\lambda^2$ in Eq. (\ref{eqn:s3_level}) will changed correspondingly. The procedure is somewhat involved.}
\label{fig:ggg16how}
\end{figure}

Perhaps the most important change we made to McDonald's methods~\cite{kn:sinh2} is using the quadruple precision instead of double precision in the calculation (and this is not a luxury on mordern computers). The accuracy problem mentioned by McDonald in this method is overcome. As a result, if we define the root-mean-square error (RMS):
\[
\epsilon = {(\sum_{i,j} f^2/\sum_{i,j} g^2)}^{1/2}
\]
where
\[
f = \Delta \Psi(i,j) + \lambda^2 \sinh(\Psi(i,j))
\]
and
\[
g = \left| \Delta \Psi(i,j) \right| + \left| \lambda^2 \sinh(\Psi(i,j)) \right|.
\]
 Our solutions can at least reach an accuracy of $10^{-7}$.

    We may construct higher-order multipole solutions by putting monopole  solutions side by side with alternating signs, and achieve doubly-periodic boundary conditions in this way, in a ``checkerboard'' solution, so that dipole (Figs. \ref{fig:ggg16how}) and higher-multipole solutions can be constructed. There is some loss of accuracy in this procedure, and we have made sure that the accuracy of all solutions is of the order of $10^{-6}$ (RMS) in our calculations.


\begin{thebibliography}{0}
\expandafter\ifx\csname natexlab\endcsname\relax\def\natexlab#1{#1}\fi
\expandafter\ifx\csname bibnamefont\endcsname\relax
  \def\bibnamefont#1{#1}\fi
\expandafter\ifx\csname bibfnamefont\endcsname\relax
  \def\bibfnamefont#1{#1}\fi
\expandafter\ifx\csname citenamefont\endcsname\relax
  \def\citenamefont#1{#1}\fi
\expandafter\ifx\csname url\endcsname\relax
  \def\url#1{\texttt{#1}}\fi
\expandafter\ifx\csname urlprefix\endcsname\relax\def\urlprefix{URL }\fi
\providecommand{\bibinfo}[2]{#2}
\providecommand{\eprint}[2][]{\url{#2}}

\end{thebibliography}


\begin{thebibliography}{1}
\bibitem{kn:q1}
 W.H.Matthaeus, W.T. Stribling, D. Martinez, S. Oughton, and D. Montgomery, ``Selective decay and coherent vortices in two-dimensional incompressible turbulence,''  Phys. Rev. Lett. \textbf{66}, 2731 (1991).
\bibitem{kn:q2}
W.H. Matthaeus, W.T. Stribling, D. Martinez, S. Oughton, and D. Montgmery,
``Decaying two-dimensional turbulence at very long times,'' Physica D \textbf{51}, 531 (1991).
\bibitem{kn:q3}
 D. Montgomery, W.H. Matthaeus, W.T. Stribling, D. Martinez, and S. Oughton, ``Relaxation in two dimensions and the `sinh-Poisson' equation,'' Phys. Fluids A \textbf{2}, 4 (1992).
\bibitem{kn:q4}
 G.R Joyce and D. Montgomery, ``Negative temperature states for a two-dimensional guiding-centre plasma,'' J. Plasma Phys. \textbf{10}, 107 (1973).
\bibitem{kn:q5}
D. Montgomery and G.R. Joyce, ``Statistical mechanics of negative temperature states,'' Phys. Fluids \textbf{17}, 1139 (1973).
\bibitem{kn:sinh2}
B.E. McDonald, ``Numerical calculation of nonunique solutions of a two-dimensional sinh-Poisson equation,'' J. Comp. Phys. \textbf{16}, 630 (1974).
\bibitem{kn:q6}
D.L. Book, S. Fisher, and B.E. McDonald, ``Steady-state distributions of interacting discrete vortices,'' Phys. Rev. Lett. \textbf{34}, 4 (1975).
\bibitem{kn:q7}
Y.B. Pointin and T.S. Lundgren, ``Statistical mechanics of two-dimensional vortices in a bounded container,'' Phys. Fluids \textbf{19}, 459 (1976).
\bibitem{kn:q8}
J.H. Williamson, ``Statistical mechanics of a guiding-centre plasma,'' J. Plasma Phys. \textbf{17}, 85 (1977).
\bibitem{kn:q9}
A.C. Ting, H.H. Chen, and Y.C. Lee, ``Exact solutions of nonlinear boundary value problem: the vortices of the two-dimensional sinh-Poisson equation,'' Physica D \textbf{26}, 37 (1987).
\bibitem{kn:q10}
R.A. Smith, ``Phase transition behavior in a negative-temperature guiding-center plasma,'' Phys. Rev. Lett. \textbf{63}, 1479 (1989). 
\bibitem{kn:q11}
R.A. Smith and T. O'Neil, ``Nonaxisymmetric thermal equilibria of a cylindrically-bounded guiding-center plasma or discrete vortex system,'' Phys. Fluids B \textbf{2},  2961 (1990).
\bibitem{kn:q12}
R.A. Smith, ``Maximization of vortex entropy as an organizing principle of intermittent, decaying, two-dimensional turbulence,'' Phys. Rev. A \textbf{43}, 1126 (1991).
\bibitem{kn:q13}
L.J. Campbell and K. O'Neil, ``Statistics of 2-D point vortices and high energy vortex states,'' J. Stat. Phys. \textbf{65}, 495 (1991).
\bibitem{kn:q14}
R. H. Kraichnan and D. Montgomery, ``Two-dimensional turbulence,'' Rep. Prog. Phys. \textbf{43}, 547 (1980).
\bibitem{kn:q15}
M. K.-H. Kiessling, ``Statistical mechanics of classical particles with logarithmic interactions,'' Comm. Pure \& Appl. Math. \textbf{46}, 2108 (1993).
\bibitem{kn:q16}
G.L. Eyink and H. Spohn, ``Negative temperature states and large-scale, long-lived vortices in two-dimensional turbulence,'' J. Stat. Phys.  \textbf{70}, 833 (1993).
\bibitem{kn:q17}
A.J. Chorin, \textit{Vorticity and Turbulence}, (Springer-Verlag, New York, 1994).
\bibitem{kn:q18}
K.W. Chow, N.W.M. Mo, and S.K. Tang, ``Exact solutions of a non-linear boundary value problem: The vortices of the two dimensional sinh-Poisson equation,'' Phys. Fluids \textbf{10}, 1111 (1998).
\bibitem{kn:sinh1}
B.N. Kuvshinov and T.J. Schep, ``Double-periodic arrays of vortices,'' Phys. Fluids \textbf{12}, 3282 (2000).
\bibitem{kn:q19}
L. Onsager, ``Statistical Hydrodynamics,'' Nuovo Cimento Suppl. \textbf{6}, 279 (1949).
\bibitem{kn:q20}
C.C. Lin, \textit{On the motion of vortices in two dimensions}, (University of Toronto Press, Toronto, 1943).
\bibitem{kn:q21}
D. Montgomery, X. Shan, and W.H. Matthaeus, ``Navier-Stokes relaxation to sinh-Poisson states at finite Reynolds numbers,'' Phys. Fluids A \textbf{5}, 2207 (1993).
\bibitem{kn:q22}
R. Robert and J. Sommeria, ``Statistical equilibrium states for two-dimensional flow,'' J. Fluid Mech. \textbf{229}, 291 (1991).
\bibitem{kn:q23}
R. Robert and J. Sommeria, ``Relaxation towards a statistical equilibrium in two-dimensional perfect fluid dynamics,'' Phys. Rev. Lett. \textbf{69}, 2776 (1992).
\bibitem{kn:q24}
P.H. Chavanis and J. Sommeria, ``Classification of self-organized vortices in two-dimensional turbulence: the case of a bounded domain,'' J. Fluid Mech. \textbf{314}, 267 (1996).
\bibitem{kn:q25}
P.H. Chavanis and J. Sommeria, ``Classification of robust isolated vortices in two-dimensional hydrodynamics,'' J. Fluid Mech. \textbf{356}, 259 (1998).
\bibitem{kn:q26}
J. Miller, P.C. Weichman, and M.C. Cross, ``Statistical Mechanics, Euler's Equation, and Jupiter's Red Spot,'' Phys. Rev. A \textbf{45}, 2328 (1992), and references therein.
\bibitem{kn:pq26}
H. Brands, S.R. Maassen and H.J.H. Clercx, ``Statistical mechanical predictions and Navier-Stokes dynamics of two-dimensional flows on a bounded domain,'' Phys. Rev. E \textbf{60}, 2864 (1999).
\bibitem{kn:q27}
D. Lynden-Bell, ``Statistical mechanics of violent relaxation in stellar systems,'' M. Not. R. Astron. Soc. \textbf{136}, 101 (1967).
\bibitem{kn:q28}
S.A. Orszag, ``Accurate solution of the Orr-Sommerfeld stability equation,'' J. Fluid Mech. \textbf{50}, 689 (1971).
\bibitem{kn:q29}
G.S. Patterson and S.A. Orszag, ``Spectral calculations of isotropic turbulence: Efficient removal of aliasing interaction,''
Phys. Fluids \textbf{14}, 2538 (1971).
\bibitem{kn:q30}
W.H. Matthaeus and D. Montgomery, ``Selective decay hypothesis at high mechanical and magnetic Reynolds numbers,'' Ann. N.Y. Acad. Sci. \textbf{357}, 203 (1980).
\bibitem{kn:sinh3}
A.C. Ting,  W.H.  Matthaeus, and D. Montgomery, ``Turbulent relaxation processes in magnetohydrodynamics,'' Phys. Fluids \textbf{29}, 3261 (1986). See Sec. VI.
\bibitem{kn:q31}
R.H. Kraichnan, ``Inertial ranges of two-dimensional turbulence,'' Phys. Fluids \textbf{10}, 1417 (1967).
\bibitem{kn:q33}
E. Segre and S. Kida, ``Late states of incompressible 2D decaying vorticity field,'' Fluid Dyn. Res. \textbf{23}, 89 (1998).
\bibitem{kn:sinh4}
P. Dmitruk, D. Gomez, A. Costa and S.P. Dawson, ``Asymptotic states of decaying turbulence in two-dimensional imcompressible flows,'' Phys. Rev. E \textbf{54}, 2555 (1996).
\bibitem{kn:q32}
J.C. McWilliams, ``The emergence of isolated coherent vortices in turbulent flow,'' J. Fluid Mech. \textbf{146}, 21 (1984).

\end{thebibliography}
\end{document}